\documentclass[12pt]{article}
\usepackage{color}
\usepackage{latexsym}
\usepackage{epsfig,amssymb,euscript, mathrsfs}
\usepackage{amsmath}
\newsavebox{\ns}
\newsavebox{\dbrane}
\newsavebox{\dbshort}

\def\be{\begin{equation}}
\def\ee{\end{equation}}
\def\bea{\begin{eqnarray}}
\def\eea{\end{eqnarray}}

\newcommand{\nn}{\nonumber}

\newcommand\R{\mathbb{R}}
\newcommand\Z{\mathbb{Z}}

\newcommand\C{\mathbb{C}}

\newcommand\diff{\mathrm{d}}
\newcommand{\de}{\partial}

\newcommand{\dd}{\mathrm{d}}

\newcommand{\ii}{\mathrm{i}}

\newcommand{\ex}{\mathrm{e}}
\newcommand{\vol}{\mathrm{vol}}

\newcommand\g{\gamma}

\newcommand{\rcut}{\varrho}
\newcommand{\Reeb}{\xi}

\newlength{\sswidth}

\newcommand{\s}{s}

\numberwithin{equation}{section}       

\begin{document}

\begin{titlepage}

\begin{center}

\today

\vskip 2.3 cm 

\vskip 5mm

{\Large \bf The supersymmetric NUTs and bolts of holography} 

\vskip 15mm

{Dario Martelli$^1$, Achilleas Passias$^1$ and James Sparks$^2$\\}

\vskip 1cm

$^1$\textit{Department of Mathematics, King's College London, \\
The Strand, London WC2R 2LS,  United Kingdom\\}

\vskip 0.8cm

$^2$\textit{Mathematical Institute, University of Oxford,\\
24-29 St Giles', Oxford OX1 3LB, United Kingdom\\}

\end{center}

\vskip 1 cm

\begin{abstract}

\noindent  
We show that a given conformal boundary can have a rich and intricate space of 
supersymmetric supergravity solutions filling it, focusing on 
the case where this conformal boundary is a biaxially squashed Lens space.  
Generically we find that the biaxially squashed Lens space $S^3/\Z_p$
admits Taub-NUT-AdS fillings, with topology $\R^4/\Z_p$, as well as 
smooth Taub-Bolt-AdS fillings with non-trivial topology.
We show that the Taub-NUT-AdS solutions always lift to 
solutions of M-theory, and correspondingly that the gravitational free energy then
agrees with the  large $N$ limit of the dual field theory free energy, 
obtained from the localized partition function of a class of ${\cal N}=2$ Chern-Simons-matter theories. 
However, the solutions of Taub-Bolt-AdS type only lift to M-theory for appropriate 
classes of internal manifold, meaning that these solutions 
exist only for corresponding classes of three-dimensional  $\mathcal{N}=2$ field  theories.

\end{abstract}

\end{titlepage}

\pagestyle{plain}
\setcounter{page}{1}
\newcounter{bean}
\baselineskip18pt
\tableofcontents

\section{Introduction}

It has recently been appreciated that putting supersymmetric field theories on curved Euclidean manifolds allows one 
to perform exact non-perturbative computations, using the technique of supersymmetric localization
\cite{Pestun:2007rz,Kapustin:2009kz,Jafferis:2010un}.
This motivates the study of rigid supersymmetry on curved manifolds, see \emph{e.g.}
\cite{Festuccia:2011ws}  -- \cite{newcyril}.
Thus, when a field theory defined on (conformally) flat  space  admits a gravity dual,
it is natural to extend the holographic duality to cases where this field theory can be put on a non-trivial curved background. 
There are currently only a few explicit examples of such constructions, which arise for classes of ${\cal N}=2$ Chern-Simons-matter 
theories put on certain squashed three-spheres \cite{Martelli:2011fu,Martelli:2011fw}. In the latter two references we presented supersymmetric 
gravity duals  for the cases of (a cousin of) the elliptically squashed three-sphere studied in \cite{Hama:2011ea} and the biaxially squashed three-sphere
studied in \cite{Imamura:2011wg}. One of the results of the present paper will be the construction of the gravity dual to field theories on a biaxially 
squashed three-sphere considered in \cite{Hama:2011ea}. What distinguishes this  from the set-up 
studied in \cite{Imamura:2011wg} is a different choice 
of background R-symmetry gauge field.

More generally, in this paper we will perform an exhaustive study of supersymmetric
 asymptotically locally AdS$_4$ solutions whose conformal boundary is given 
by a biaxially squashed \emph{Lens space} $S^3/\Z_p$. We will first work within (Euclidean) minimal gauged supergravity in four dimensions, 
determining the general local form of  the supersymmetric solutions with $SU(2)\times U(1)$ symmetry, and then we will discuss in detail 
the global properties of these solutions, both in four dimensions and in eleven-dimensional supergravity. Despite the high degree of 
symmetry of the problem, we uncover
a surprisingly intricate web of supersymmetric solutions. One of our main findings is that generically 
a given  conformal  boundary can be ``filled'' with more than one supersymmetric solution, with different topology.
More specifically, we will show that for a given choice of conformal class of metric and
gauge field  there  exist supersymmetric solutions with the topology of $\R^4$ (or $\Z_p$ orbifolds of this) -- the \emph{NUTs} -- 
and different supersymmetric  solutions with the topology  of ${\cal M}_p \equiv$ total space of ${\cal O}(-p)\to S^2$ -- the \emph{bolts}.\footnote{In particular, $H_2({\cal M}_p,\Z)\cong\Z$ and 
there is hence a non-trivial two-cycle, which is referred to as a ``bolt''. } The discussion of these Taub-Bolt-AdS 
solutions is subtle: they typically exist only in certain ranges of the squashing parameter, depending on $p$ and the amount of supersymmetry 
preserved, and moreover typically they have globally different boundary conditions to
the corresponding $\Z_p$ quotient of a Taub-NUT-AdS solution (related to the addition of a flat Wilson line 
at infinity for the gauge field). 
Appealing to a conjecture \cite{anderson} that the (conformal) isometry group of the conformal boundary extends to the isometry of 
the bulk,\footnote{See also Appendix B of \cite{Papadimitriou:2005ii}.}
 we will have found \emph{all} possible supersymmetric fillings of a given boundary, at least in the context of four-dimensional minimal gauged supergravity. 

The results we find have interesting implications  for the AdS/CFT correspondence. 
Recall that when there exist inequivalent fillings of a fixed boundary one should sum over 
all the contributions in the saddle point approximation to the path integral. 
Equivalently, the partition function of the dual field theory (in the large $N$ limit) is given by the \emph{sum} of the 
exponential of minus the supergravity action, evaluated on each solution with a fixed boundary. 
If different solutions dominate the path integral 
(have smallest free energy) in different regimes of the parameters, 
then passing from one solution to another is interpreted as a phase transition between vacua of the  theory. 
In the example of the Hawking-Page phase transition   \cite{Hawking:1982dh}, 
discussed in \cite{Witten:1998zw},  the two gravity solutions with  the same boundary are
  thermal AdS$_4$ and the Schwarzchild-AdS$_4$ solution, and the paramater being dialled is the 
temperature of the black hole (or equivalently of the dual field theory).
The more sophisticated examples  discussed  in \cite{Hawking:1998ct,Chamblin:1998pz} share a number of
similarities with the results presented here, but there are some crucial differences. 
The latter references studied 
Taub-NUT-AdS and Taub-Bolt-AdS solutions,  whose conformal boundary metric is precisely the biaxially squashed 
three-sphere. 
However,  these are all non-supersymmetric Einstein solutions,
and do not possess any gauge field.\footnote{As we shall discuss, the Taub-NUT-AdS metric
has  self-dual Weyl tensor, and hence it can be made supersymmmetric by 
adding particular instanton fields \cite{Dunajski:2010zp}. 
 The Taub-Bolt-AdS metric
in \cite{Hawking:1998ct,Chamblin:1998pz} is not self-dual, and cannot be made supersymmetric by adding any instanton.}
On the other hand, the solutions in this paper will all have a non-trivial gauge field turned on, which is necessary in order to 
preserve supersymmetry. We will therefore refrain from interpreting the squashing parameter as the inverse temperature. 
Whether or not one should sum over our Taub-Bolt-AdS solutions, in the saddle point approximation to 
quantum gravity, depends on whether they are interpreted as different vacua of the same theory,  
or rather as vacua of (subtly) different field theories. This in turn depends on the uplifting 
of the solutions to M-theory, discussed briefly in the next paragraph, but we shall argue that,
at least in some cases, the Taub-Bolt-AdS solutions have (subtly) different boundary conditions to the Taub-NUT-AdS 
solutions. 

An interesting aspect of the supersymmetric Taub-Bolt-AdS solutions (with topology ${\cal M}_p$) is that these 
can be uplifted to solutions ${\cal M}_p \, \widetilde{\times}\, Y_7$ of M-theory only for particular internal Sasaki-Einstein manifolds $Y_7$. Indeed, the key issue here is that $Y_7$ is necessarily 
fibred over $\mathcal{M}_p$, which we have denoted with the tilde.
As we shall explain,  for all these solutions the free energy of the field theory has not   yet been studied  in the literature, 
and therefore we cannot compare our gravity results with an existing field theory calculation.
However, for both classes of  solutions of Taub-NUT-AdS type (1/2 BPS and 1/4 BPS), 
where the dual field theories are  placed on squashed three-spheres, 
we obtain a precise matching between our gravity results and the results from localization in field theory.

The rest of the paper is structured as follows.  In section \ref{gaugedsex} we derive the general local form of the 
 solutions of interest. Section \ref{selfdual}
 is devoted to a discussion of regular self-dual Einstein solutions.
In sections \ref{1/2} and \ref{1/4} we discuss global properties of the solutions preserving 1/2 and 1/4 supersymmetry, respectively. In these sections the analysis is carried out in four dimensions. In section \ref{liftingit}
we discuss the subtleties associated
to embedding the solutions in M-theory, and make some comments on the holographic dual field theories. 
Section \ref{sowhat} concludes with a discussion. Seven appendices contain technical material  complementing the main 
body of the paper. 


\section{$SU(2)\times U(1)$-invariant solutions of gauged supergravity}
\label{gaugedsex}

We begin by presenting all Euclidean supersymmetric solutions of $d=4$, 
$\mathcal{N}=2$ gauged supergravity with $SU(2)\times U(1)$ symmetry. 
The action for the bosonic sector of 
this theory \cite{Freedman:1976aw} reads
\bea\label{4dSUGRA}
S &=& -\frac{1}{16\pi G_4}\int \diff^4x\sqrt{g}\left(R + 6\ell^{-2} - F^2 \right)~.
\eea 
Here $R$ denotes the Ricci scalar of the metric $g_{\mu\nu}$ and we have defined $F^2 \equiv F_{\mu\nu} F^{\mu\nu}$.
$G_4$ is the four-dimensional Newton constant and $\ell$ is a parameter with dimensions of length, related to the cosmological constant  via $\Lambda=-3\ell^{-2}$. The graviphoton is an Abelian gauge field $A$ with field strength  $F=\diff A$.

The equations of motion derived from (\ref{4dSUGRA}) read
\bea\label{EOM}
R_{\mu\nu} &=& -3\ell^{-2}g_{\mu\nu} +2\left(F_\mu^{\ \rho}F_{\nu\rho}-\frac{1}{4}F^2 g_{\mu\nu}\right)~,\nonumber\\
\diff * F &=&0~.
\eea
In Euclidean signature the gauge field may in principle be complex, although for the solutions in this paper the field strength $F$ will in fact be either real or purely imaginary.\footnote{In principle the metric may also be complex, 
although we will not consider that possibility here.} 

A solution is supersymmetric if there is a non-trivial Dirac 
spinor $\epsilon$ satisfying the Killing spinor equation
\bea\label{KSE}
\left( \nabla_\mu - \ii \ell^{-1} A_\mu + \frac{1}{2} \ell^{-1} \Gamma_\mu  + \frac{\ii}{4} F_{\nu\rho} \Gamma^{\nu\rho} \Gamma_\mu \right) \epsilon &=& 0~.
\eea
This takes the same form as in Lorentzian signature, except that here 
 $\Gamma_\mu$, $\mu=1,2,3,4$, generate the Clifford algebra $\mathrm{Cliff}(4,0)$, so 
$\{\Gamma_\mu,\Gamma_\nu\}=2g_{\mu\nu}$.
It was shown in \cite{Gauntlett:2007ma, Gauntlett:2009zw}  that any such solution uplifts (locally) to a supersymmetric solution of eleven-dimensional  
supergravity. As we will see, global aspects of this uplift can be subtle, and 
we will postpone a detailed discussion of these issues until  section \ref{liftingit}. In the remainder of 
this section all computations will be \emph{local}. In what follows we set $\ell =1$; factors of $\ell$ may be restored by dimensional analysis.

\subsection{General solution to the Einstein equations}

Our aim is to find, in explicit form, \emph{all} asymptotically locally AdS$_4$ solutions in Euclidean signature with boundary a biaxially squashed Lens space. Recall that the round metric on $S^3$ has $SU(2)_l\times SU(2)_r$ isometry. A biaxially squashed 
Lens space is described by an $SU(2)_l\times U(1)_r$-invariant metric on $S^3/\Z_p$, 
where $\Z_p\subset SU(2)_r$. Given a (conformal) Killing vector field on a compact 
three-manifold $\mathcal{M}^{(3)}$, a theorem of  Anderson \cite{anderson} shows that
this extends to a Killing vector  for 
any asymptotically locally AdS$_4$ Einstein metric on $\mathcal{M}^{(4)}$ with conformal boundary 
$\mathcal{M}^{(3)}=\partial \mathcal{M}^{(4)}$, provided $\pi_1(\mathcal{M}^{(4)},\mathcal{M}^{(3)})=0$.  
In particular, this result 
applies directly to the class of \emph{self-dual} solutions that we will discuss momentarily. 
Anderson also conjectures that this result 
extends to more general asymptotically locally AdS$_4$  solutions
to the Einstein-Maxwell equations. Assuming this conjecture holds, we may hence
restrict our  search to $SU(2)\times U(1)$-invariant solutions.\footnote{This result should be contrasted with the 
corresponding situation for asymptotically locally Euclidean metrics, where Killing vector 
fields on the boundary do not necessarily extend inside. The canonical examples 
are the Gibbons-Hawking multi-centre solutions \cite{Gibbons:1979zt}.}

The general ansatz for the  metric and gauge field takes the form 
\bea
\diff s^2_4 & = & \alpha^2 (r) \dd r^2 + \beta^2( r) (\sigma_1^2 + \sigma_2^2)  + \gamma^2(r) \sigma_3^2~, \nn \\
A & = & h(r) \sigma_3 ~,\label{metricansatz}
\eea 
where $\sigma_1,\sigma_2,\sigma_3$ are $SU(2)$ left-invariant one-forms, which may be written in terms of Euler angular variables as
\bea\label{leftinvariant}
\sigma_1 + \ii \sigma_2 & = & \ex^{-\ii\psi}(\diff\theta + \ii \sin\theta \diff\varphi)~,~~~ \sigma_3 \ = \ \diff\psi + \cos\theta\diff\varphi ~.
\eea
Note that in the case 
$h(r)\equiv 0$, when the metric is necessarily Einstein,  the general form of the solutions was obtained by Page-Pope \cite{Page:1985bq}.  We are not aware of any study of the equations in the most general Einstein-Maxwell case. 
In appendix  \ref{app:Einstein} we show that the general solution to (\ref{EOM}) 
with the ansatz (\ref{metricansatz}) is given by
\bea
\diff s^2_4 &=& \frac{r^2-s^2}{\Omega(r)}\diff r^2 + (r^2-s^2)(\sigma_1^2 + \sigma_2^2)+\frac{4s^2\Omega(r)}{r^2-s^2} \sigma_3^2~,\nn \\
A &=& \left(P\frac{r^2+s^2}{r^2-s^2}-Q\frac{2rs}{r^2-s^2}\right) \sigma_3~,\label{solution}
\eea
where 
\bea\label{poly}
\Omega(r) &=& (r^2-s^2)^2+(1-4s^2)(r^2+s^2)-2Mr+P^2-Q^2~.
\eea
Here $s$, $M$, $P$ and $Q$  are integration constants.
This  coincides with an analytic continuation of the Reissner-Nordstr\"om-Taub-NUT-AdS (RN-TN-AdS) solutions originally found in
 \cite{Carter:1968ks} and \cite{plebe}, and reduces to the Page-Pope metrics  for $P^2-Q^2=0$. The supersymmetry properties of the Lorentzian 
 solutions were 
 studied in \cite{AlonsoAlberca:2000cs} and \cite{Caldarelli:2003pb}.

It is a simple matter to check that the metric (\ref{solution}) is asymptotically 
locally AdS$_4$ as $|r|\rightarrow\infty$.
 At large $|r|$ the metric is to leading 
order
\bea\label{asymptoticmetric}
\diff s^2_4 & \approx & \frac{\diff r^2}{r^2} + r^2\left(\sigma_1^2+\sigma_2^2+ 
4s^2\sigma_3^2\right)~, 
\eea
so that the conformal boundary at $r=\pm \infty$ is (locally) 
a biaxially squashed $S^3$. 

\subsection{BPS equations}
\label{bpsses}

The requirement of supersymmetry imposes constraints on the four parameters $s, M, P$ and $Q$. 
In appendix \ref{app:BPS} we show that the integrability condition of \eqref{KSE} implies
\bea
D \ = \ 0~, \qquad B_+B_- \ = \ 0~,\label{integrabb}
\eea
where
\bea
D &\equiv & 2\left[MP-sQ(1-4s^2)\right]~,\nonumber\\
B_\pm & \equiv & (M\pm sQ)^2 - s^2(1\pm P-4s^2)^2 - (1\pm 2P-5s^2)(P^2-Q^2) ~.\label{BPSeqns}
\eea
We emphasize that these are \emph{necessary} but \emph{not sufficient} conditions for supersymmetry, and indeed we shall find examples of non-supersymmetric 
solutions satisfying both the integrability conditions (\ref{integrabb}).
One can show that solutions to the algebraic equations (\ref{BPSeqns}) fall into three classes:
\bea
\mbox{Class I}: \qquad M &= &  \pm 2sQ ~,~~~~~~~~~~~~~~P \ = \ \mp\frac{1}{2}(4s^2-1) ~,\nn\\
\mbox{Class II}: \qquad M &= &  \pm Q\sqrt{4s^2-1}~,~~~~~P\  = \ \mp s\sqrt{4s^2-1} ~,\nn\\
\mbox{Class III}: \qquad M &=& \mp s(4s^2-1)~,~~~~~~P \ = \ \pm Q~.\label{classes}
\eea
As we will show in the next section by explicitly solving the Killing spinor equation 
(\ref{KSE}), Class I corresponds to 1/4 BPS solutions while Class II corresponds to 1/2 BPS solutions. 
Class III are Einstein but in general \emph{not} supersymmetric, 
although both Classes II and III satisfy $D=B_+=B_-=0$. 
The upper and lower signs in (\ref{classes}) in fact lead to the same (local) solutions for the metric 
and gauge field:
in Class II the upper and lower signs are exchanged by sending $\{r\rightarrow -r,s\rightarrow -s\}$, 
while for Class I the upper and lower signs are exchanged by sending $\{r\rightarrow -r,\psi\rightarrow -\psi, 
\varphi\rightarrow -\varphi\}$. Thus, after a change of variable, the solutions for the metric and gauge 
field are in fact identical.
Without loss of generality we will thus focus 
on the following two cases:
\bea\label{1/4BPS}
\mbox{1/4 BPS}: \qquad M &= &  2sQ ~,~~~~~~~~~~~~~~P \ = \ -\frac{1}{2}(4s^2-1) ~,\\
\label{1/2BPS}
\mbox{1/2 BPS}: \qquad M &= & Q\sqrt{4s^2-1}~,~~~~~P\  = \ -s\sqrt{4s^2-1} ~.
\eea

\subsection{Killing spinors}\label{KSsection}

In this section we solve the Killing spinor equation \eqref{KSE}. We will do so separately for the two classes of BPS constraints (\ref{1/4BPS}), (\ref{1/2BPS}). 
In this section we will only derive the form of the Killing spinors in a convenient local orthonormal frame; 
\emph{global} aspects of these spinors will be addressed later in the paper, and in 
particular in appendix \ref{spinc}.
The Einstein metrics in Class III will be discussed further 
in section \ref{selfdual}. 

We work in the local orthonormal frame
\begin{equation}
\begin{aligned}
&e^1 \ = \ \sqrt{r^2-s^2}\, \sigma_1~,~~
&&e^2 \ =\  \sqrt{r^2-s^2}\, \sigma_2~,  \\
&e^3 \ = \ 2s\sqrt{\frac{\Omega(r)}{r^2-s^2}}\sigma_3~,~~
&&e^4 \ = \ \sqrt{\frac{r^2-s^2}{\Omega(r)}}\diff r~,
\end{aligned}\label{frame}
\end{equation}
and write $\Omega(r)$ as
\bea\label{Omegaroots}
\Omega(r) & =&  (r-r_1)(r-r_2)(r-r_3)(r-r_4)~.
\eea
We take the following basis of four-dimensional gamma matrices:
\bea\label{gamma}
\Gamma_\alpha &=& \left(\begin{array}{cc}0 & \tau_\alpha \\ \tau_\alpha  & 0\end{array}\right)~,\qquad \Gamma_4 \ = \ \left(\begin{array}{cc}0 & \ii \mathbb{I}_2 \\ 
-\ii \mathbb{I}_2 & 0\end{array}\right)~,
\eea
where $\tau_\alpha$, $\alpha =1,2,3$ are the Pauli matrices. Accordingly, 
\bea
\Gamma_5 & \equiv & \Gamma_1\Gamma_2\Gamma_3\Gamma_4\  =\  \left(\begin{array}{cc} \mathbb{I}_2 & 0 \\ 0 &  -\mathbb{I}_2 \end{array}\right)~.
\eea
We decompose the Dirac spinor $\epsilon$ into positive and negative chirality parts as
\bea
\epsilon & = &  \left(\begin{array}{c}\epsilon_+ \\ \epsilon_- \end{array} \right)~,
\eea
and further denote the components of $\epsilon_\pm$ as
\bea
\epsilon_\pm & = & \left(\begin{array}{c}\epsilon^{(+)}_\pm \\ \epsilon^{(-)}_\pm \end{array} \right)~.
\eea

\subsubsection{1/2 BPS solutions}

In this section we solve the Killing spinor equation (\ref{KSE}) for the second class of BPS constraints \eqref{1/2BPS}. 
We first obtain an algebraic relation between $\epsilon_+$ and $\epsilon_-$ by using the integrability condition \eqref{int}. In particular, by decomposing \eqref{int} into chiral parts using the \eqref{gamma} basis of gamma matrices we derive
\bea
\epsilon_-^{(+)} &=& \ii\sqrt{\frac{r-s}{r+s}}\sqrt{\frac{(r-r_1)(r-r_2)}{(r-r_3)(r-r_4)}}\epsilon_+^{(+)}~,\nn\\
\epsilon_-^{(-)} &=& \ii\sqrt{\frac{r-s}{r+s}}\sqrt{\frac{(r-r_3)(r-r_4)}{(r-r_1)(r-r_2)}}\epsilon_+^{(-)}~.\label{algebII}
\eea
Here we have identified the roots of $\Omega(r)$ in (\ref{Omegaroots}) as 
\bea\label{1/2roots}
\left\{\begin{array}{c}r_4 \\ r_3\end{array}\right\} &=& \frac{1}{2}\left[-\sqrt{4s^2-1}\pm 
\sqrt{8s^2-4Q-1}\right]~,\nn\\
\left\{\begin{array}{c}r_2 \\ r_1\end{array}\right\} &=& \frac{1}{2}\left[\sqrt{4s^2-1}\pm 
\sqrt{8s^2+4Q-1}\right]~.
\eea
We continue by looking at the $\mu = r$ component of the Killing spinor equation. Decomposing this into chiral parts we obtain
\bea
\partial_r \, \epsilon_+  &=& -  \frac{\ii}{2} \sqrt{\frac{r^2-s^2}{\Omega (r)}}\, \epsilon_- 
- \ii \sqrt{\frac{r^2-s^2}{\Omega (r)}} \cdot \frac{s\sqrt{4s^2-1}+Q}{2(r-s)^2} \, \tau_3 \,\epsilon_-  \nn ~, \\
\partial_r \, \epsilon_-  &=& + \frac{\ii }{2}  \sqrt{\frac{r^2-s^2}{\Omega (r)}}\, \epsilon_+ 
+\ii \sqrt{\frac{r^2-s^2}{\Omega (r)}} \cdot \frac{s \sqrt{ 4 s^2 -1}-Q}{2(r+s)^2}\,\tau_3 \, \epsilon_+  ~.
\eea
Using the relations \eqref{algebII} it is straightforward to solve the above first order ODEs. The general solution is
\bea\label{spinor1/2}
\epsilon_+ &=& \left(\begin{array}{c} \sqrt{\frac{(r-r_3)(r-r_4)}{r-s}} \chi^{(+)} \\ \sqrt{\frac{(r-r_1)(r-r_2)}{r-s}}\chi^{(-)}  \end{array}\right)~,~~~~ 
\epsilon_- \ =\  \ii \left(\begin{array}{c} \sqrt{\frac{(r-r_1)(r-r_2)}{r+s}}  \chi^{(+)} \\ \sqrt{\frac{(r-r_3)(r-r_4)}{r+s}} \chi^{(-)}  \end{array}\right)~,
\eea
where the components $\chi^{(\pm)}$ depend only on the angular coordinates. 
We may then form the $r$-independent two-component spinor
\bea\label{chipm}
\chi & \equiv & \left(\begin{array}{c}\chi^{(+)} \\ \chi^{(-)}\end{array}\right)~,
\eea
The remaining components of the Killing spinor equation (\ref{KSE}) then
reduce to the following Killing spinor equation for $\chi$:
\bea\label{3dKSE1/2}
\left(\nabla_\alpha^{(3)}  - 
\ii A^{(3)}_\alpha - \frac{\ii s}{2}\gamma_\alpha - \frac{\ii}{2}\sqrt{4s^2-1}\gamma_\alpha\gamma_3\right)\chi &=& 0~.
\eea
Indeed, this is a particular instance of the new minimal rigid supersymmetry equation \cite{Klare:2012gn,newcyril}, 
which in turn is (locally) equivalent to  the charged conformal Killing spinor equation \cite{Klare:2012gn}.
Here $\nabla^{(3)}$ denotes the spin connection for the three-metric
\bea\label{3metric}
\diff s^2_3 &=& \sigma_1^2+ \sigma_2^2 + 4s^2\sigma_3^2~,
\eea
with $\gamma_{\alpha}=\tau_\alpha$, $\alpha=1,2,3$ generating the corresponding 
Cliff$(3,0)$ algebra in an orthonormal frame,
and 
\bea\label{3gaugefield}
A^{(3)} &=& \lim_{r\rightarrow\infty} A  \ = \ P\sigma_3 \ = \ - s\sqrt{4s^2-1}\sigma_3~.
\eea
The three-metric (\ref{3metric}) and gauge field (\ref{3gaugefield}) 
are in fact the conformal boundary of (\ref{solution}) at $r=\infty$.  
It is important to stress here that, in general, the expression (\ref{3gaugefield}) 
is valid only \emph{locally}, that is in a coordinate patch. The precise global form of the gauge field, and how this 
interacts with the spin structure, will be discussed later in the paper, and in particular in appendix \ref{spinc}.

The general solution to (\ref{3dKSE1/2}) in the orthonormal  frame
\bea\label{tildeframe}
\tilde{e}^1 \ = \ \sigma^1~,~~~~~\tilde{e}^2 \ = \ \sigma^2~,~~~~~\tilde{e}^3 \ = \ 2s\sigma^3
\eea
induced from the $r\rightarrow\infty$ limit of the frame (\ref{frame}) ($\tilde{e}^a=\lim_{r\rightarrow\infty} e^a/r$) is
\bea\label{chi}
\chi &=& \left(\begin{array}{cc}\cos\tfrac{\theta}{2}\ex^{\ii(\psi+\varphi)/2} & 
-\sin\tfrac{\theta}{2}\ex^{\ii(\psi-\varphi)/2} \\ \gamma\sin\tfrac{\theta}{2}\ex^{-\ii(\psi-\varphi)/2} & \gamma\cos\tfrac{\theta}{2}\ex^{-\ii(\psi+\varphi)/2} \end{array}\right)\chi_{(0)}~,
\eea
where $\chi_{(0)}$ is any constant two-component spinor and we have defined
\bea
\gamma &\equiv & \ii (2s+\sqrt{4s^2-1})~.
\eea
The Killing spinors in this 1/2 BPS class are thus given explicitly by 
(\ref{spinor1/2}), with $\chi$ given by (\ref{chipm}), (\ref{chi}).

\subsubsection{1/4 BPS solutions}

In this section we solve the Killing spinor equation (\ref{KSE}) for the first class of BPS constraints  \eqref{1/4BPS}.
We again obtain an algebraic relation between $\epsilon_+$ and $\epsilon_-$ by using the integrability condition \eqref{int}:
\bea\label{algebI}
\epsilon_-^{(+)} &=& \epsilon_+^{(+)} \ = \ 0~,\nonumber \\
\epsilon_-^{(-)} &=& \ii\sqrt{\frac{r-s}{r+s}}\cdot \sqrt{\frac{(r-r_1)(r-r_2)}{(r-r_3)(r-r_4)}}\epsilon_+^{(-)}~.
\eea 
Here we have identified the roots of $\Omega(r)$ in (\ref{Omegaroots}) as
\bea\label{1/4roots}
\left\{\begin{array}{c}r_4 \\ r_3\end{array}\right\} &=& s\pm \sqrt{\frac{2Q+4s^2-1}{2}}~,\nn\\
\left\{\begin{array}{c}r_2 \\ r_1\end{array}\right\} &=&-s\pm \sqrt{\frac{-2Q+4s^2-1}{2}} ~.
\eea
The $\mu = r$ component of the Killing spinor equation reads
\bea
\partial_r \, \epsilon_+ &=& -\frac{\ii}{2}\sqrt{\frac{r^2-s^2}{\Omega(r)}}\, \epsilon_- - \ii \sqrt{\frac{r^2-s^2}{\Omega(r)}} \cdot \frac{1- 2Q-4s^2}{4(r- s)^2}\, \tau_3\, \epsilon_-~,\nonumber\\
\partial_r \, \epsilon_- &=& +\frac{\ii}{2}\sqrt{\frac{r^2-s^2}{\Omega(r)}}\, \epsilon_+ + \ii \sqrt{\frac{r^2-s^2}{\Omega(r)}}  \cdot \frac{1+ 2Q-4s^2}{4(r+ s)^2}\,\tau_3 \, \epsilon_+~.
\eea
Using the relations \eqref{algebI} the general solution is
\bea\label{spinor1/4}
\epsilon &=&  \left(\begin{array}{c} \sqrt{\frac{(r-r_3)(r-r_4)}{r-s}} \\  \ii \sqrt{\frac{(r-r_1)(r-r_2)}{r+s}} \end{array} \right) \otimes \chi~,
\eea
where again $\chi$ is a two-component spinor independent of $r$.
The remaining components of equation (\ref{KSE}) reduce to the following Killing spinor equation for $\chi$:
\bea\label{3dKSE1/4}
\left(\nabla_\alpha^{(3)} - \ii A^{(3)}_\alpha + \frac{\ii s}{2}\gamma_\alpha\right)\chi &=& 0~.
\eea
This is another instance of the new minimal rigid supersymmetry equation \cite{Klare:2012gn,newcyril}, which in \cite{Klare:2012gn} was shown to arise generically on the boundary of 
supersymmetric 
solutions of minimal gauged supergravity.
Here $\nabla_\alpha^{(3)}$ and $\gamma_\alpha$ are the spin connection and gamma matrices 
for the same biaxially squashed three-sphere metric (\ref{3metric}), while (locally)
the gauge field is now
\bea\label{3gaugefield1/4}
A^{(3)} &=&  \lim_{r\rightarrow\infty} A  \ = \ P\sigma_3 \ = \ - \frac{1}{2}(4s^2-1)\sigma_3~.
\eea
Notice that (\ref{3dKSE1/4}) is \emph{different} to the 1/2 BPS equation (\ref{3dKSE1/2}). The general solution to (\ref{3dKSE1/4}) in the orthonormal  frame (\ref{tildeframe}) is
\bea\label{1/4BPSchi}
\chi &=&  \left(\begin{array}{c}0 \\ \chi_{(0)}^{(-)} \end{array} \right)  ~,
\eea
where $\chi_{(0)}^{(-)}$ is a constant.


\section{Regular self-dual Einstein solutions}
\label{selfdual}

Having completed the local analysis, 
in this section we continue by finding all globally regular supersymmetric Einstein solutions. 
These are necessarily self-dual, meaning that the Weyl 
tensor is self-dual, with the gauge field being an instanton, 
{\it i.e.} with self-dual field strength $F$.\footnote{Of course, a change of orientation 
replaces self-dual by anti-self-dual in these statements.}
The condition of regularity means
requiring that the local metric given in (\ref{solution}) extends to a smooth complete metric on a four-manifold $\mathcal{M}^{(4)}$, and that the gauge field $A$ and Killing spinor are non-singular. 
Here it is important to specify globally precisely what are the gauge transformations of the gauge field $A$, and we shall 
find, throughout the whole paper, that regularity of the metric automatically implies that $A$ satisfies the quantization 
condition for a \emph{spin$^c$} gauge field on $\mathcal{M}^{(4)}$, and that the Killing spinors are correspondingly then smooth
spin$^c$ spinors.\footnote{\label{newfoot}In section \ref{liftingit} we shall discuss how uplifiting these solutions to \emph{eleven dimensions} imposes further 
conditions, in particular it will turn out that $\lambda A$ is a \emph{bona fide} connection, for some rational number $\lambda$ that we will determine. 
Correspondingly, the eleven-dimensional metric and Killing spinors will be globally defined only for certain choices of $p$, related to $\lambda$.}
We shall find two Einstein metrics in this class, both of which 
are known in the literature: the Taub-NUT-AdS solution, with the topology 
$\mathcal{M}^{(4)}=\R^4$ \cite{peder}, and the Quaternionic-Eguchi-Hanson solutions, with 
topology the total space of the complex line bundle $\mathcal{M}^{(4)}=\mathcal{M}_p\equiv\mathcal{O}(-p)\rightarrow S^2$, 
for $p\geq 3$ \cite{lebru,hitch}. 
In fact these both derive from the \emph{same} local 
solution in (\ref{solution}). These  are not supersymmetric 
without the addition of an instanton gauge field. We recover 
the instanton found by the authors in \cite{Martelli:2011fw}, and also find new 
regular supersymmetric solutions in both the 1/2 BPS and 1/4 BPS classes.

\subsection{BPS equations}

It is straightforward to show that the metric in (\ref{solution}) is Einstein
 if and only if $P^2-Q^2=0$. The field strength $F$ is then 
self-dual, meaning that the gauge field $A$ is an instanton. Thus, as commented in the previous section, 
the metrics in Class III are all Einstein. Recall that in this case
\bea\label{Mselfdual}
M=\mp s(4s^2-1)~,
\eea
and the metric function $\Omega(r)$ in (\ref{poly}) simplifies to
\bea\label{OmegaSD}
\Omega(r) &=& (r\mp s)^2\left[1+(r\mp s)(r\pm 3s)\right]~.
\eea
For the 1/2 BPS Class II, setting $P=\pm Q$  
the BPS condition (\ref{1/2BPS})  implies 
\bea\label{Q1/2}
\mbox{1/2 BPS}: \qquad Q &=& \mp s\sqrt{4s^2-1}~,
\eea
and hence again $M$ is given by (\ref{Mselfdual}). For the 1/4 BPS Class I, instead the 
BPS condition (\ref{1/4BPS}) gives
\bea\label{Q1/4}
\mbox{1/4 BPS}: \qquad Q &=& \mp \frac{1}{2}(4s^2-1)~,
\eea
which means that yet again $M$ is given by (\ref{Mselfdual}).

Thus for all cases with $P^2=Q^2$ the metric is given by the \emph{same} Einstein metric, with 
the metric function $\Omega(r)$ given by (\ref{OmegaSD}), but 
the gauge field instantons for the 1/2 BPS (\ref{Q1/2}) and 1/4 BPS (\ref{Q1/4}) classes are different. Class III clearly contains these supersymmetric solutions, 
but allows for an arbitrary rescaling of the instanton, described by the free 
parameter $P=\pm Q$. In fact we prove in appendix \ref{classIII} that the only 
\emph{supersymmetric} solutions in Class III are the solutions above in 
Class I and II. We may thus henceforth discard Class III.

\subsection{Einstein metrics}\label{Einstein}

The Einstein metric described in the previous subsection is
\bea\label{metricSD}
\diff s^2_4 &=& \frac{r^2-s^2}{\Omega(r)}\diff r^2 + (r^2-s^2)(\sigma_1^2 + \sigma_2^2)+\frac{4s^2\Omega(r)}{r^2-s^2} \sigma_3^2~,
\eea
where 
\bea\label{OmegaSDagain}
\Omega(r) &=& (r\mp s)^2\left[1+(r\mp s)(r\pm 3s)\right]~.
\eea
One can check that  the Weyl tensor of this metric is 
self-dual. Notice that without loss of generality we may consider only the case 
$r\rightarrow +\infty$ for the asymptotic boundary (\ref{asymptoticmetric}). Due to the $\pm$ signs in (\ref{OmegaSDagain})
we may also without loss of generality assume that $s\geq 0$.\footnote{At this point it might 
look more convenient to fix a choice of sign and simply take $s\in\R$. However, this 
choice of parametrization turns out to be inconvenient when comparing to the 
non-Einstein solutions discussed in later sections.}

It will be useful to note that the four roots of $\Omega(r)$ in (\ref{OmegaSDagain})  in this case may
be written as
\bea\label{Einsteinroots}
\left\{\begin{array}{c}r_4 \\ r_3\end{array}\right\} &=& \pm s~,\nn\\
\left\{\begin{array}{c}r_2 \\ r_1\end{array}\right\} &=& \left\{\begin{array}{c} \mp s + \sqrt{4s^2-1}\\ \mp s - \sqrt{4s^2-1} \end{array}\right\} ~.
\eea
In particular, $r_1$ and $r_2$ are complex for $0\leq s<\tfrac{1}{2}$. 
Notice these agree with the corresponding limits of the general roots 
in (\ref{1/4roots}); the relation to the roots in (\ref{1/2roots}) is more complicated, 
and will be discussed in section \ref{1/2}.

\subsubsection{Taub-NUT-AdS}

We begin by considering the upper signs in (\ref{Einsteinroots}). In this case $r_3=r_4=s$ is the largest root of 
$\Omega(r)$, so that $\Omega(r)>0$ for $r>s$. This case was discussed in \cite{Martelli:2011fw}, and 
 the metric is automatically regular at the double root $r=s$ provided 
the Euler angle $\psi$ has  period $4\pi$, so that the surfaces of constant 
$r>s$ are diffeomorphic to $S^3$.
  Then $\{r=s\}$ is a NUT-type coordinate singularity, and the metric is a 
  smooth and complete metric on $\mathcal{M}^{(4)}=\R^4$, with the origin of $\R^4$ being 
  naturally identified with $\{r=s\}$. In fact the metric is the metric 
  on AdS$_4$ for the particular value $s=\tfrac{1}{2}$, with the limit $s=0$ 
  being singular.  The conformal boundary is correspondingly 
  the round three-sphere for $s=\tfrac{1}{2}$, with $s>\tfrac{1}{2}$ and $0<s<\tfrac{1}{2}$ either  ``stretching'' or  ``squashing''
  the size of the Hopf fibre $S^1$ relative to the $S^2$ base.

\subsubsection{Quaternionic-Eguchi-Hanson}

We next consider the lower signs in (\ref{Einsteinroots}). In this case 
it is not possible to make the metric regular for $0<s<\tfrac{1}{2}$, since 
 in this range the largest root is at $r=-s<0$, and the 
coefficient of $\sigma_3^2$ then blows up at $r=s>0$, which leads to a singular metric. 
However, for $s>\tfrac{1}{2}$ the largest root is now at $r_2=s+\sqrt{4s^2-1}$, 
and thus we might obtain a regular metric by taking $r\geq r_2$.  To 
examine this possibility, we note that near to $r=r_2$ the 
 metric is to leading order
\bea
\diff s^2_4 & \approx & \frac{r_1+s}{2(r-r_2)}\diff r^2 + 
(r_2^2-s^2)(\sigma_1^2+\sigma_2^2)+\frac{8s^2(r-r_2)}{(r_1+s)}\sigma_3^2~.
\eea
Changing coordinate to
\bea
R^2 &=& 2(r_1+s)(r-r_2)~,
\eea
the metric is to leading order near $R=0$ given by
\bea\label{localmet}
\diff s^2_4 & \approx & \diff R^2 + \left(\frac{2s}{r_1+s}\right)^2R^2(\diff\psi + 
\cos\theta\diff\varphi)^2 + (r_2^2-s^2)(\sigma_1^2+\sigma_2^2)~.
\eea
We obtain a smooth metric on the $S^2$ at $R=0$ provided that 
$\theta\in [0,\tfrac{\pi}{2}]$ and $\varphi$ has period $2\pi$. 
On surfaces $r>r_2$ we must then take $\psi$ to have period $4\pi/p$, 
so that these three-manifolds are biaxially squashed Lens spaces 
$S^3/\Z_p$. The collapse of the metric (\ref{localmet}) at $R=0$ is smooth 
if and only if the period $\Delta\psi=4\pi/p$ of $\psi$ satisfies
\bea
\frac{2s}{r_1+s}\Delta\psi = 2\pi~.
\eea
We thus conclude that the squashing parameter is fixed to be
\bea\label{sp}
s &= & s_p \ \equiv \ \frac{p}{4\sqrt{p-1}}~.
\eea
Since $s>\tfrac{1}{2}$, this implies that for each integer $p\geq 3$ there 
exists a unique smooth Quaternionic-Eguchi-Hanson metric on the total 
space $\mathcal{M}_p$ of the complex line bundle $\mathcal{O}(-p)\rightarrow S^2$. 
In particular, the conformal boundary is then the biaxially squashed 
Lens space $S^3/\Z_p$, with squashing parameter fixed 
in terms of $p$ via (\ref{sp}).

The Quaternionc-Eguchi-Hanson metric is often presented in a different 
coordinate system. The change of variable
\bea
r(\rho)^2 &=& s^2 + \frac{\rho-a^2}{(1-\rho)^2}~,\nn\\
s^2 &=& \frac{1}{2(1-a^2)}~,
\eea 
leads to the metric
\bea
\diff s^2_4 &=& \frac{1}{(1-\rho)^2}\left[\frac{(\rho-a^2)\diff\rho^2}{\rho^2-a^2}+(\rho-a^2)(\sigma_1^2+\sigma_2^2)+\frac{\rho^2-a^2}{\rho-a^2}\sigma_3^2\right]~.
\eea
In these coordinates the conformal boundary is at $\rho=1$,  and $a=a_p\equiv 1 - \frac{8(p-1)}{p^2}$.

\subsection{Instantons}

As already commented, the Taub-NUT-AdS and Quaternionic-Eguchi-Hanson 
manifolds are, by themselves, not supersymmetric. However, they become 
1/2 BPS and 1/4 BPS solutions by turning on the instanton gauge field in 
(\ref{solution}) with $P=\pm Q$ and $Q$ fixed in terms of $s$ via 
(\ref{Q1/2}) and (\ref{Q1/4}), respectively. This is clear locally. 
In the remainder of this section we examine global issues. 
In particular, the instantons for the Quaternionic-Eguchi-Hanson solution 
will turn out to be automatically spin$^c$ connections in general, with the corresponding 
Killing spinor $\epsilon$ also being a spin$^c$ spinor. This is clearly necessary in order to have a 
smooth, globally-defined four-dimensional solution, 
since $\mathcal{M}_p\equiv\mathcal{O}(-p)\rightarrow S^2$ is a spin manifold if and only if $p$ is 
even, while it is spin$^c$ for all $p\in \mathbb{Z}$. We emphasize that 
in this section we are treating the solutions as purely four-dimensional. When we 
uplift to eleven-dimensional solutions in section \ref{liftingit} we will need to reconsider 
the gauge field $A$; in particular, what gauge transformations it inherits from 
eleven dimensions, and just as importantly whether it is $A$ that is ``observable'', or 
rather some multiple of it -- \emph{cf.} footnote \ref{newfoot}.

We begin by noting that with $P=\pm Q$ the \emph{local} gauge field (\ref{solution}) is
\bea\label{Ainst}
A &=&  Pf_\pm(r)\sigma_3~,
\eea
where we have defined
\bea
f_\pm(r) &\equiv & \frac{r\mp s}{r\pm s}~.
\label{fplusminus}
\eea
The corresponding field strength is thus
\bea\label{Finst}
F &=& \diff A \ = \ Pf_\pm'(r)\diff r\wedge \sigma_3 - Pf_\pm(r)\sigma_1\wedge \sigma_2~.
\eea
The value of $P$ is fixed to be
\bea\label{Pvals}
\mbox{1/4 BPS}: \qquad P  & = & -\frac{1}{2}(4s^2-1) ~,\nn\\
\mbox{1/2 BPS}: \qquad P & =  & -s\sqrt{4s^2-1} ~.
\eea

\subsubsection{Taub-NUT-AdS}

Recall that for the Taub-NUT-AdS solution we must take the upper signs 
in (\ref{Ainst}). Then this gauge field  is a globally well-defined 
one-form on $\{r>s\}\cong \R_+\times S^3$. Crucially, at $r=s$ the function $f_+(s)=0$. In fact near to this point $f_+(r)$ vanishes as $\rho^2$ 
as $\rho\rightarrow 0$, where $\rho$ denotes geodesic distance from the origin 
of $\R^4$ at $r=s$. It follows that $A$ is a global smooth one-form 
on the whole of $\mathcal{M}^{(4)}=\R^4$, and that the instanton is everywhere smooth 
and exact. This is true for either value of $P$ in (\ref{Pvals}).
 It follows that 
for all $s>0$ we get a 1/2 BPS and a 1/4 BPS smooth 
Euclidean supersymmetric supergravity solution on $\R^4$. 
The 1/2 BPS solution was found in \cite{Martelli:2011fw}, while the 1/4 BPS solution is new.

\subsubsection{Quaternionic-Eguchi-Hanson}
\label{QEHinstanton}

Recall that for the Quaternionic-Eguchi-Hanson solution we must take the lower signs 
in (\ref{Ainst}). 
In this case the latter gauge field is not defined at $r=r_2$, 
where the vector field $\partial_\psi$ has zero length. However, 
the field stength (\ref{Finst}) is manifestly a smooth global two-form 
on the four-manifold $\mathcal{M}_p= \mathcal{O}(-p)\rightarrow S^2$. 
It is straightforward to compute the flux through the 
$S^2\subset \mathcal{M}_p$ at $r=r_2$:
\bea
\int_{S^2} \frac{F}{2\pi} &=& -2Pf_-(r_2) \ = \ \begin{cases} \ 4s^2-1+2s\sqrt{4s^2-1} & \quad \mbox{1/4 BPS} \\ 
\ 4s^2+2s\sqrt{4s^2-1} & \quad \mbox{1/2 BPS}\end{cases}~,
\eea
where we have used (\ref{Pvals}). However, now using the fact that 
$s=s_p$ is fixed in terms of $p\geq 3$ via (\ref{sp}), we find the remarkable result
\bea\label{QEHflux}
\int_{S^2} \frac{F}{2\pi} &=&  \begin{cases} \ \frac{p}{2}-1 & \quad \mbox{1/4 BPS} \\ 
\ \frac{p}{2} & \quad \mbox{1/2 BPS}\end{cases}~.
\eea
In particular, for $p$ even we see that $F/2\pi$ defines 
an integral cohomology class in $H^2(\mathcal{M}_p,\Z)\cong\Z$, while 
for $p$ odd instead $F/2\pi$ has \emph{half-integer} period. 
This is precisely the condition that $A$ is a spin$^{c}$ connection. 
Recall that the curvature $F$ of a spin$^c$ connection $A$ on a manifold $\mathcal{M}$ satisfies the quantization condition
\bea\label{qu}
2\int_{\Sigma}\frac{F}{2\pi} &=& \int_{\Sigma} w_2(\mathcal{M}) \ \mathrm{mod}\ 2~,
\eea
where $\Sigma\subset \mathcal{M}$  runs over all two-cycles in $\mathcal{M}$. Here 
$w_2(\mathcal{M})\in H^2(\mathcal{M},\Z_2)$ denotes the second Stiefel-Whitney class 
of (the tangent bundle of) $\mathcal{M}$. For $\mathcal{M}_p=\mathcal{O}(-p)\rightarrow S^2$, it is straightforward to 
compute that $w_2(\mathcal{M}_p)= p$ mod $2\in \Z_2\cong H^2(\mathcal{M}_p,\Z_2)$. 
Thus for both 1/2 BPS and 1/4 BPS cases in (\ref{QEHflux}) we see 
that $A$ is a spin$^c$ connection for \emph{all} values of $p\geq 3$. 

This is also clearly necessary for the Killing spinors in section \ref{KSsection} 
to be globally well-defined. For $p$ an odd integer, the manifolds $\mathcal{M}_p$ 
are \emph{not} spin manifolds, so it is not possible to globally 
define a spinor $\epsilon$ on $\mathcal{M}_p$. However, 
from the Killing spinor equation (\ref{KSE}) we see that $\epsilon$ 
is charged under the gauge field $A$. This precisely defines a spin$^c$ spinor, with spin$^c$ gauge field $A$, provided that the 
curvature $F=\diff A$ satisfies the quantization condition (\ref{qu}). 
Thus the Killing spinors, in both 1/2 BPS and 1/4 BPS cases, are globally 
spin$^c$ spinors on $\mathcal{M}_p$. This is discussed in 
detail in appendix \ref{spinc}.
The upshot is that 
both the 1/2 BPS and 1/4 BPS Quaternionic-Eguchi-Hanson 
solutions on $\mathcal{M}_p=\mathcal{O}(-p)\rightarrow S^2$ 
lead to globally defined Euclidean supersymmetric supergravity solutions, 
for all $p\geq 3$.  Specifically, the four-component Dirac (spin$^c)$ spinors 
$\epsilon$ in the two cases are smooth sections of the bundles
\bea\label{nigel}
\begin{cases} \ \pi^*\left[\mathcal{O}(p-2)\oplus \mathcal{O}(0)\oplus\mathcal{O}(-2)\oplus\mathcal{O}(p)\right]& \quad \mbox{1/4 BPS} \\ \ \pi^*\left[\mathcal{O}(p-1)\oplus \mathcal{O}(1)\oplus\mathcal{O}(-1)\oplus\mathcal{O}(p+1)\right] & \quad \mbox{1/2 BPS}\end{cases}~,
\eea
where $\pi:\mathcal{M}_p\rightarrow S^2$ denotes projection onto the bolt/zero-section.

We refer the reader to appendix \ref{spinc} for a detailed discussion, but we conclude this section 
with some comments about the global form of the above Killing spinors and gauge field.  In fact these comments 
will apply equally to all the four-dimensional solutions in this paper.
The conformal
 boundary of the Quaternionic-Eguchi-Hanson solutions is a squashed $S^3/\Z_p$, 
with particular squashing fixed in terms of $p$ by (\ref{sp}). 
In the 1/2 BPS case the three-dimensional Killing spinor $\chi$ in (\ref{chi}) 
on constant $r>r_2$ hypersurfaces
appears to depend on the coordinate 
$\psi$, but this is an artifact of the \emph{frame} not being invariant
 under $\partial_\psi$. One can check that $\mathcal{L}_{\partial_\psi}\chi=0$, and 
one is then free to take the $\Z_p$ quotient along $\psi$ and preserve supersymmety. 
When $p$ is odd  the bulk spinors are necessarily spin$^c$ spinors, 
and these restrict to the unique spin bundle on the surfaces 
$\{r>r_2\}\cong S^3/\Z_p$. When $p$ is even the bulk is a spin manifold, 
and the  surfaces 
$\{r>r_2\}\cong S^3/\Z_p$ have \emph{two inequivalent} spin structures, 
which we refer to as ``periodic'' and ``anti-periodic'' in appendix \ref{spinc}.\footnote{This 
is by analogy with the two spin structures on $S^1$, but it is not meant to indicate 
any particular periodicity properties of the spinors.} 
The spinor bundle of the bulk in fact restricts to the anti-periodic 
spinor bundle on $S^3/\Z_p$, but the spin$^c$ bundle in (\ref{nigel}) that our Killing spinors 
are sections of restricts to the \emph{periodic} spinor bundle on $S^3/\Z_p$. 
The $\frac{p}{2}$ units of flux in (\ref{QEHflux}) play a crucial role 
in this discussion. 

The 1/4 BPS case is essentially the same, but with one small difference. 
The three-dimensional Killing spinor $\chi$ in (\ref{1/4BPSchi}) appears to be independent 
of $\psi$, but now the rotating frame in fact means that $\mathcal{L}_{\partial_\psi}\chi = 
\frac{\ii}{2}\chi$,  introducing an overall $\psi$-dependence of $\ex^{\ii \psi/2}$ in $\chi$. Thus the 1/4 BPS spinors on $\{r>r_2\}\cong S^3/\Z_p$ hypersurfaces 
apparently depend on $\psi$, which would seem to prevent one from 
quotienting by $\Z_p$ and preserving supersymmetry. However, 
in solving the Killing spinor equation in section \ref{KSsection} we did not 
take into account the \emph{global} form of the gauge field $A^{(3)}$. 
The full gauge field is
\bea\label{flateric}
A^{(3)} &=& A^{(3)}_{\mathrm{global}} + A^{(3)}_{\mathrm{flat}} 
\ =  \ P \sigma_3 + A^{(3)}_\mathrm{flat}~,
\eea
where $A^{(3)}_{\mathrm{flat}}$ is a flat connection. The factor of $-1$ 
in the flux (\ref{QEHflux}), relative to the 1/2 BPS case, precisely 
induces on $S^3/\Z_p$ a flat connection on the torsion line bundle 
$\mathscr{L}^{-1}$ with $c_1(\mathscr{L})=1\in\Z_p\cong H^2(S^3/\Z_p,\Z)$. 
The concrete effect of this is to introduce (locally) a phase 
$\ex^{-\ii \psi/2}$ into the Killing spinor $\chi$, cancelling the phase 
$\ex^{\ii\psi/2}$ described above, and meaning that the  
 correct global form of the Killing spinor $\chi$ is in fact independent of 
$\psi$. 
Thus the $-1$ factor in (\ref{QEHflux}), relative to the 1/2 BPS case, is crucial in order that these 1/4 BPS solutions are globally supersymmetric. 
We refer the interested reader to appendix \ref{spinc} for a detailed discussion of these issues.

Finally, let us comment further on the global form of the boundary gauge field in (\ref{flateric}).
 The gauge field at infinity $A^{(3)}$ is 
naively given by (\ref{Ainst}) restricted to $r=\infty$, which is 
\bea
A^{(3)}_{\mathrm{global}} &\equiv & P\sigma_3~,
\eea
where $\sigma_3$ is a globally defined one-form on $S^3/\Z_p$ (it is the global 
angular form for the fibration $S^3/\Z_p\rightarrow S^2$). Thus at first 
sight the gauge field at infinity is a global one-form, and thus is a connection on a 
trivial line bundle. However, this conclusion is false in general. The above argument 
is incorrect -- the gauge field in (\ref{Ainst}) is defined only \emph{locally} on $\mathcal{M}_p$, 
since it is ill-defined on the bolt at $r=r_2$, 
and for $p$ odd is not even globally a gauge field.  This is discussed carefully in appendix 
\ref{spinc}. If 
\bea
\int_{S^2}\frac{F}{2\pi} &=& \frac{n}{2}~,
\eea
then the upshot is that the gauge field at conformal infinity is
(\ref{flateric})
where $A^{(3)}_{\mathrm{flat}}$ is a certain flat connection. Using the result of 
appendix \ref{spinc}, we compute the first Chern class of the latter (which determines it uniquely) as
\bea\label{QEHtorsion}
\Z_p\ \cong \ H^2(S^3/\Z_p,\Z) \ \ni \ c_1(A^{(3)}_{\mathrm{flat}}) &=& \begin{cases} \  \begin{cases}\  \frac{p}{2}-1 & \quad \mbox{$p$ even} \\ \ p-1 & \quad \mbox{$p$ odd}\end{cases} & \quad  \mbox{1/4 BPS} \\ \  \begin{cases}\  \frac{p}{2} & \quad  \quad \ \  \mbox{$p$ even} \\ \ 0 & \quad  \quad \ \  \mbox{$p$ odd} \end{cases} & \quad \mbox{1/2 BPS}  \end{cases}~.
\eea
Notice that the integers on the right hand side are defined only mod $p$. The term $P\sigma_3$ thus gives only the globally defined  part of the gauge field, in general.

We conclude by emphasizing again that when we lift these solutions to eleven dimensions, 
in some cases we will need to re-examine the global form of the gauge transformations of $A$ 
inherited from eleven dimensions, to determine which solutions have the ``same'' boundary 
data. In particular, a flat gauge field such as $A^{(3)}_{\mathrm{flat}}$ is always locally trivial, and the \emph{only} information it contains 
is therefore global. 
 

\section{Regular 1/2 BPS solutions}
\label{1/2}

In this section we find all globally regular supersymmetric solutions satisfying the 1/2 BPS condition
(\ref{1/2BPS}). 
For all such solutions the (conformal class of the) boundary three-manifold 
will be $S^3/\Z_p$ with biaxially squashed metric
\bea\label{boundary3}
\diff s^2_3 &=& \sigma_1^2+\sigma_2^2+4s^2\sigma_3^2~,
\eea
where $\sigma_3=\diff\psi+\cos\theta\diff\varphi$ and $\psi$ has period $4\pi/p$, while the boundary 
gauge field is
\bea\label{A3infinity}
A^{(3)} &=& P \sigma_3  + A^{(3)}_{\mathrm{flat}}  \ = \ - s\sqrt{4s^2-1}\sigma_3 + A^{(3)}_{\mathrm{flat}} ~.
\eea
The flat gauge field $A^{(3)}_{\mathrm{flat}}$ is present for precisely the same global reasons 
discussed at the end of section \ref{selfdual}.
The boundary Killing spinor equation 
is (\ref{3dKSE1/2}), which we reproduce here for convenience
\bea\label{barry}
\left(\nabla_\alpha^{(3)}  - 
\ii A^{(3)}_\alpha - \frac{\ii s}{2}\gamma_\alpha - \frac{\ii}{2}\sqrt{4s^2-1}\gamma_\alpha\gamma_3\right)\chi &=& 0~.
\eea
The solution $\chi$ is given by (\ref{chi}). 
It will be important to note that a solution to the above boundary data with 
given $s$ is diffeomorphic to the same solution with $s\rightarrow -s$. 
\emph{Thus it is only $|s|$ that is physically meaningful at infinity.}
This is completely obvious for the metric (\ref{boundary3}). 
We may effectively change the sign of $s$ in the gauge field (\ref{A3infinity}) 
by the change of coordinates $\{\psi\rightarrow -\psi, \varphi\rightarrow -\varphi\}$, which 
sends $\sigma_3\rightarrow -\sigma_3$. Similarly, we may effectively change the sign of $s$ in the 
Killing spinor equation  (\ref{barry}) by sending $\gamma_\alpha\rightarrow -\gamma_\alpha$, which 
generate the same Clifford algebra $\mathrm{Cliff}(3,0)$.  

As we shall see, and perhaps surprisingly, for fixed 
conformal boundary data we sometimes find more than one smooth supersymmetric 
filling, with different topologies.
This moduli space will be described in 
section \ref{1/2moduli}.

\subsection{Self-dual Einstein solutions}\label{sd2}

The 1/2 BPS Einstein solutions were described in section \ref{selfdual}. 
For \emph{any} choice of conformal boundary data, meaning for all $p\in\mathbb{N}$ and
all choices of squashing parameter $s>0$, 
there exists the 1/2 BPS Taub-NUT-AdS$/\Z_p$ solution on $\R^4/\Z_p$. 
This has metric (\ref{metricSD}), (\ref{OmegaSDagain}) and 
$\psi$ is taken to have period $4\pi/p$. This solution then has an isolated $\Z_p$ orbifold 
singularity at $r=s$ for $p>1$, or, removing the singularity, the topology is 
$\R_{>0}\times S^3/\Z_p$. Although $\R^4/\Z_p$ is (mildly) singular for $p>1$, there is 
evidence that this solution is indeed an appropriate gravity dual \cite{Alday:2012au}. 
In the latter reference the large $N$ limit of the free energy of the ABJM theory on the unsquashed ($s=\frac{1}{2}$) $S^3/\Z_p$
was computed, and found to agree with the free energy of AdS$_4/\Z_p$. 

On the other hand, for each $p\geq 3$ and specific squashing parameter 
$s= s_p=\tfrac{p}{4\sqrt{p-1}}$ we also have the Quaternionic-Eguchi-Hanson solution. 
Thus for each $p\geq 3$ and $s=s_p$ there exist two supersymmetric self-dual Einstein
fillings of the same boundary data: the Taub-NUT-AdS solution on $\R^4/\Z_p$  and the 
Quaternionic-Eguchi-Hanson solution on $\mathcal{M}_p=\mathcal{O}(-p)\rightarrow S^2$. 
However,  in concluding this we must be careful about the \emph{global} boundary data 
in the two cases.  As discussed around equation (\ref{QEHtorsion}), the 
1/2 BPS Quaternionic-Eguchi-Hanson solution has a gauge field on the conformal  
boundary $S^3/\Z_p$ with torsion first Chern class $c_1=\frac{p}{2}$ mod $p$ when 
$p$ is even. That is, globally $A^{(3)}$ is a connection on the torsion line bundle 
$\mathscr{L}^{\frac{p}{2}}$ when $p$ is even, where 
$c_1(\mathscr{L})=1\in \Z_p\cong H^2(S^3/\Z_p,\Z)$ (notice $c_1=0$ mod $p$ when $p$ is odd). 
 However, at the same time, the spinors in the bulk restrict to sections of the spin bundle 
$\mathscr{S}_1$ on the boundary. As discussed in detail in appendix  \ref{spinc},  in fact the latter bundle is isomorphic 
to $\mathscr{S}_0\otimes \mathscr{L}^{\frac{p}{2}}  \cong \mathscr{S}_1$, therefore the net effect of the non-trivial flat
connection on the torsion line bundle 
$\mathscr{L}^{\frac{p}{2}}$ is to turn the boundary spinor into sections of  $\mathscr{S}_0 \cong \mathscr{S}_1\otimes \mathscr{L}^{\frac{p}{2}} $, the periodic spin bundle, 
precisely as for the spinors on the Taub-NUT-AdS solutions. Effectively, 
the additional flat gauge field induced from the bulk then cancels against the corresponding difference
 in the spin connection. 

\subsection{Non-self-dual Bolt solutions}

\subsubsection{Regularity analysis}
\label{12regularity}

We begin by analysing when the general metric in (\ref{solution}) is regular, where for the 1/2 BPS class
the metric function $\Omega(r)=(r-r_1)(r-r_2)(r-r_3)(r-r_4)$ has roots\footnote{Notice that this parametrization 
of the roots is \emph{different} to the self-dual Einstein limit in section \ref{Einstein}. For example, 
setting $Q=-s\sqrt{4s^2-1}$ we have from (\ref{rpm}) that $r_\pm = s$ for $s>0$, which thus match onto the 
roots $r_3,r_4$ of section \ref{Einstein}, while $r_\pm = -s\pm\sqrt{4s^2-1}$ for $s\leq-\frac{1}{2}$, which
thus match onto the roots $r_1,r_2$ of section \ref{Einstein}. }
\bea\label{1/2rootsagain}
\left\{\begin{array}{c}r_4 \\ r_3\end{array}\right\} &=& \frac{1}{2}\left[-\sqrt{4s^2-1}\pm 
\sqrt{8s^2-4Q-1}\right]~,\nn\\
\left\{\begin{array}{c}r_2 \\ r_1\end{array}\right\} &=& \frac{1}{2}\left[\sqrt{4s^2-1}\pm 
\sqrt{8s^2+4Q-1}\right]~.
\eea
Again, without loss of generality we may take the conformal boundary to be at $r=+\infty$.
A complete metric will then necessarily close off at the largest root $r_0$ of $\Omega(r)$, which 
must satisfy $r_0\geq s$ (if $r_0<s$ then the metric (\ref{solution}) is singular at $r=s$). 
Given (\ref{1/2rootsagain}), 
the largest root is thus either $r_0=r_+$ or $r_0=r_-$, where
\bea\label{rpm}
r_\pm &\equiv &  \frac{1}{2}\left[\pm\sqrt{4\s^2-1}+ \sqrt{8\s^2\pm 4Q-1}\right]~.
\eea

We first note that $r_0=r_\pm = s$ leads only to the $Q=\mp s\sqrt{4s^2-1}$ Taub-NUT-AdS 
solutions considered in the previous section. Thus $r_0>s$ and if $\psi$ has period $4\pi/p$ 
then the only possible topology is $\mathcal{M}_p=\mathcal{O}(-p)\rightarrow S^2$.  
Regularity of the metric near to the $S^2$ zero section at $r=r_0$ requires 
\bea\label{regularity}
\left|\frac{r_0^2-s^2}{s\Omega'(r_0)}\right| &=& \frac{2}{p}~.
\eea
This condition ensures that near to  $\rho=0$, where $ \rho\equiv \lambda\sqrt{r-r_0}$ is geodesic distance near the bolt (for appropriate constant $\lambda>0$), the metric (\ref{solution}) takes the form 
\bea\label{nearthebolt}
\diff s^2_4 &\approx & \diff\rho^2 + \rho^2\left[\diff \left(\frac{p\psi}{2}\right)+\frac{p}{2}\cos\theta\diff\varphi\right]^2 + (r_0^2-s^2)(\diff\theta^2+\sin^2\theta\diff\varphi^2)~.
\eea
Here $p\psi/2$ has period $2\pi$. 
Imposing (\ref{regularity}) at $r_0=r_\pm$ gives
\bea\label{Qvalues}
Q &=& Q_\pm(\s) \ \equiv \ \mp \frac{128s^4-16s^2-p^2}{64\s^2}~.
\eea
In turn, one then finds that the putative largest root is
\bea\label{rootschoice}
r_\pm(Q=Q_\pm(\s)) &=& \frac{1}{8}\left[\frac{p}{|\s|}\pm 4\sqrt{4\s^2-1}\right]~.
\eea

At this point we should pause to notice that a solution with given $s>0$ will 
be equivalent to the corresponding solution with $s\rightarrow -s<0$. 
This is because $Q_\pm(s)=Q_\pm(-s)$ in (\ref{Qvalues}), which then 
leads to exactly the same set of roots in (\ref{1/2rootsagain}), and thus the same local metric, 
while $P(-s)=-P(s)$. However, from the explicit form of the gauge field in (\ref{solution})
 we see that the diffeomorphism $\{\psi\rightarrow -\psi,\varphi\rightarrow-\varphi\}$ maps $\sigma_3\rightarrow -\sigma_3$, 
 which together with $s\rightarrow -s$ then leaves the gauge field invariant. 
Thus our parametrization of the roots in (\ref{1/2rootsagain})  is such that 
we need only consider $s>0$, which we henceforth assume.

Recall that in order to have a smooth metric, we require $r_0>s$. Imposing this for $r_0=r_\pm(Q_\pm(\s))$ gives
\bea
r_\pm(Q_\pm(\s)) - s &=& f_p^\pm (s)~,
\eea
where we must then determine the range of $s$ for which the function
\bea
f_p^\pm(s) &\equiv & \frac{1}{2}\left[\frac{p}{4s}-2s \pm\sqrt{4s^2-1}\right]
\eea
is strictly positive, in order to have a smooth metric. In addition, we must 
verify that (\ref{rootschoice}) really \emph{is} the largest root. We thus 
define
\bea
r_\pm(Q_\pm(s))-r_\mp(Q_\pm(s)) &=& h^\pm_p(s)~,
\eea
where as in all other formulae in this paper the signs are read entirely along the top or the bottom, 
and one finds
\bea
h^\pm_p(s) & \equiv & \frac{1}{2}\left[\frac{p}{4s} \pm 2\sqrt{4s^2-1}-\sqrt{16s^2-2-\frac{p^2}{16s^2}}\right]~.
\eea
Then (\ref{rootschoice}) is indeed the largest root provided also $h^\pm_p(s)$ is positive, or is 
complex.\footnote{If $h_p^\pm(s)$ is negative, one cannot then simply take the larger root
$r_\mp(Q_\pm(s))$ to be $r_0$, as the regularity condition (\ref{regularity}) does not hold.}

We are thus reduced to determining the subset of $\{s>0\}$ for which 
$f_p^\pm(s)$ is strictly positive, and $h_p^\pm(s)$ is either strictly positive or complex 
(since then the putative larger root is in fact complex). We refer to the two sign choices 
as positive and negative branch solutions.
The behaviour for $p=1$ and $p=2$ is qualitatively different from that with $p\geq 3$, so 
we must treat these cases separately. 

\subsection*{$p=1$}

It is straightforward to see that $f_1^\pm(s)<0$ for $s\in [\frac{1}{2},\infty)$, 
so that the metric cannot be made regular for $s$ in this range.  
Specifically, $f_1^\pm(\frac{1}{2})=-\frac{1}{4}$:
since $f_1^-(s)$ is monotonic decreasing, this rules out taking $r_0=r_-(Q_-(s))$ given by (\ref{rootschoice}); 
on the other hand $f_1^+(s)$ monotonically increases to zero 
from below as $s\rightarrow\infty$, and we thus also rule out $r_0=r_+(Q_+(s))$ in (\ref{rootschoice}). 
For $s\in (0,\frac{1}{2})$ the putative largest root is complex, 
so this range is also not allowed. 
We thus conclude that there are no additional 1/2 BPS 
solutions with $p=1$. This proves that the \emph{only} 1/2 BPS solution with $S^3$ 
boundary is Taub-NUT-AdS.

\subsection*{$p=2$}

We have $f_2^\pm(\frac{1}{2})=0$. Since $f_2^-(s)$ is monotonic decreasing on $s\in (\frac{1}{2},\infty)$ we rule out the  branch
$r_0=r_-(Q_-(s))$ for $s\in [\frac{1}{2},\infty)$.
 On the other hand, one can check that $\frac{\diff}{\diff s}f^+_2(\frac{1}{2})=+\infty$, 
$f_2^+(s)$ has a single turning point on $s\in (\frac{1}{2},\infty)$ at $s=\frac{1}{4}\sqrt{2+2\sqrt{5}}$, and 
$f_2^+(s)\rightarrow 0$ from above as $s\rightarrow\infty$. In particular 
for all $s>\frac{1}{2}$ we may take $Q=Q_+(s)$ and $r_0(s)=r_+(Q_+(s))$, since we have shown that then $r_0(s)>s$ for all $s>\frac{1}{2}$.  We must then check that $r_0(s)$ really is the largest root of $\Omega(r)$ in this range. This follows since $h^+_2(s)>0$ holds for all $s$ in this range, and thus this 
positive branch exists for all $s>\frac{1}{2}$. 
Again, the roots are complex for $s\in(0,\frac{1}{2})$. 
In conclusion, we have shown that for all 
$s\in(\frac{1}{2},\infty)$ we have 
a regular 1/2 BPS solution on $\mathcal{M}_2=\mathcal{O}(-2)\rightarrow S^2$.\footnote{Notice that the $s=\frac{1}{2}$ limiting solution
fills a \emph{round} Lens space $S^3/\Z_2$. 
We shall discuss this further in section \ref{roundLens}.}

\subsection*{$p\geq 3$}

\subsubsection*{Positive branches}

One can check that for all $p>2$ we have $f^+_p(\frac{1}{2})> 0$, $\frac{\diff}{\diff s}f^+_p(\frac{1}{2})=+\infty$, and
$f^+_p(s)$ has a single
turning point on $s\in (\frac{1}{2},\infty)$ given by 
\bea
\frac{\diff}{\diff s}\, f_p^+(s) &=& 0 \qquad \Longrightarrow \qquad s \ = \ \sqrt{\frac{p(4-p+\sqrt{p(p+8)})}{32(p-1)}}~.
\eea
Moreover, then $f_p^+(s)\rightarrow 0$ from above as $s\rightarrow\infty$. Setting $Q=Q_+(s)$, 
we must check that $r_+(Q_+(s))$ is the largest root. 
In fact $r_-(Q_+(s))$, and hence $h^+_p(s)$, is real here only for $s\geq \frac{1}{4}\sqrt{1+\sqrt{p^2+1}}$. In this range (which notice is automatic when $p=2$) one can check that $h_p^+(s)$ is strictly 
positive. In conclusion, taking $Q=Q_+(s)$ one finds that $r_0(s)=r_+(Q_+(s))$ is indeed the largest 
root of $\Omega(r)$ and satisfies $r_0(s)>s$ for all $s\geq \frac{1}{2}$. Thus the metric  is regular.
In conclusion, we have shown that for all 
 $s\in[\frac{1}{2},\infty)$ we have 
a regular 1/2 BPS solution on $\mathcal{M}_p=\mathcal{O}(-p)\rightarrow S^2$.  

\subsubsection*{Negative branches}

For $p\geq 3$ we also have regular solutions from the negative branch. 
Indeed, we now have $f_p^-(\frac{1}{2})>0$. Since $f_p^-(s)$ is monotonic decreasing, it follows that 
$f_p^-(s)$ is positive on precisely $[\frac{1}{2},{s}_p)$ for some ${s}_p>1$. One easily finds
\bea
{s}_p &=& \frac{p}{4\sqrt{p-1}}~.
\eea
Again, notice here that $p=2$ is special, since ${s}_2=\frac{1}{2}$. There is thus potentially another branch 
of solutions for $s$ in the range
\bea\label{srange}
\frac{1}{2}& \leq&  s\  < \ \frac{p}{4\sqrt{p-1}} \ = \ s_p~.
\eea
To check this is indeed the case, we note
that  $h_p^-(s)$ is real only for $s\geq \frac{1}{4}\sqrt{1+\sqrt{p^2+1}}$, and one can 
check that provided also $s<{s}_p$ then $h_p^-(s)$ is positive. Thus $r_-(Q_-(s))$ is indeed 
the largest root of $\Omega(r)$ for $Q=Q_-(s)$ and $s$ satisfying (\ref{srange}).
In conclusion, we have shown that for all 
 $s\in[\frac{1}{2},s_p)$ we have 
a regular 1/2 BPS solution on $\mathcal{M}_p=\mathcal{O}(-p)\rightarrow S^2$.  
The limiting solutions for $s=s_p$, which notice are where the 
roots $r_\pm(Q_-(s))$ are equal, will be discussed later. 

\subsubsection{Gauge field and spinors}

Having determined this rather intricate branch structure of solutions, 
let us now turn to analysing the global properties of the gauge field.
After a suitable gauge transformation, the latter can be written \emph{locally} as
\bea\label{noninstanton}
A &=&  \frac{s}{r^2-s^2}\left[-2Qr-(r^2+s^2)\sqrt{4\s^2-1}\right]\, \sigma_3~.
\eea
In particular, this gauge potential is \emph{singular} on the $S^2$ at $r=r_0$, 
but is otherwise globally defined on the complement  $\mathcal{M}_p\setminus S^2$ of the bolt. 
The field strength $F=\diff A$ is easily verified to be a
globally defined smooth two-form on $\mathcal{M}_p$, with non-trivial flux 
through the  $S^2$ at $r=r_0$. Indeed, for $Q=Q_\pm(s)$ one 
computes the period through the $S^2$ at $r_0(s)=r_\pm(Q_\pm(s))$ (respectively) to be 
\bea\label{fluxy}
\int_{S^2} \frac{F}{2\pi} &=& -\frac{2s}{r_0(s)^2-s^2}\left[-2Q_\pm(s)r_0(s)-(r_0(s)^2+s^2)\sqrt{4\s^2-1}\right]\nonumber\\
&=& \pm \frac{p}{2}~,
\eea
the last line simply being a remarkable identity satisfied by the largest root $r_0(s)$. 
Thus the positive/negative branch solutions have a gauge field flux 
$\pm \frac{p}{2}$ through the bolt, respectively. Following appendix \ref{spinc}, 
and precisely as for the 1/2 BPS Quaternionic-Eguchi-Hanson solutions in section \ref{QEHinstanton}, 
both branches then induce the \emph{same} spinors and global gauge field at conformal infinity, for fixed $p$ and $s$ (the crucial point here being that $\frac{p}{2}\equiv - \frac{p}{2}$ mod $p$, so that the torsion line bundles
on the boundary are the same for the positive and negative branches). 
Again, in eleven dimensions we will  need to reconsider this conclusion, 
as the physically observable gauge field is not necessarily $A$, but rather a multiple of it.

 For completeness we note that the Dirac spin$^c$ spinors are smooth sections of the following 
bundles:
\bea\label{henry}
\begin{cases} \ \pi^*\left[\mathcal{O}(p-1)\oplus \mathcal{O}(1)\oplus\mathcal{O}(-1)\oplus\mathcal{O}(p+1)\right]  & \quad \mbox{positive branch} \\ \ \pi^*\left[\mathcal{O}(-1)\oplus \mathcal{O}(-p+1)\oplus\mathcal{O}(-p-1)\oplus\mathcal{O}(1)\right] & \quad \mbox{negative branch}\end{cases}~,
\eea
and that when $p$ is even the boundary gauge field $A^{(3)}$ is a connection 
on $\mathscr{L}^{\frac{p}{2}}$.

\subsubsection{Special solutions}\label{roundLens}

For $p\geq2$ the positive branches described in section \ref{12regularity} 
all terminate at $s=\frac{1}{2}$, while for $p\geq 3$ 
the negative branches terminate at $s= \frac{1}{2}$ 
and $s=s_p$. In this section we consider these special 
limiting solutions. 

\subsubsection*{Positive branches}

When $s= \frac{1}{2}$ note firstly that the conformal boundary 
$S^3/\Z_p$ is \emph{round}, and secondly that 
the \emph{global} part of the gauge field $A^{(3)}_{\mathrm{global}}$ 
on the conformal boundary is
 identically zero. 
Indeed, notice that $P=0$ when $s=\frac{1}{2}$, while
\bea\label{Q12}
Q \ = \ Q_+\left(\tfrac{1}{2}\right) &=& \frac{(p-2)(p+2)}{16}~.
\eea
Thus for $p=2$ in particular we see that $P=0=Q$ and thus this solution is self-dual, but 
with a round $S^3/\Z_2$ boundary. It is not surprising, therefore, to discover 
that $s=\tfrac{1}{2}$ is simply AdS$_4/\Z_2$ in this case. However, 
due to the single unit of gauge field flux through the bolt (which in this singular limit 
has collapsed to zero size), the
global gauge field on the boundary is the unique non-trivial flat $U(1)$ connection on $S^3/\Z_2$.\footnote{Correspondingly,  the spinors inherited from  the bulk are sections of $\mathscr{S}_1$, so that altogether the boundary spinors
are sections of  $\mathscr{S}_0$. } 

For $p\geq 3$ we also have $P=0$, but now $Q>0$ in (\ref{Q12}). Thus the gauge field 
in the bulk is \emph{not} an instanton, and correspondingly we obtain 
a non-trivial smooth non-self-dual solution on $\mathcal{M}_p=\mathcal{O}(-p)\rightarrow S^2$. 
We will refer to all these solutions as \emph{round Lens filling} solutions -- 
locally, the conformal boundary is equivalent to the round three-sphere.  

Although this branch does not terminate at $s=s_p$, we note that 
at this point $Q_+(s_p)=s_p\sqrt{4s_p^2-1} = -P$ so that the solution 
is \emph{self-dual}. In fact this solution is precisely the Quaternionic-Eguchi-Hanson 
solution! Thus although this was isolated as a self-dual solution, we see that it 
exists as a special case of a family of non-self-dual solutions.

\subsubsection*{Negative branches}

The discussion for the limit $s=\frac{1}{2}$ is similar to that for the positive branches above. 
The only difference is that now
\bea
Q \ = \ Q_-\left(\tfrac{1}{2}\right) &=& -\frac{(p-2)(p+2)}{16}~.
\eea
However, since $P=0$ and $r_+(Q_+(\tfrac{1}{2}))=r_-(Q_-(\tfrac{1}{2}))$, we see 
that these are actually the \emph{same} round Lens filling solutions as on the positive 
branch. Thus the positive and negative branches actually \emph{join together} at this point.
 
Finally, recall that the $s=s_p$ limit has $h_p^-(s_p)=0$, implying that 
we have a \emph{double root}. It follows that this must locally be
a Taub-NUT-AdS solution, and indeed one can check that 
this negative branch joins onto Taub-NUT-AdS$/\Z_p$ 
with squashing parameter $s=s_p$. 

\subsection{Moduli space of solutions}\label{1/2moduli}

\begin{figure}[ht!]
\centering
\includegraphics[width=0.85\textwidth]{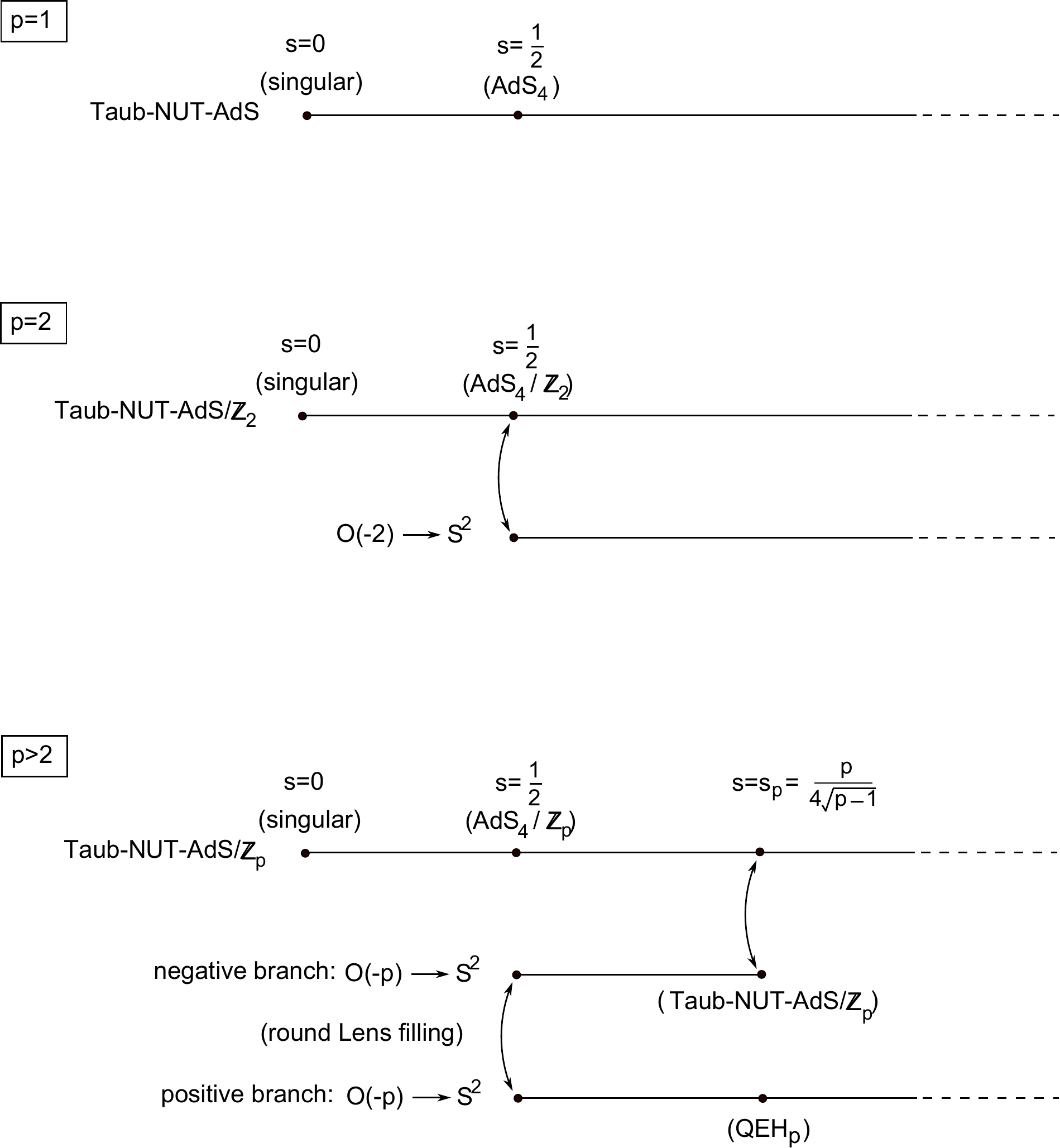}
\caption{The moduli space of 1/2 BPS solutions with biaxially squashed $S^3/\Z_p$ boundary, with squashing parameter 
$s$. The arrows denote
identification of solutions on different branches. Notice that these moduli spaces 
are connected for each $p$, but that for $p\geq 2$ the space multiply covers the $s$-axis. 
The self-dual Quaternionic-Eguchi-Hanson solution QEH$_p$ appears as a special point on the positive branch for $p\geq 3$.}\label{12BPS}
\end{figure}

We have summarized the intricate branch structure of solutions in Figure \ref{12BPS}.
In general the conformal boundary has biaxially squashed $S^3/\Z_p$ metric (\ref{boundary3}), 
with squashing parameter $s>0$, and boundary gauge field given by (\ref{A3infinity}). 
The 1/2 BPS fillings of this boundary may then be summarized as follows:

\begin{itemize} 
\item For $p=1$, the boundary $S^3$ 
with arbitrary squashing parameter $s>0$ 
has a unique 1/2 BPS filling, namely the Taub-NUT-AdS solution. 
For $s=\frac{1}{2}$ one obtains the AdS$_4$ metric as a special case. 
The gauge field curvature is real for $s>1/2$ and imaginary for $s<1/2$.

\item For $p\geq 2$ and arbitrary squashing parameter $s>0$ we always have the (mildly singular) 
Taub-NUT-AdS$/\Z_p$ solution. Thus for all boundary data there always 
exists a gravity filling, provided one allows for orbifold singularities. 
However, starting with $p=2$ there can exist other 1/2 BPS solutions, leading 
to non-unique supersymmetric fillings of the same boundary:

\item For $p=2$ and $s>\frac{1}{2}$
there is also a 1/2 BPS filling with the topology $\mathcal{M}_2=\mathcal{O}(-2)\rightarrow S^2$.
This degenerates to AdS$_4/\Z_2$ in the $s\rightarrow \frac{1}{2}$ limit, but with a non-trivial 
flat connection. This solution 
was first found in \cite{Martelli:2011fw},  where it was dubbed supersymmetric Eguchi-Hanson-AdS.
 Notice that for $p=2$ and $s=\frac{1}{2}$ there then exists 
a unique filling of the round $S^3/\Z_2$, which is the singular AdS$_4/\Z_2$ solution.

\item For all $p>2$ and $s>\frac{1}{2}$ there is an even more intricate 
structure. There is always a positive branch filling with topology $\mathcal{M}_p=\mathcal{O}(-p)\rightarrow S^2$, 
which includes the Quaternionic-Eguchi-Hanson solution at the specific value 
$s=s_p=\frac{p}{4\sqrt{p-1}}$. In the $s=\frac{1}{2}$ limit (which is non-singular) this branch joins onto a negative 
branch set of solutions, with the same topology. However, this negative branch then exists 
only for $s<s_p$, and joins onto the Taub-NUT-AdS$/\Z_p$ general solutions  in the 
$s\rightarrow s_p$ limit.  In particular, notice that this moduli space is connected, 
but multiply covers the $s$-axis.
\end{itemize}

\subsection{Holographic free energy}
\label{free1/2}

In this subsection we compute the holographic free energy of the 1/2 BPS solutions 
summarized above, using standard holographic renormalization methods \cite{Emparan:1999pm, Skenderis:2002wp}. 
Further details can be found in appendix \ref{app:FreeEnergy}.  A
subtlety for $p>1$ is how to calculate the holographic free 
energy of the \emph{singular} Taub-NUT-AdS$/\Z_p$ solutions, that we shall discuss later. 

The total on-shell action is
\bea
I &=&  I^{\text{grav}}_{\text{bulk}} + I^F + I^{\text{grav}}_{\text{ct}} + I^{\text{grav}}_{\text{bdry}}~.
\label{alli}
\eea
Here the first two terms are the bulk supergravity action (\ref{4dSUGRA}) 
\bea
I^{\text{grav}}_{\text{bulk}} + I^F &\equiv & -\frac{1}{16\pi G_4}\int \diff^4x\sqrt{g}\left(R + 6 - F^2 \right)~,
\eea
evaluated on a particular solution.
This is divergent, but we may regularize it using holographic renormalization. Introducing 
a cut-off at some large value of $r=\rcut$, with corresponding hypersurface $\mathcal{S}_\rcut
=\{r=\rcut\}$, we then add the following boundary terms
\bea
I^{\text{grav}}_{\text{ct}} + I^{\text{grav}}_{\text{bdry}} &=&  \frac{1}{8\pi G_4} \int_{\mathcal{S}_\rcut} \diff^3x \sqrt{\gamma} \left( 2 + \tfrac{1}{2} R(\gamma) - K\right) ~.
\eea
Here $R(\gamma)$ is the Ricci scalar of the induced metric $\gamma_{\mu\nu}$ on $\mathcal{S}_\rcut$, and $K$ is the trace of
 the second fundamental form of $\mathcal{S}_\rcut$, the latter being the Gibbons-Hawking boundary term. 

In all cases  the manifold closes off at $r=r_0$, the largest root of $\Omega(r)$, and we compute
\bea
I^{\text{grav}}_{\text{bulk}} &=& \frac{1}{8\pi G_4} \frac{16\pi^2}{p} \left.( 2 s r^3 - 6 s^3 r) \right |_{r_0}^\rcut ~,\\
I^{\text{grav}}_{\text{ct}} + I^{\text{grav}}_{\text{bdry}} &=&  \frac{1}{8\pi G_4} \frac{16\pi^2}{p}  [2 Q s \sqrt{4s^2-1}   -2 s \rcut^3 + 6s^3 \rcut +{O}(\rcut^{-1})]~.
\eea
As expected, the divergent terms cancel as $\rcut \to \infty$. 
The contribution to the action of the gauge field is finite in all cases and does not need regularization. 
For the Taub-NUT-AdS case $r_0=s$ and we compute
\bea
I^F_{\mathrm{NUT}} &=&  \frac{16\pi^2}{8\pi G_4} Q^2 \ = \  \frac{2\pi}{G_4} s^2(4s^2-1)  \quad  \qquad  (p=1)~,
\eea
while for the Taub-Bolt-AdS cases $r_0=r_\pm>s$ and we compute
\bea
I^F_{\mathrm{Bolt}} \ = \ \frac{1}{8\pi G_4}\frac{16\pi^2}{p}
\frac{2 s r_0\left[\left(Qr_0+s^2\sqrt{4s^2-1}\right)^2 + \left(Qs + sr_0 \sqrt{4s^2-1}\right)^2\right]}{(r_0^2-s^2)^2}~.
\eea
Combining all the above contributions to the action we obtain the following simple expressions 
\begin{equation}
\begin{aligned}
I_{\mathrm{NUT}}   &\ = \ 2s^2\frac{\pi}{G_4}~ \quad \qquad \qquad  (p=1)~,   \\[2mm]
I_{\mathrm{Bolt_\pm}} &\ = \ \left[\frac{1}{2} \pm \frac{\sqrt{4s^2-1}}{sp}\left(s^2-\frac{p^2}{16}\right)\right]\frac{\pi}{G_4}~.\\
\end{aligned}
\end{equation}
Here $I_{\mathrm{Bolt_\pm}}$ refers to the actions of the positive and negative 
branch solutions, respectively. Recall that  $I_{\mathrm{Bolt_+}}$ exists\footnote{For $p=2$ this free 
energy was computed in  \cite{Martelli:2011fw}.} 
for any $p\geq 2$, while $I_{\mathrm{Bolt_-}}$ exists for any $p\geq 3$.

For any $p\geq 2$ we can always fill the boundary squashed Lens space $S^3/\Z_p$
with the mildly singular Taub-NUT-AdS/$\Z_p$ solution, where  $\Z_p$ acts on the coordinate $\psi$. 
In these cases one may be concerned that the supergravity approximation breaks down and the classical 
on-shell gravity action (\ref{alli}) does not reproduce the correct free energy of the holographic dual field theories.  
In particular, the fact that the Taub-Bolt-AdS solutions smoothly reduce to the Taub-NUT-AdS/$\Z_p$ solutions at the special 
points $(p=2,s=\tfrac{1}{2})$ and $(p\geq 3,s=s_p)$ (see Figure \ref{12BPS}) 
implies that the holographic free energies of these orbifold solutions must be given by the limits 
\bea
\lim_{s\to \tfrac{1}{2}} I_{\mathrm{Bolt_+}} & = & \frac{1}{2}  \frac{\pi}{G_4}  \qquad \qquad  ~~~(p=2)~, \nn\\
 \lim_{s\to s_p} I_{\mathrm{Bolt_-}} & = & \frac{p^2}{8(p-1)} \frac{\pi}{G_4}   \qquad  (p\geq 3 )~, 
\eea
respectively. These differ from the naive values $\frac{1}{p}I_{\mathrm{NUT}}$ of the Taub-NUT-AdS/$\Z_p$ solutions
by a contribution that can be understood as associated to  flux  trapped  at the $\Z_p$ singularity \cite{Martelli:2011fw}. 
In turn, this trapped flux is related directly to the fact that the Taub-NUT-AdS$/\Z_p$ limits 
of the Taub-Bolt-AdS solutions necessarily have an additional flat gauge field $A^{(3)}_{\mathrm{flat}}$ turned on, 
relative to the simple $\Z_p$ quotient of the $p=1$ Taub-NUT-AdS solution.
  In similar circumstances  (\emph{e.g.} in singular 
ALE Calabi-Yau two-folds), a method for computing the contribution of this flux is to resolve the space. However, presently we cannot resolve the space while preserving supersymmetry (and $SU(2)\times U(1)$ isometry), 
as such geometries would  contain two parameters and their existence is precluded by our general analysis. 
It is natural to assume that, by continuity, 
the free energy of the orbifold Taub-NUT-AdS$/\Z_p$ branch onto which the bolt solutions join  contains
 the contribution of this trapped flux for generic values of $s$. 
One way to compute the free energies of these solutions 
is to  resolve the NUT orbifold singularity, replacing it with a non-vanishing two-sphere $S^2_\varepsilon$, while not preserving supersymmetry. 
Using this method, further discussed in  appendix \ref{proofising}, we find that for a gauge field with $\frac{n}{2}$ units of flux at the singularity
the contribution to the free energy is given by 
\bea
I_\mathrm{sing} & = & \frac{n^2}{8p}\cdot \frac{\pi}{G_4}~.
\eea
The total free energy of the orbifold solutions with $\pm \frac{p}{2}$ units of  flux is then given by
\bea
 I_\mathrm{NUT+flux}^\mathrm{orb} & = &  \frac{1}{p} I_{\mathrm{NUT}} +  I_\mathrm{sing}  \ = \  \left( \frac{2s^2}{p} + \frac{p}{8} \right)   \frac{\pi}{G_4}~.
\eea

\begin{figure}[t!]
\begin{tabular}{cc}
\includegraphics[width=0.45\textwidth]{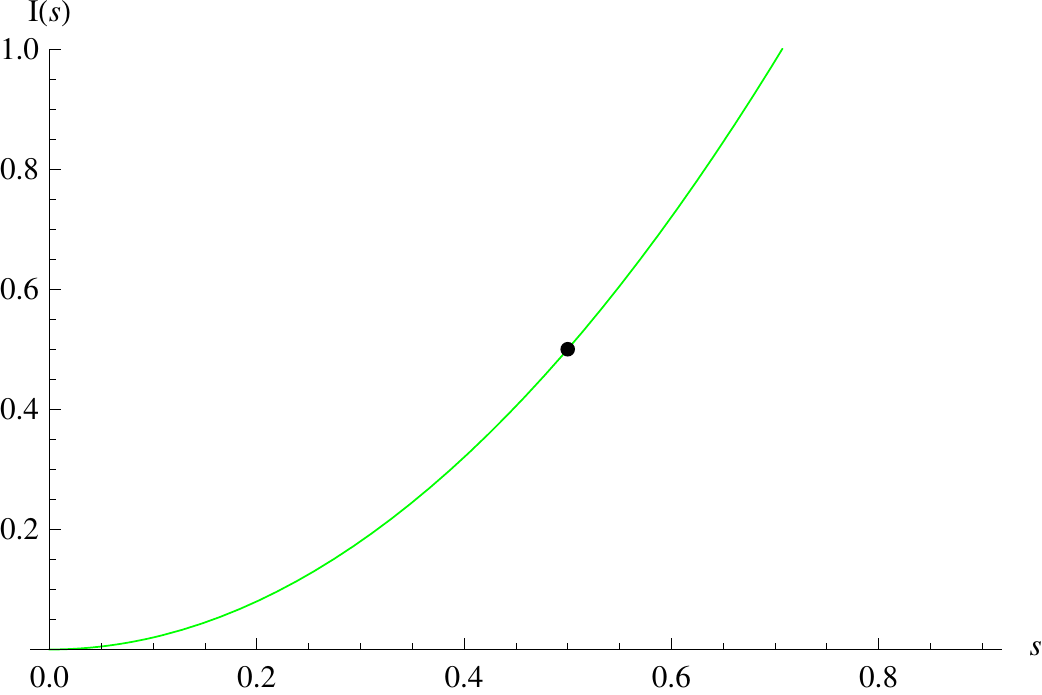} & \includegraphics[width=0.45\textwidth]{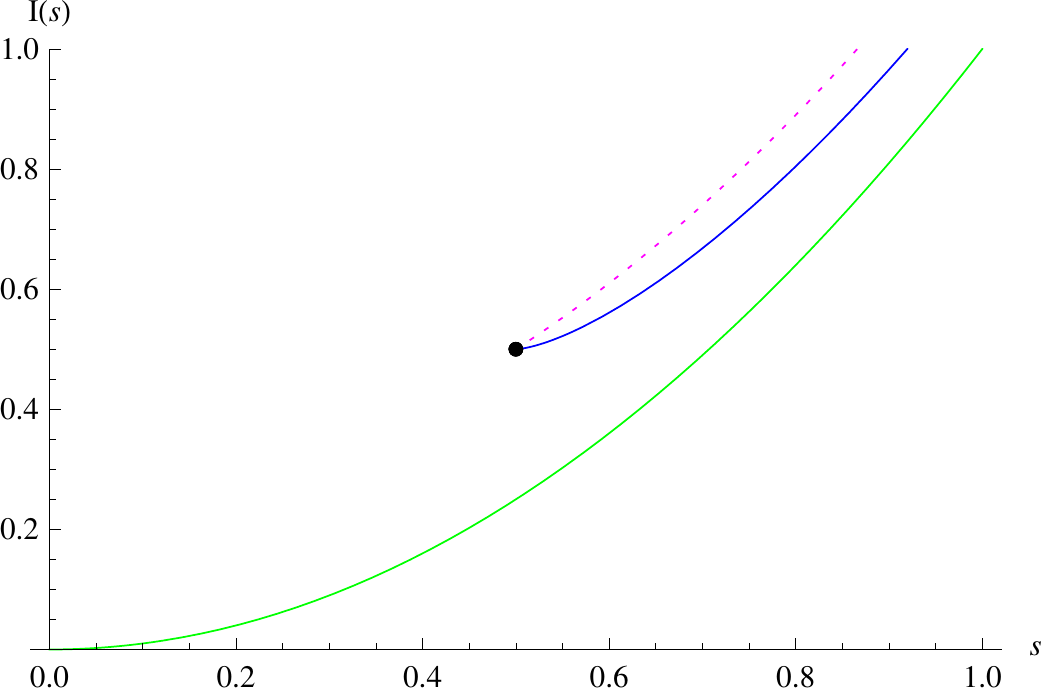} \\
\includegraphics[width=0.45\textwidth]{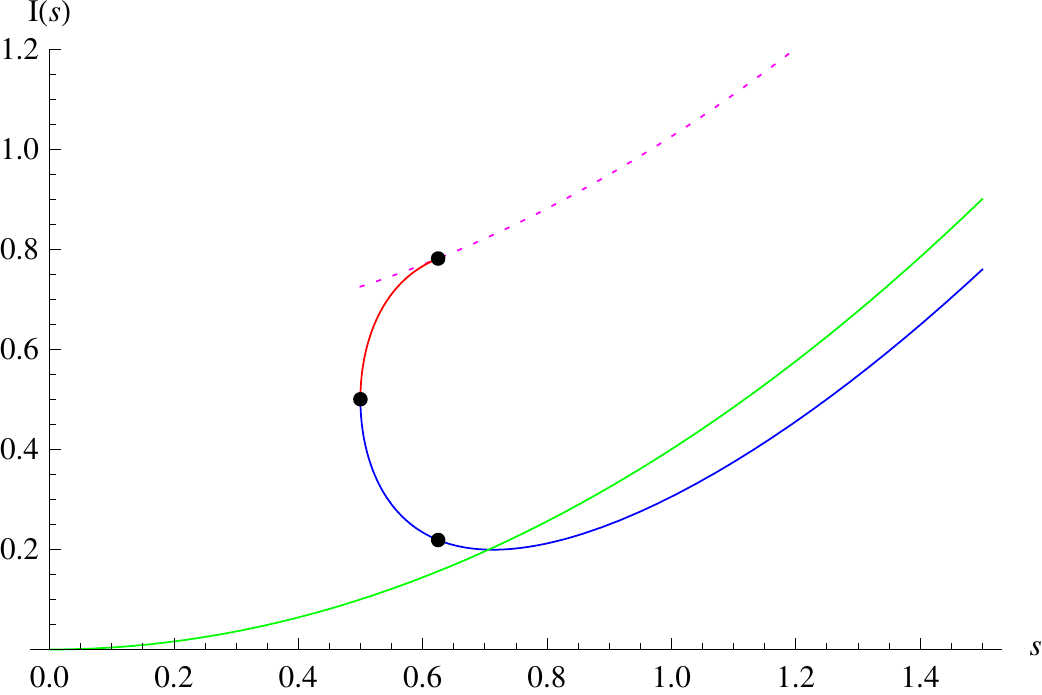} & \includegraphics[width=0.45\textwidth]{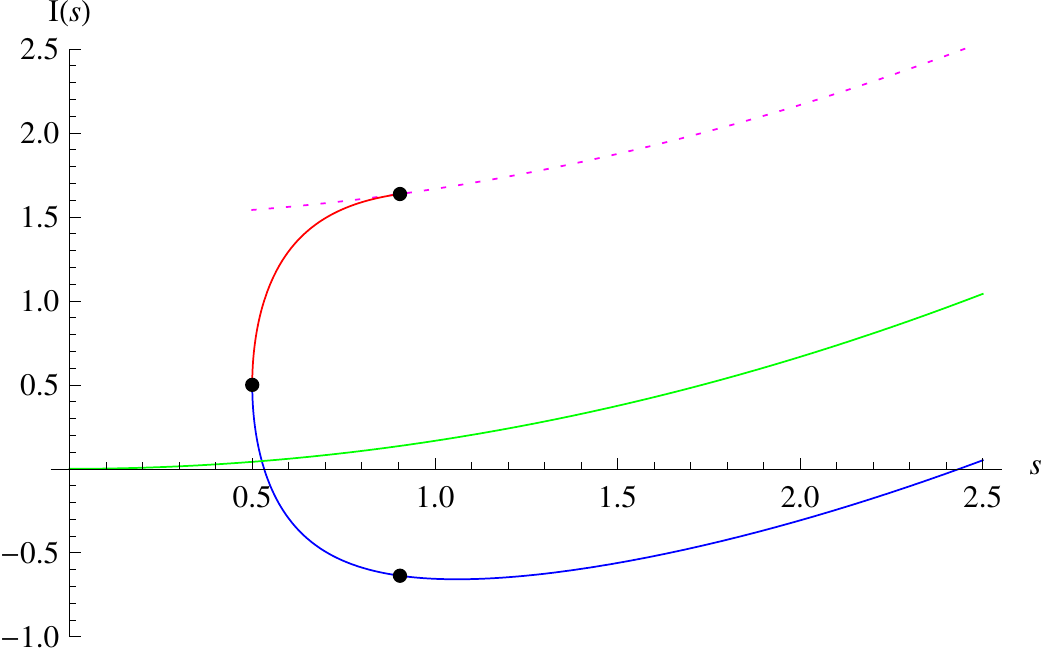} 
\end{tabular}
\caption{Plots of the free energies $I(s)$ of the different branches for $p=1,2,5,12$, respectively. 
The first plot is the free energy of the 1/2 BPS Taub-NUT-AdS solution. In the other plots the green curve is the  free energy 
$\tfrac{1}{p}I_\mathrm{NUT}$  of the Taub-NUT-AdS$/\Z_p$ solution, while the dotted line in magenta is the free energy $I_\mathrm{NUT+flux}^\mathrm{orb}$,
 including the contribution of $\pm\frac{p}{2}$ units of flux at the orbifold singularity. 
The red curve is the free energy  $I_{\mathrm{Bolt_-}}$ of the negative branch. The blue curve is the free energy $I_{\mathrm{Bolt_+}}$  of the positive branch. 
The free energies of the special solutions are marked with points.}
\label{freeplots2}
\end{figure}

In Figure \ref{freeplots2} we have plotted the holographic free energies for various values of $p$. 
The first plot is the free energy of the unique 1/2 BPS filling of the squashed $S^3$, with the marked point 
being the AdS$_4$ solution without gauge field. In the second plot $p=2$ and we see that the free energy 
of the positive branch bolt solution joins at $s=\frac{1}{2}$ to the free energy 
of  the orbifold Taub-NUT-AdS$/\Z_2$ solution with $1$ unit of flux at the singularity, as observed in \cite{Martelli:2011fw}. 
On the same plot the green curve is the  free energy of Taub-NUT-AdS$/\Z_2$, without any trapped flux.  
In the remaining two plots ($p=5$ and $p=12$ respectively) the negative branch bolt solutions appear. 
The curve of the free energy $I_{\mathrm{Bolt_-}}$ connects the free energy $I_\mathrm{NUT+flux}^\mathrm{orbifold}$ of the orbifold branch   with
the free energy  $I_{\mathrm{Bolt_+}}$ of the positive branch  at the values $s=s_p$ and $s=\frac{1}{2}$, respectively. 


\section{Regular 1/4 BPS solutions}
\label{1/4}

In this section we find all regular supersymmetric solutions satisfying the 1/4 BPS condition
(\ref{1/4BPS}). 
For all solutions the (conformal class of the) boundary three-manifold 
is again a biaxially squashed $S^3/\Z_p$ with metric (\ref{boundary3}), but now the 
boundary gauge field is given by
\bea\label{A3infinity1/4}
A^{(3)} &=& P \sigma_3 + A^{(3)}_{\mathrm{flat}} \ = \ - \frac{1}{2} (4s^2-1) \sigma_3 + A^{(3)}_{\mathrm{flat}}~,
\eea
where $A^{(3)}_{\mathrm{flat}}$ is again a certain flat connection. The latter is particularly important in order 
to globally have supersymmetry on the boundary in this case, precisely as for the 1/4 BPS Quaternionic-Eguchi-Hanson
solutions in section \ref{QEHinstanton}.
The boundary Killing spinor equation 
is (\ref{3dKSE1/4}), which we reproduce here for convenience:
\bea\label{barry1/4}
\left(\nabla_\alpha^{(3)} - \ii A_\alpha^{(3)} + \frac{\ii s}{2}\gamma_\alpha \right)\chi &=& 0~.
\eea
Again as in  section \ref{1/2}, a solution 
to the above boundary data with given $s$ is diffeomorphic to the same solution with $s\rightarrow -s$. 

As for the case of 1/2 BPS solutions, for fixed 
conformal boundary data we find more than one smooth supersymmetric 
filling, with different topologies. What is exceptional in the 1/4 BPS class of solutions
is that for an $S^3$ boundary the Taub-NUT-AdS solution is \emph{not} the unique filling, as one might expect, 
but rather there is also a filling with an $\mathcal{M}_1 = \mathcal{O}(-1) \rightarrow S^2$ topology. 
The full moduli space will be summarized in section \ref{1/4moduli}.

\subsection{Self-dual Einstein solutions}

The 1/4 BPS Einstein solutions were described in section \ref{selfdual}. 
For \emph{any} choice of conformal boundary data, meaning for all $p\in\mathbb{N}$ and
all choices of squashing parameter $s>0$, 
there exists the 1/4 BPS Taub-NUT-AdS solution on $\R^4/\Z_p$. 
This has metric (\ref{metricSD}), (\ref{OmegaSDagain}) and 
$\psi$ is taken to have period $4\pi/p$. This solution then has an isolated $\Z_p$ orbifold 
singularity at $r=s$ for $p>1$, or, removing the singularity, the topology is 
$\R_{>0}\times S^3/\Z_p$. In taking the $\Z_p$ quotient in this 
1/4 BPS case notice that in order to preserve supersymmetry we must also 
turn on an additional flat gauge field which is a connection 
on $\mathscr{L}^{-1}$. Here recall that $\mathscr{L}$ is the line bundle 
on $\R_{>0}\times S^3/\Z_p$ with torsion first Chern class 
$c_1(\mathscr{L})=1\in H^2(\R_{>0}\times S^3/\Z_p,\Z)\cong \Z_p$. 
The reason for this is as discussed for the Quaternionic-Eguchi-Hanson solutions 
in section \ref{QEHinstanton} -- the Killing spinors for the 1/4 BPS 
Taub-NUT-AdS solution are not invariant under $\mathcal{L}_{\partial_\psi}$, 
and the additional torsion gauge field is \emph{required} in order to have supersymmetry on the 
quotient space.

On the other hand, for each $p\geq 3$ and specific squashing parameter 
$s= s_p=\tfrac{p}{4\sqrt{p-1}}$ we also have the 1/4 BPS Quaternionic-Eguchi-Hanson solution. 
Thus for each $p\geq 3$ and $s=s_p$ there exist two supersymmetric self-dual Einstein
fillings of the same boundary data: the Taub-NUT-AdS solution on $\R^4/\Z_p$ and the 
Quaternionic-Eguchi-Hanson solution on $\mathcal{M}_p=\mathcal{O}(-p)\rightarrow S^2$. 
Again, the  boundary gauge field is important in comparing the 
\emph{global} boundary data for these two solutions, and the discussion is essentially 
the same as for the 1/2 BPS case in section \ref{sd2}. In fact the only difference
between the two cases is the additional contribution of $\mathscr{L}^{-1}$ described 
in the previous paragraph.

\subsection{Non-self-dual Bolt solutions}

\subsubsection{Regularity analysis}
\label{reg1/4}

We begin by analysing when the general metric in \eqref{solution} is regular, where for the 1/4 BPS class
the metric function $\Omega(r)=(r-r_1)(r-r_2)(r-r_3)(r-r_4)$ has roots
\bea\label{1/4rootsagain}
\left\{\begin{array}{c}r_4 \\ r_3\end{array}\right\} &=&   \s \pm \sqrt{\frac{-1+2Q+4\s^2}{2}} \nn~, \\
\left\{\begin{array}{c}r_2 \\ r_1\end{array}\right\} &=&  -\s \pm \sqrt{\frac{-1-2Q+4\s^2}{2}} ~.
\eea
Again, without loss of generality we may take the conformal boundary to be at $r=+\infty$.
A complete metric will then necessarily close off at the largest root $r_0$ of $\Omega(r)$, which 
must satisfy $r_0 \geq \s$. 
Given (\ref{1/4rootsagain}), the largest root is thus either $r_0=r_+$ or $r_0=r_-$, where
\bea
r_\pm &\equiv&  \pm \s + \sqrt{\frac{-1\pm2Q+4\s^2}{2}}~.
\eea

We first note that $r_0=r_\pm = \s$ leads only to the $Q=\mp \frac{1}{2}(4\s^2-1)$ Taub-NUT-AdS 
solutions considered in the previous section. Thus $r_0>\s$ and if $\psi$ has period $4\pi/p$ 
then the only possible topology is $\mathcal{M}_p=\mathcal{O}(-p)\rightarrow S^2$.  
Regularity of the metric near to the $S^2$ zero section at $r=r_0$ requires, as in the previous section,  
\bea\label{regularity1/4}
\left|\frac{r_0^2-\s^2}{s\Omega'(r_0)}\right| &=& \frac{2}{p}~.
\eea
Imposing \eqref{regularity1/4} at $r_0=r_+$ gives
\bea\label{Qvalue1/4plus}
Q  \ = \ \begin{cases} \ Q_+(\s)~,& \s>0  \\ \\ -Q^\pm_-(\s)~,&\s<0 \end{cases}~,
\eea
while for $r_0=r_-$ imposing \eqref{regularity1/4} gives
\bea\label{Qvalue1/4minus}
Q  \ = \ \begin{cases} \ Q^\pm_-(s)~,& \s>0   \\  \\ - Q_+(\s)~,&\s<0 \end{cases}~.
\eea
Here  we have defined
\bea
Q_+(\s) & \equiv & \frac{p^2 - (16\s^2-p)\sqrt{f^+_p(\s)}}{128\s^2}~, \nn \\
Q^\pm_-(\s) & \equiv & - \frac{p^2\mp(16s^2+p)\sqrt{f^-_p(\s)}}{128\s^2}~,
\eea
and have introduced the polynomials 
\bea
f^\pm_p(\s) \ \equiv \  (16 s^2 \pm p)^2-128 \s^2  ~.
\eea

Similarly to the 1/2 BPS solutions, notice that a solution with given $s>0$ will 
be equivalent to the corresponding solution with $s\rightarrow -\s<0$. 
This is because $r_+(\s) = r_-(-\s)$, which then 
leads to exactly the same set of roots in (\ref{1/4rootsagain}), and thus the same local metric. In addition, $P(-\s)=P(\s)$ and $Q_\pm(\s) = Q_\pm(-\s)$ and hence the gauge field in (\ref{solution}) is also invariant. Thus our parametrization of the roots in (\ref{1/4rootsagain})  is such that 
we need only consider $\s>0$, which we henceforth assume.

The putative largest root for \eqref{Qvalue1/4plus} and \eqref{Qvalue1/4minus}, respectively, is
\bea\label{rootschoice1/4}
r_+(Q=Q_+(\s)) \ = \ \frac{p+\sqrt{f^+_p(\s)}}{16\s} ~, \nn \\
r_-(Q=Q^\pm_-(\s)) \ = \ \frac{p\mp\sqrt{f^-_p(\s)}}{16\s}~.
\eea
The above expressions are real provided $f_p^\pm(\s)$ are positive semidefinite.

Recall that in order to have a smooth metric we require $r_0>\s$. Imposing this for $r_0=r_+(Q_+(\s))$ and $r_0=r_-(Q_-^\pm(\s))$ is equivalent to 
determining the range of $\s$ for which the functions 
\bea
a_p(\s) &\equiv& (p-16\s^2) + \sqrt{f_p^+(\s)}~, \nn \\
b^\pm_p(\s) &\equiv& (p-16\s^2)\mp\sqrt{f_p^-(\s)}~,
\eea
are strictly positive, respectively. In addition, we must 
verify that \eqref{rootschoice1/4} really \emph{is} the largest root. We thus define 
\bea
c_p(\s) & \equiv & r_+(Q_+(\s)) - r_-(Q_+(\s))~, \nn \\
d^\pm_p(\s) & \equiv & r_-(Q^\pm_-(\s)) - r_+(Q^\pm_-(\s))~,
\eea
and one finds
\bea
c_p(\s) \ = \ \frac{p + 16\s^2 + \sqrt{f_p^+} - \sqrt{(p-16\s^2 - \sqrt{f_p^+})^2 -4p^2}}{16\s}~, \\
d^\pm_p(\s) \ = \ \frac{p  - 16\s^2 \mp \sqrt{f_p^-} - \sqrt{(p+16\s^2 \pm \sqrt{f_p^-})^2 -4p^2}}{16\s}~.
\eea
Then \eqref{rootschoice1/4} is indeed the largest root provided also $c_p(\s)$ or $d_p^\pm(\s)$, respectively, is positive or complex. 

We are thus reduced to determining the subset of $\{\s>0\}$ for which $f_p^\pm(\s)$ is real and non-negative, and, respectively as appropriate, $a_p(s)$, $b^\pm_p(\s)$ are strictly positive and $c_p(s)$, $d_p^\pm(\s)$ are either strictly positive or complex. We refer to the two sign choices in $r_\pm$ as positive and negative branch solutions.
The behaviour for $p=1$ and $p=2$ is again qualitatively different from that with $p\geq 3$.

\subsection*{$p=1$}

\subsubsection*{Positive branch}

The polynomial $f_1^+(s)$ is positive semidefinite for $s \in (0,\frac{\sqrt{2}-1}{4}] \cup [\frac{\sqrt{2}+1}{4},\infty)$ but $a_1(s)$ is positive only for $s\in(0,\frac{\sqrt{2}-1}{4}]$. In this range $(1-16\s^2 - \sqrt{f_1^+})^2-4$ is negative and so $c_1(\s)$ is complex; hence $r_+(Q_+(\s))$ is indeed the largest root of $\Omega(r)$. In conclusion, for $s\in(0,\frac{\sqrt{2}-1}{4}]$ and $Q=Q_+(s)$ we have a regular 1/4 BPS solution on $\mathcal{M}_1 = \mathcal{O}(-1) \rightarrow S^2$.

\subsubsection*{Negative branches}

The polynomial $f_1^-(s)$ is positive semidefinite for $s \in (0,\frac{\sqrt{3}-\sqrt{2}}{4}] \cup [\frac{\sqrt{2}+\sqrt{3}}{4},\infty)$ but $b^\pm_1(s)$ is positive only for $\s\in(0,\frac{\sqrt{3}-\sqrt{2}}{4}]$. In this range $(1+16\s^2 \pm \sqrt{f_1^-})^2 -4$ is negative and so $d^\pm_1(\s)$ is complex; hence $r_-(Q^\pm_-(\s))$ is indeed the largest root of $\Omega(r)$. In conclusion, for $\s\in(0,\frac{\sqrt{3}-\sqrt{2}}{4}]$ and $Q=Q^\pm_-(s)$ we have two regular 1/4 BPS solutions on $\mathcal{M}_1 = \mathcal{O}(-1) \rightarrow S^2$.

\subsection*{$p=2$}

\subsubsection*{Positive branch}
For $p=2$ the expressions for $r_+(Q_+(s))$ and $Q_+(s)$ simplify to
\bea
r_+(Q_+(s)) & = & \frac{1}{4s} - s~,~~~~~Q_+(s) \  = \  \frac{1}{16s^2} - \frac{1}{2} + 2s^2~.
\eea
The above values satisfy \eqref{regularity1/4} for $s \in (0,\frac{1}{2\sqrt{2}})$. In this range $a_2(s)$ is positive while  $c_2(s) $ is complex, {\it i.e.}\ $r_+(Q_+(s))$ is indeed the largest root of $\Omega(r)$. In conclusion, for $s\in(0,\frac{1}{2\sqrt{2}})$ and $Q=Q_+(s)$ we have a regular 1/4 BPS solution on $\mathcal{M}_2 = \mathcal{O}(-2) \rightarrow S^2$. In the limit $s=\frac{1}{2\sqrt{2}}$, the root $r_+(Q_+(s)) = s = \frac{1}{2\sqrt{2}}$ which corresponds to a Taub-NUT solution.

\subsubsection*{Negative branches}
The polynomial $f_2^-(s)$ is positive semidefinite for $s \in (0,\frac{2-\sqrt{2}}{4}] \cup [\frac{\sqrt{2}+2}{4},\infty)$ but $b^\pm_2(s)$ is positive only for $s\in(0,\frac{2-\sqrt{2}}{4}]$.  In this range $(2+16\s^2 \pm \sqrt{f_2^-})^2 -16$ is negative and so $d^\pm_2(s)$ is complex. In conclusion, for $s\in(0,\frac{2-\sqrt{2}}{4}]$ and $Q=Q^\pm_-(s)$ we have two regular 1/4 BPS solutions on $\mathcal{M}_2 = \mathcal{O}(-2) \rightarrow S^2$.

\subsection*{$p\geq3$}

\subsubsection*{Positive branch}

The polynomial $f_p^+(s)$ is positive definite for all $s>0$ since it has imaginary roots and $a_p(s)$ is also positive for all $s>0$. In this range $c_p(s)$ is positive and hence $r_+(Q_+(s))$ is indeed the largest root of $\Omega(r)$. In conclusion, for $s>0$ and $Q=Q_+(s)$ we have a regular 1/4 BPS solution on $\mathcal{M}_p = \mathcal{O}(-p) \rightarrow S^2$.

\subsubsection*{Negative branches}

The polynomial $f_p^-(s)$ is positive semidefinite for $s \in (0,\frac{\sqrt{2+p}-\sqrt{2}}{4}] \cup [\frac{\sqrt{2}+\sqrt{2+p}}{4},\infty)$ but $b^\pm_p(s)$ is positive only for $s\in(0,\frac{\sqrt{2+p}-\sqrt{2}}{4}]$. In this range $d^\pm_2(s)$ is positive and hence $r_-(Q^\pm_-(s))$ is indeed the largest root of $\Omega(r)$. In conclusion, for $s\in(0,\frac{\sqrt{2+p}-\sqrt{2}}{4}]$ and $Q=Q^\pm_-(s)$ we have two regular 1/4 BPS solutions on $\mathcal{M}_p = \mathcal{O}(-p) \rightarrow S^2$.

\

It is important to remark that these various branches of solutions really \emph{are} distinct 
solutions. In particular, one should verify that the two negative branch solutions 
are not diffeomorphic. We have  checked this is this case by comparing 
the value of the square of the Ricci tensor $R_{\mu\nu}R^{\mu\nu}$
evaluated on the bolt $S^2$ at $r=r_0$ (this may be defined in a coordinate-independent manner 
as the fixed point set of $U(1)_r$, generated by $\partial_\psi$).  Indeed, one easily computes the general expression
\bea\label{Rsquare}
R_{\mu\nu}R^{\mu\nu} &=& 36 + \frac{4(P^2-Q^2)^2}{(r^2-s^2)^4}~.
\eea
It is a simple exercise to compute this at $r=r_0$ for the various cases, and 
check that the solutions we claim are distinct give distinct values of this 
curvature invariant on the bolt.

\subsubsection{Gauge field and spinors}

Let us now turn to analysing the global properties of the gauge field.
After a suitable gauge transformation, the latter can be written \emph{locally} as
\bea\label{noninstanton1/4}
A &=&  \frac{s}{r^2-s^2}\left[-2Qr-\frac{1}{2s}(r^2+s^2)(4s^2-1)\right]\, \sigma_3~.
\eea
In particular, this gauge potential is \emph{singular} on the $S^2$ at $r=r_0$, 
but is otherwise globally defined on the complement  $\mathcal{M}_p\setminus S^2$ of the bolt. 
The field strength $F=\diff A$ is easily verified to be a
globally defined smooth two-form on $\mathcal{M}_p$, with non-trivial flux 
through the  $S^2$ at $r=r_0$. Indeed, for $Q=Q_\pm(s)$ one 
computes the period through the $S^2$ at $r_0(s)=r_\pm(Q_\pm(s))$ (respectively) to be 
\bea\label{fluxy1/4}
\int_{S^2} \frac{F}{2\pi} &=& -\frac{2s}{r_0(s)^2-s^2}\left[-2Q_\pm(s)r_0(s)-\frac{1}{2}(r_0(s)^2+s^2)(4\s^2-1)\right]\nonumber\\
&=& \pm \frac{p}{2}-1~.
\eea
Thus the positive/negative branch solutions have a gauge field flux 
$\pm \frac{p}{2}-1$ through the bolt, respectively. 
Both branches then induce the \emph{same} spinors and global gauge field at conformal infinity, for fixed $p$ and $s$.
The factor of $-1$ in the quantization condition (\ref{fluxy1/4}) is precisely the same as for the 1/4 BPS Quaternionic-Eguchi-Hanson 
solutions (\ref{QEHflux}) in section \ref{QEHinstanton}, and its relation to having globally well-defined spinors 
on the conformal boundary, invariant under $\mathcal{L}_{\partial_\psi}$, is precisely 
the same as the discussion around equation (\ref{flateric}).

We note that the Dirac spin$^c$ spinors are smooth sections of the following 
bundles:
\bea\label{albert}
\begin{cases} \ \pi^*\left[\mathcal{O}(p-2)\oplus \mathcal{O}(0)\oplus\mathcal{O}(-2)\oplus\mathcal{O}(p)\right]& \quad \mbox{positive branch} \\ \ \pi^*\left[\mathcal{O}(-2)\oplus \mathcal{O}(-p)\oplus\mathcal{O}(-p-2)\oplus\mathcal{O}(0)\right] & \quad \mbox{negative branch}\end{cases}~.
\eea
When $p$ is even the boundary gauge field $A^{(3)}$ is a connection 
on $\mathscr{L}^{\frac{p}{2}-1}$, while when $p$ is odd it is a connection on 
$\mathscr{L}^{-1}$. The three-dimensional boundary spinors are 
correspondingly sections of $\mathscr{S}_0\otimes \mathscr{L}^{-1}$ 
and $\mathscr{S}\otimes\mathscr{L}^{-1}$, respectively (see appendix \ref{spinc}).

\subsubsection{Special solutions}

For $p<3$ the positive branches described in section \ref{reg1/4} terminate at $s=\frac{\sqrt{2}-\sqrt{2-p}}{4}$ while for $p\geq3$ the 
positive branch exists for all $s>0$, but there are special solutions at $s=\tfrac{1}{2}$ and $s=s_p$. The negative branches  terminate at $\frac{\sqrt{p+2}-\sqrt{2}}{4}$ for all $p$. In this section we describe these various special and/or limiting solutions.

\subsubsection*{Positive branches}

For $p=1$ the positive branch exists for $s\in (0,\tfrac{\sqrt{2}-1}{4}]$. As usual the 
$s=0$ limit is singular, but the terminating solution with $s=\tfrac{\sqrt{2}-1}{4}$ 
is a regular solution. At this value of $s$ we have $f^+_1(s)=0$, although 
we have not found an invariant geometric interpretation of this characterization of the solution. 
For $p=2$ the positive branch exists for  $s\in (0,\tfrac{1}{2\sqrt{2}})$, but 
here the terminating solution in the limit $s\rightarrow \tfrac{1}{2\sqrt{2}}$ 
 degenerates to the Taub-NUT-AdS$/\Z_2$ solution, 
which of course has an orbifold singularity. Thus for $p=2$ the positive branch 
joins onto the Taub-NUT-AdS$/\Z_2$ solutions. Notice that, in contrast
to the 1/2 BPS case, here the limiting Taub-NUT-AdS$/\Z_2$ solution has 
zero torsion, since $\frac{p}{2}-1=0$ when $p=2$.

For $p\geq 3$ the positive branch exists for all $s>0$, but there are some notable 
special solutions on this branch. Firstly, $s=\frac{1}{2}$ leads to a round 
metric on $S^3/\Z_p$, and thus this solution is a ``round Lens filling solution'', 
as dubbed in section \ref{1/2}. However, while for the 1/2 BPS solutions the 
round Lens filling solutions were terminating solutions that joined together the positive
and negative branches, here it appears as a special point on the positive branch. 
Of course, it is not a surprise to see the self-dual Quaternionic-Eguchi-Hanson solution 
arise from the special value $s=s_p=\frac{p}{4\sqrt{p-1}}$, and this is another 
special solution on the $p\geq 3$ 1/4 BPS  positive branch.

\subsubsection*{Negative branches}

The negative branches  terminate at $s=\frac{\sqrt{p+2}-\sqrt{2}}{4}$ for all $p\geq 1$. 
At this value of $s$ we have $f^-_p(s)=0$, and in fact the two negative branches 
become identical at this point, and thus \emph{join together}. Again, 
we have not found a geometrical characterization of the condition that 
$f^-_p(s)=0$. Notice that for $p\geq 10$ we have
 $\tilde s_p\equiv (\sqrt{p+2}-\sqrt{2})/4>1/2$, and therefore there exist 
two additional  round Lens filling solutions on the negative branches. These are 
distinct solutions, as follows by comparing the curvature invariant (\ref{Rsquare}) on the bolt $S^2$.

\subsection{Moduli space of solutions}
\label{1/4moduli}

We have summarized the even more intricate branch structure of the 1/4 BPS solutions in Figure \ref{14BPS}.
In general the conformal boundary has biaxially squashed $S^3/\Z_p$ metric (\ref{boundary3}), 
with squashing parameter $s>0$, and boundary gauge field given by (\ref{A3infinity1/4}). 
The 1/4 BPS fillings of this boundary may then be summarized as follows:

\begin{itemize} 
\item For $p=1$, the boundary $S^3$ 
with arbitrary squashing parameter $s>0$ always has the Taub-NUT-AdS 
solution as filling, but for $s\in (0,\frac{\sqrt{2}-1}{4}]$ 
there is also a smooth positive branch solution with topology 
$\mathcal{M}_1=\mathcal{O}(-1)\rightarrow S^2$, while for 
$s\in (0,\frac{\sqrt{3}-\sqrt{2}}{4}]$ there are \emph{two} negative 
branch solutions (which are connected to each other) of the same topology. The Taub-NUT, positive, and negative branch solutions 
are disconnected from each other; this in fact had to be the case, as we shall see in the next 
section that they have different constant free energy. Notice that the 
$s=\frac{1}{2}$  AdS$_4$ metric sits on the Taub-NUT-AdS 
branch. 

\item For $p\geq 2$ and arbitrary squashing parameter $s>0$ we always have the (mildly singular) 
Taub-NUT-AdS$/\Z_p$ solution. Thus for all boundary data there always 
exists a gravity filling, provided one allows for orbifold singularities. 

\item For $p=2$
there is a positive branch filling for $s\in (0,\frac{1}{2\sqrt{2}})$ with 
topology $\mathcal{M}_2=\mathcal{O}(-2)\rightarrow S^2$. This joins 
onto the Taub-NUT-AdS$/\Z_2$ branch at $s=\frac{1}{2\sqrt{2}}$, and 
we shall indeed see that these have the same free energy. Notice that, since 
$\frac{p}{2}-1=0$ for $p=2$, the gauge field is a connection on a trivial line bundle.
For $s\in (0,\frac{2-\sqrt{2}}{4}]$  there are again two negative 
branch solutions. These are connected to each other, but disconnected 
from the positive branch and Taub-NUT-AdS branch. 

\item For all $p>2$ and $s>0$  there exists a positive 
branch   filling with topology $\mathcal{M}_p=\mathcal{O}(-p)\rightarrow S^2$. 
This includes  the Quaternionic-Eguchi-Hanson solution at the specific value 
$s=s_p=\frac{p}{4\sqrt{p-1}}$, and the round Lens filling solution at $s=\frac{1}{2}$. 
However, this positive branch is disconnected from the Taub-NUT-AdS branch. 
For $s\in (0,\frac{\sqrt{p+2}-\sqrt{2}}{4}]$  there are again two negative 
branch solutions, which are connected to each other but disconnected 
from the positive branch and Taub-NUT-AdS branch. For $p\geq 10$ there exist 
two additional distinct round Lens filling solutions on the negative branches.

\begin{figure}[ht!]
\centering
\includegraphics[width=0.8\textwidth]{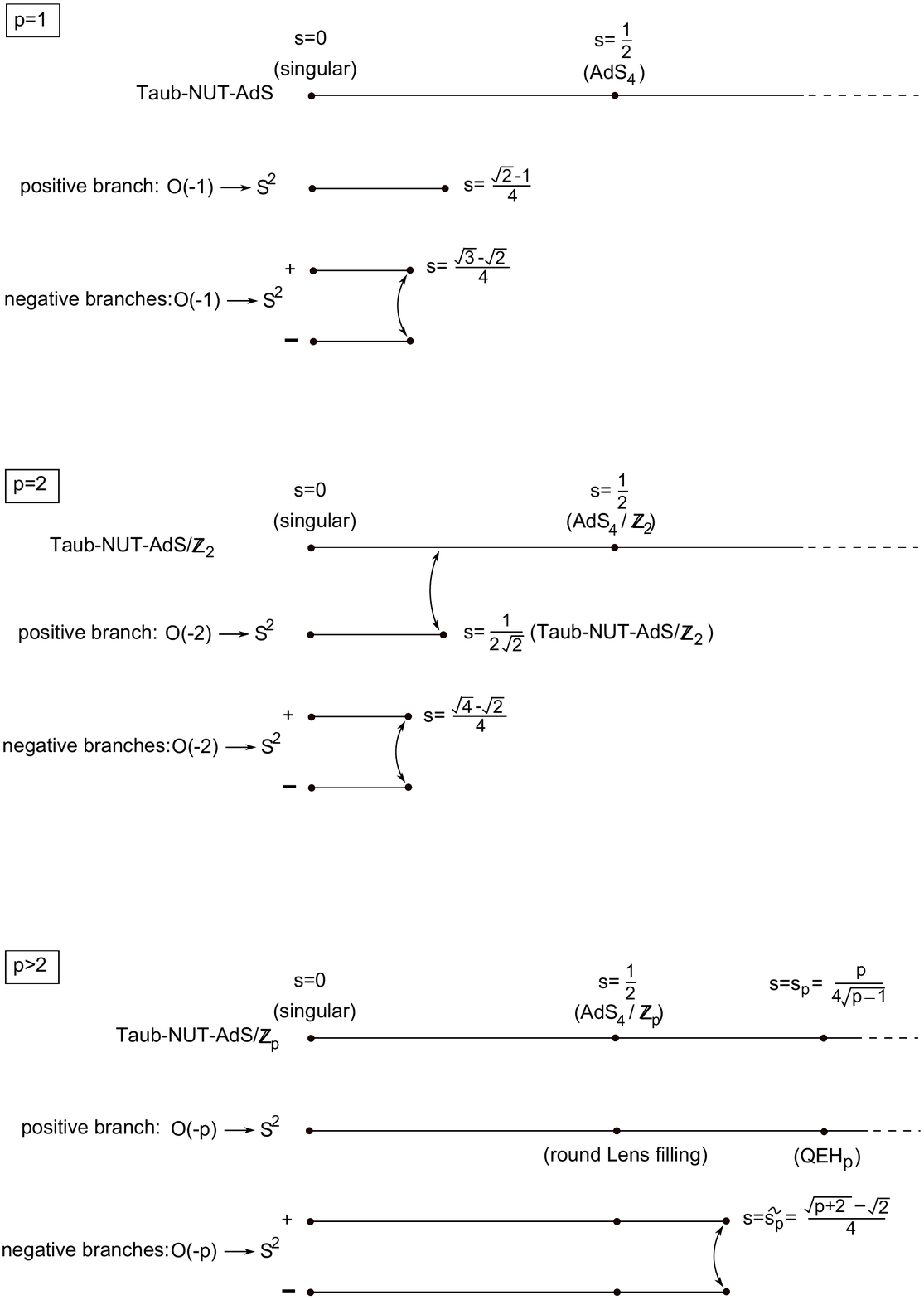}
\caption{The moduli space of 1/4 BPS solutions with biaxially squashed $S^3/\Z_p$ boundary, with squashing parameter 
$s$. The arrows denote identification of solutions on different branches. Notice that these moduli spaces 
are generally disconnected, as follows from the fact that the free energies 
are different. Note also that the negative branches extend past 
the round Lens filling solutions at $s=\frac{1}{2}$ only when $p\geq 10$ (which is the case plotted).}
\label{14BPS}
\end{figure}

\end{itemize}

\subsection{Holographic free energy}
\label{free1/4}

In this subsection we compute the holographic free energy of the 1/4 BPS solutions summarized above. 
This follows similarly section \ref{free1/2}, thus we will be more brief. 
Again we refer to appendices \ref{app:FreeEnergy} and \ref{proofising} for  further details.
We compute
\bea
I^{\text{grav}}_{\text{bulk}} &=& \frac{1}{8\pi G_4}\frac{16\pi^2}{p}\left.( 2 s r^3 - 6 s^3 r) \right |_{r_0}^\rcut ~, \\
I^{\text{grav}}_{\text{ct}} + I^{\text{grav}}_{\text{bdry}} &=&  \frac{1}{8\pi G_4} \frac{16\pi^2}{p} [4 Q s^2  -2 s \rcut^3 + 6s^3 \rcut + \mathcal{O}(\rcut^{-1})] ~,
\eea
where $r_0=r_\pm$ is the appropriate  largest root of $\Omega(r)$, where the manifold closes off. 
Removing the cut-off $\rcut \to \infty$ 
the divergent terms cancel. The contribution to the action from the bulk gauge field is as follows. 
For the NUT case $r_0=s$ and we have
\bea
I^F_{\mathrm{NUT}} \ = \ \frac{16\pi^2}{8\pi G_4} Q^2 \ = \ \frac{2\pi}{G_4} \frac{(1-4s^2)^2}{4}~,
\eea
while for the Taub-Bolt-AdS cases $r_0>s$ and we have
\bea
I^F_{\mathrm{Bolt}} \ = \ \frac{1}{8\pi G_4}\frac{16\pi^2}{p}  
                   \frac{s r_0\left[-4Q(1-4s^2)(r_0+s)^2 + (r_0^2+s^2)(2Q + 1-4s^2)^2\right]}{2(r_0^2-s^2)^2}~.
\eea
Combining all the above contributions to the action we obtain 
the following remarkably simple expressions
\begin{equation}\label{14free}
\begin{aligned}
I_{\mathrm{NUT}}   &\ = \ \frac{1}{2}\frac{\pi}{G_4} \qquad  \qquad  (p=1)~, \\
I_{\mathrm{Bolt\pm}} &\ = \ \frac{4\mp p}{8}\frac{\pi}{G_4}  \qquad  ~(p\geq 2)~.\\
\end{aligned}
\end{equation}
Again,  $I_{\mathrm{Bolt_\pm}}$ refers to the free energies of the positive and negative 
branch solutions, respectively. In particular, the two distinct (non-diffeomorphic) 
negative branches  in fact have the \emph{same} free energy, that we denote $I_{\mathrm{Bolt_-}}$.

\begin{figure}[t!]
\begin{tabular}{cc}
\includegraphics[width=0.45\textwidth]{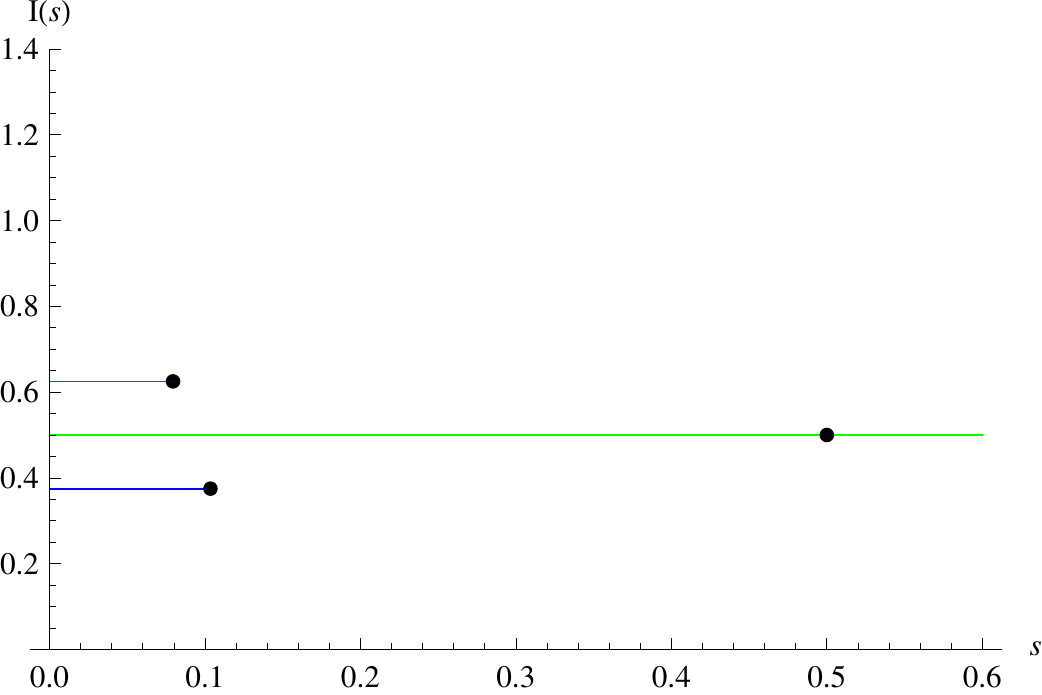} & \includegraphics[width=0.45\textwidth]{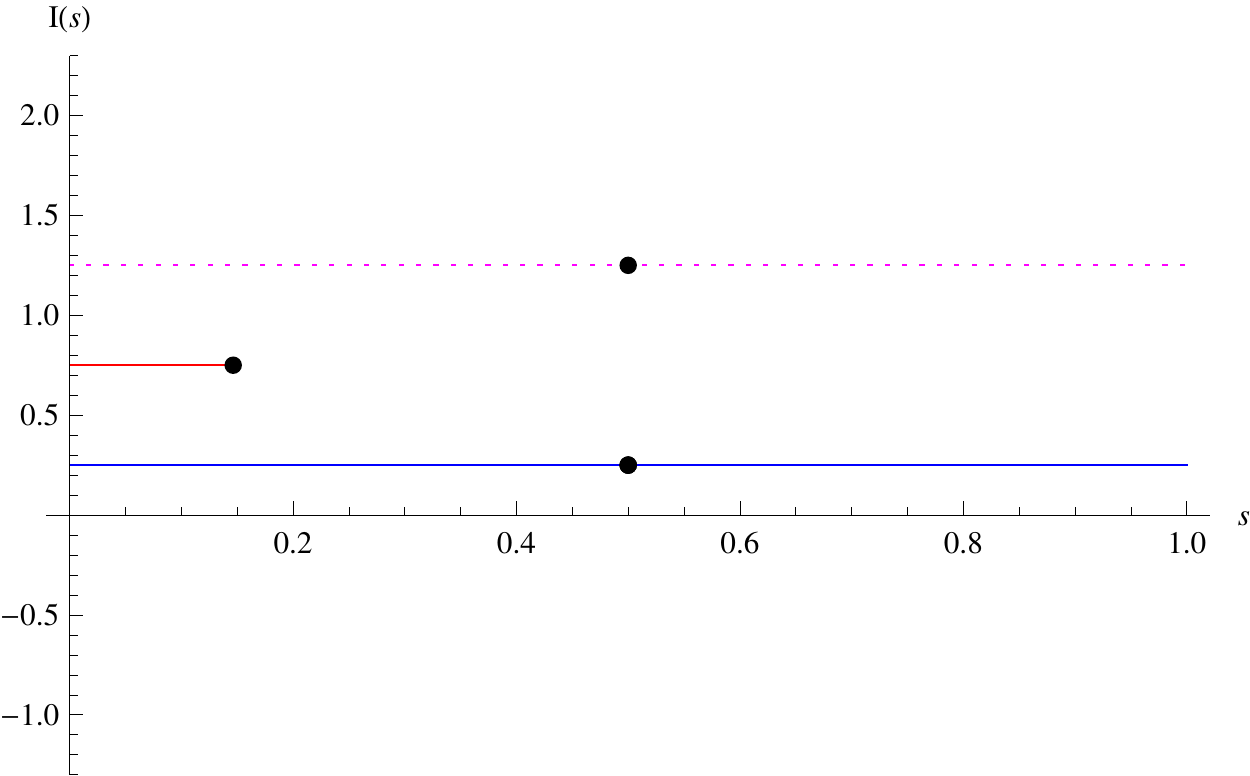} \\
\includegraphics[width=0.45\textwidth]{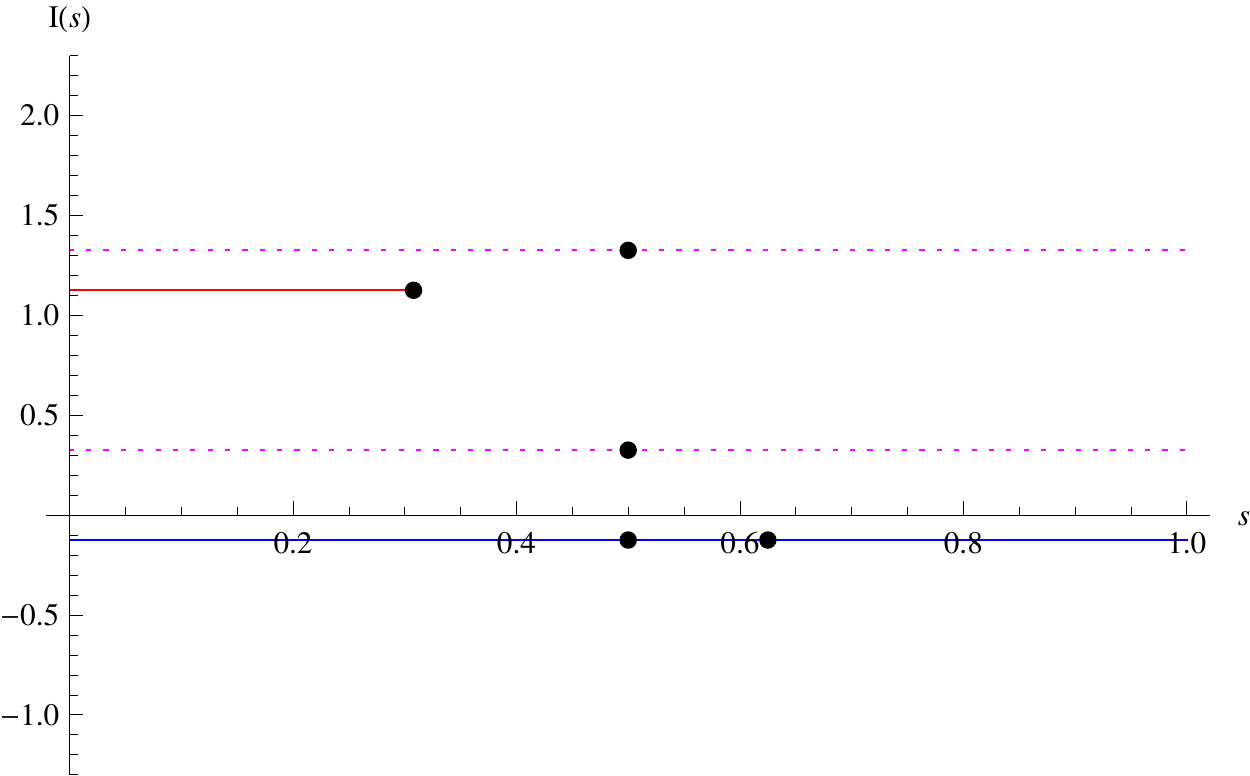} & \includegraphics[width=0.45\textwidth]{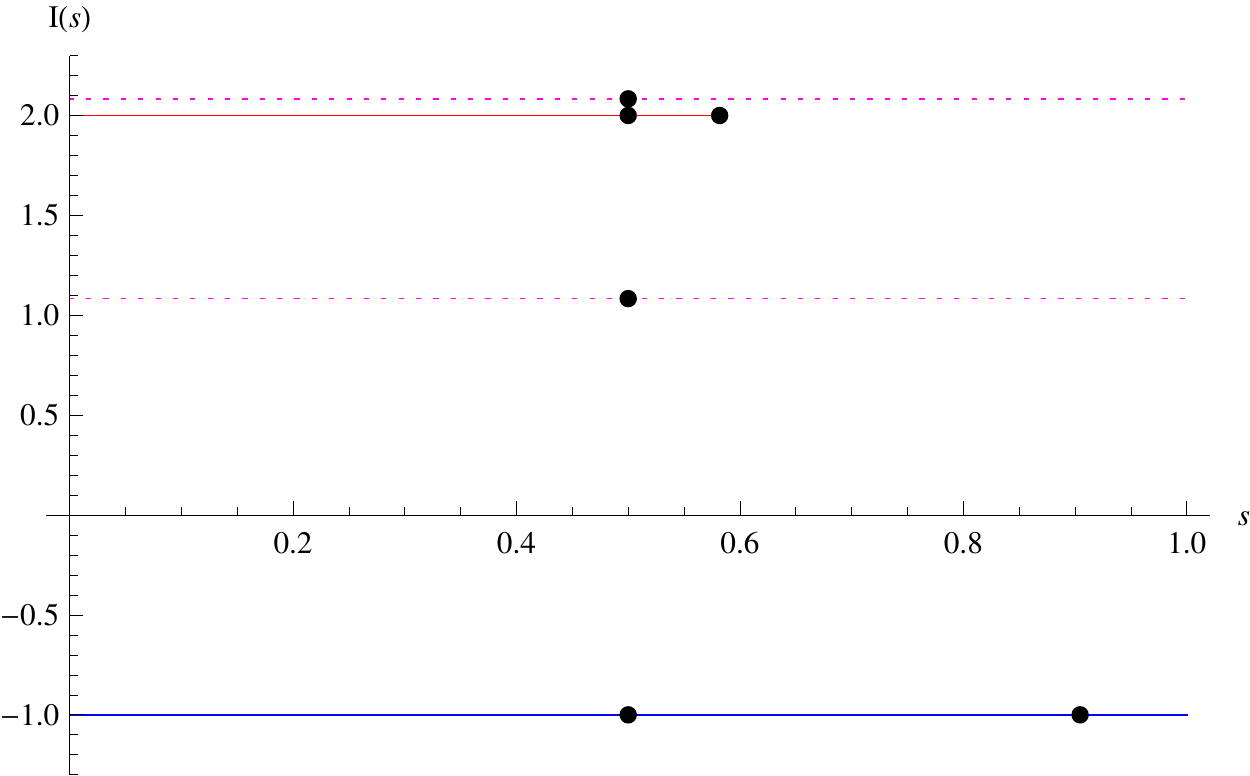} 
\end{tabular}
\caption{Plots of the free energies $I(s)$ of the different branches for $p=1,2,5,12$, respectively. 
The dotted lines in magenta are the free energies 
$I_\mathrm{NUT+flux_\pm}^\mathrm{orb}$,  including the contribution of $\pm\frac{p}{2}-1$ units of flux at the orbifold singularity. 
The red lines are the free energies  $I_{\mathrm{Bolt_-}}$ of the negative branches. The blue lines are the free energies 
$I_{\mathrm{Bolt_+}}$  of the positive branches. 
The  special solutions are marked with points.}
\label{freeplots4}
\end{figure}

As for the 1/2 BPS solutions, for any $p\geq 2$ we can fill the boundary squashed Lens space $S^3/\Z_p$
with the 1/4 BPS Taub-NUT-AdS/$\Z_p$ solution, where $\Z_p$ acts on the coordinate $\psi$.
Here we must consider more specifically the orbifold NUT 
solutions with $\pm \frac{p}{2} -1$ units of  flux trapped at the orbifold singularity, 
as a direct quotient of the Taub-NUT-AdS solution is not supersymmetric.
The latter solutions 
have the same global boundary data as the Taub-Bolt-AdS solutions, 
and in particular the trapped flux induces the same topological class of the gauge field 
on the conformal boundary $S^3/\Z_p$.
Using the result of appendix \ref{proofising} we compute the total action 
\bea
 I_\mathrm{NUT+flux_\pm}^\mathrm{orbifold} & = &  \frac{1}{p} I_{\mathrm{NUT}} +  I_\mathrm{sing}  \ = \  
\left( \frac{1}{2p} + \left(\frac{p}{2}\mp 1\right)^2\frac{1}{2p} \right)   \frac{\pi}{G_4}~,
\eea
where  in this case we obtain two different values 
depending on the sign of the flux. In Figure \ref{freeplots4} we plotted the holographic free energies for various values of $p$. 
The most striking feature is that we now have \emph{four} distinct smooth supergravity solutions 
filling a squashed $S^3$ boundary ($p=1$).   The corresponding free energies are shown in the first plot. 


\section{M-theory solutions and holography}
\label{liftingit}

In this section we discuss how the four-dimensional supergravity solutions uplift to 
solutions of eleven-dimensional supergravity. The full eleven-dimensional 
solution will take the form of a fibration over $\mathcal{M}^{(4)}$, where 
the fibres are copies of the internal space $Y_7$. The choice of the latter 
determines the field theory dual that is defined on the biaxially squashed 
$S^3/\Z_p$ conformal boundary of $\mathcal{M}^{(4)}$. Recall 
that for all solutions the four-dimensional gauge field $A$ satisfies 
the quantization condition for a spin$^c$ gauge field, and in particular 
$2A$ is always a connection on a line bundle $L$ over $\mathcal{M}^{(4)}$.
As we shall see, the Taub-NUT-AdS 
solutions may \emph{always} be uplifted to global supersymmetric
M-theory solutions, for any choice of internal space $Y_7$, and in this case we are able to compare 
the free energies computed in sections \ref{1/2} and \ref{1/4} to 
corresponding large $N$ field theory results, and find agreement in section \ref{agree}. 
An important point here is that the Taub-NUT-AdS solutions have 
topology $\mathcal{M}^{(4)}\cong \R^4$, so that the line bundle 
${L}$ is necessarily topologically trivial, {\it i.e.} the 
four-dimensional graviphoton $A$ is globally a one-form on $\mathcal{M}^{(4)}$.
However, as soon as $c_1({L})\in H^2(\mathcal{M}^{(4)},\Z)$ is non-zero 
this puts constraints on the possible choices of $Y_7$ -- 
this is the case for almost\footnote{\label{exception}The exception is the 1/4 BPS 
positive branch solution with $p=2$, which is the only case where 
$A$ is globally a one-form on $\mathcal{M}^{(4)}$. This then also uplifts 
for \emph{any} choice of internal space $Y_7$. However, notice that 
the free energy (\ref{14free}) of this solution is equal to the free energy of AdS$_4/\Z_2$, which 
has the same global boundary conditions.} all of the Taub-Bolt-AdS solutions, and even the Taub-NUT-AdS$/\Z_p$ 
solutions if they have non-trivial flat connections turned on.\footnote{\label{morefoot}As we shall see, 
in general the uplifting to eleven-dimensions involves not $L$, but rather $L^{\lambda/2}$ for some 
rational $\lambda\in\mathbb{Q}$. Since $c_1(L\mid_{S^3/\Z_p})\in \Z_p\cong H^2(S^3/\Z_p,\Z)$ is always torsion when restricted to the boundary $S^3/\Z_p$, this 
will be crucial when we come to ask which solutions have the same \emph{global} boundary conditions.}

This may be rephrased as follows. Given 
any supersymmetric field theory with an AdS$_4\times Y_7$ gravity dual, 
this field theory may also be put on the biaxially squashed $S^3$, preserving 
1/2 or 1/4 supersymmetry. 
Any such field theory then has a Taub-NUT-AdS filling as a gravity dual, 
of the form $\mathcal{M}^{(4)}\times Y_7$ where $\mathcal{M}^{(4)}$ 
is the Taub-NUT-AdS solution with appropriate 1/2 BPS or 1/4 BPS instanton, respectively. 
However, only a certain class of field theories, meaning only certain choices of 
$Y_7$, has in addition the 1/4 BPS Taub-Bolt-AdS 
filling of section \ref{1/4}. Similar comments apply to the case of the biaxially squashed 
Lens spaces $S^3/\Z_p$.
 We shall describe some choices of corresponding $Y_7$
in section \ref{comments}, and comment on the dual field theories. 

\subsection{Lifting NUTs}

As shown in \cite{Gauntlett:2007ma}, any supersymmetric solution to $d=4$, $\mathcal{N}=2$ gauged 
supergravity theory uplifts \emph{locally} to a supersymmetric solution of $d=11$ supergravity. More precisely, given
any Sasaki-Einstein seven-manifold $Y_7$ with contact one-form $\eta$, transverse K\"ahler-Einstein metric
$\diff s^2_T$ and with the seven-dimensional metric normalized so that $R_{ij} = 6g_{ij}$, we have
the uplifting ansatz\footnote{A caveat here is that the uplifting formulae above 
were shown in \cite{Gauntlett:2007ma} in Lorentzian signature.  Passing to 
Euclidean signature does not affect this at the level of equations of motion. Global aspects of the 
eleven-dimensional Killing spinors are discussed in appendix \ref{11spinor}.} 
\bea\label{uplift}
\diff s^2_{11} &=& R^2\left(\frac{1}{4}\diff s^2_4 + \left(\eta+\frac{1}{2}A\right)^2+\diff s^2_T\right)~,\nonumber\\
G &=& R^3\left(\frac{3}{8}\mathrm{vol}_4 -\frac{1}{4}*_4 F\wedge \diff\eta\right)~.
\eea
Here $\diff s^2_4$ is the four-dimensional gauged supergravity metric on $\mathcal{M}^{(4)}$, with volume form 
$\mathrm{vol}_4$, and the radius $R$ is
\bea
R^6 &=& \frac{(2\pi\ell_p)^6N}{6\mathrm{Vol}(Y_7)}~,
\eea
where $N$ is the number of units of flux
\bea
N &=& \frac{1}{(2\pi\ell_p)^6}\int_{Y_7} *_{11}G~.\eea
The four-dimensional Newton constant is then given by
\bea\label{Newton}
\frac{1}{16\pi G_4} &=& N^{3/2} \sqrt{\frac{\pi^2}{32\cdot 27\mathrm{Vol}(Y_7)}}~.
\eea
In fact it was more generally conjectured in \cite{Gauntlett:2007ma} that given any $\mathcal{N}=2$ warped
AdS$_4\times Y_7$ solution of eleven-dimensional supergravity there is a consistent Kaluza-Klein
truncation on $Y_7$ to $d=4$, $\mathcal{N}=2$ gauged supergravity theory. Properties of
such general solutions have recently been investigated in \cite{Gabella:2012rc, Gabella:2011sg}, and we expect the contact
structure discussed in these references to play an important role in this truncation. In particular,
it was shown in \cite{Gabella:2011sg} that (\ref{Newton}) remains true in this more general setting, provided one
replaces the Riemannian volume $\mathrm{Vol}(Y_7)$ by the contact volume.

As a specific example we may consider simply $Y_7=S^7/\Z_k$, with the $\Z_k$ action 
along the Hopf fibre of $S^7$. In this case 
$\diff s^2_T$ is the usual Fubini-Study metric on $\mathbb{CP}^3$, and $\eta=\diff\Reeb + A_{\mathbb{CP}^3}$, 
where $\Reeb$ has period $2\pi/k$ and $\diff A_{\mathbb{CP}^3}$ is the K\"ahler form on 
$\mathbb{CP}^3$, normalized to have period $2\pi$ through the linearly embedded $\mathbb{CP}^1$.
In that case $\mathrm{Vol}(S^7/\Z_k)=\pi^4/3k$. Different choices of $Y_7$ correspond to 
differerent choices of Chern-Simons-matter theory on the squashed $S^3$, 
and there are by now many examples of dual pairs, including infinite families.

The Taub-NUT-AdS solutions have topology $\mathcal{M}^{(4)}\cong \R^4$, and then 
necessarily $A$ is globally a one-form on $\R^4$. It follows immediately from the uplifting formula (\ref{uplift}) 
that we obtain a globally supersymmetric eleven-dimensional solution, again 
of the product topology $\mathcal{M}^{(4)}\times Y_7$, for any choice 
of AdS$_4\times Y_7$ solution. Specifically, because $A$ is a global one-form on 
$\mathcal{M}^{(4)}$, the twisting $\eta+\tfrac{1}{2}A$ is topologically trivial. Notice also 
that there is no flux quantization condition on $G$, since $\diff\eta$ is exact.
Thus any supersymmetric field theory on $S^3$ with an AdS$_4\times Y_7$ dual also 
has, when the theory is put on the biaxially squashed $S^3$, a supersymmetric 
$(\mbox{Taub-NUT-AdS})\times Y_7$ dual, in both the 1/2 BPS and 1/4 BPS cases.\footnote{An interesting 
subtlety here is that when the squashing parameter $s$ satisfies $0<s<1/2$ the gauge 
field is in fact \emph{complex}. One then formally obtains a complex eleven-dimensional 
metric via (\ref{uplift}). This is the only case in which we obtain a non-real 
gauge field. 
} 
We may then compare the gravitational holographic free energies of these 
solutions to corresponding exact large $N$ field theory computations, 
which we will do in the next section.

\subsection{Comparison to field theory duals}
\label{agree}

The gravitational holographic free energies of the 1/2 BPS and 1/4 BPS Taub-NUT-AdS solutions were computed in sections 
\ref{free1/2} and \ref{free1/4}, respectively. The result is
\bea
I_{\mathrm{NUT}} &=& 
\begin{cases}       
\displaystyle\ \frac{\pi}{2G_4} & \quad \mbox{1/4 BPS}  \\[4mm]
\displaystyle\ \frac{(2s)^2\pi}{2G_4} & \quad \mbox{1/2 BPS}
\end{cases}~.
\eea
Using the formula (\ref{Newton}) for the four-dimensional Newton constant, we thus obtain
\bea\label{INUT}
I_{\mathrm{NUT}} &=& 
\begin{cases}       
\displaystyle\ \frac{\sqrt{2}\pi}{3}\sqrt{\frac{\mathrm{Vol}(S^7)}{\mathrm{Vol}(Y_7)}}N^{3/2} & \quad \mbox{1/4 BPS}  \\[4mm]
\displaystyle\ (2s)^2\frac{\sqrt{2}\pi}{3}\sqrt{\frac{\mathrm{Vol}(S^7)}{\mathrm{Vol}(Y_7)}}N^{3/2} & \quad \mbox{1/2 BPS}
\end{cases}~.
\eea

In fact the 1/2 BPS case was precisely studied by the authors in \cite{Martelli:2011fw}. 
In this case the biaxially squashed $S^3$ with metric (\ref{3metric}), boundary gauge field 
(\ref{3gaugefield}) and three-dimensional Killing spinor equation (\ref{3dKSE1/2}) 
was studied in \cite{Imamura:2011wg}. In the latter reference the authors 
showed that, for a large class of $\mathcal{N}=2$ Chern-Simons-quiver gauge theories, 
the leading large $N$ free energy is precisely $(2s)^2$ times the result 
for the round sphere (see equation (148) in \cite{Imamura:2011wg}). This is precisely what we obtain from the 
1/2 BPS Taub-NUT-AdS gravity solution (\ref{INUT}), which has the same conformal boundary data! 

In the 1/4 BPS case the boundary three-metric (\ref{3metric}) is the same 
as in the 1/2 BPS case, but the boundary gauge field (\ref{3gaugefield1/4}) 
and three-dimensional Killing spinor equation (\ref{3dKSE1/4}) are different. 
General $\mathcal{N}=2$ Chern-Simons-matter theories were studied on this 
biaxially squashed $S^3$ in \cite{Hama:2011ea}, and it was found that 
the partition function is \emph{independent of the squashing parameter}. 
This is an exact statement, valid for all $N$. This then precisely agrees 
with our large $N$ gravity result in (\ref{INUT}), where we find that 
the gravitational free energy is equal to the result for the round sphere 
with $s=\frac{1}{2}$. Thus the 1/4 BPS Taub-NUT-AdS solution 
reproduces the correct large $N$ free energy.\footnote{Notice 
that it is non-trivial that the final result is independent of the squashing parameter --
each term in the action depends on $s$, with the $s$-dependence only cancelling 
when all terms are summed.} Of course, this can only be regarded as a partial 
result at this stage, because in the 1/4 BPS case there is also the Taub-Bolt-AdS 
filling, with topology $\mathcal{M}_1=\mathcal{O}(-1)\rightarrow S^2$. We 
turn to these solutions next.

\subsection{Lifting bolts}\label{liftbolt}

The Taub-Bolt-AdS solutions certainly uplift \emph{locally} to 
 eleven-dimensional supersymmetric supergravity solutions via (\ref{uplift}). 
However, globally this uplifting ansatz is inconsistent 
unless one restricts the internal space $Y_7$ appropriately. In this 
section we explain this important global subtlety. This implies 
that only a restricted class of field theories have Taub-Bolt-AdS 
fillings, in addition to the universal Taub-NUT-AdS fillings described 
in the previous section.

The discussion that follows is entirely topological, and we may in fact treat 
all of the 1/2 BPS and 1/4 BPS cases simultaneously. Specifically, 
all that we shall need to know is that the topology of the Taub-Bolt-AdS 
solutions is $\mathcal{M}^{(4)}=\mathcal{M}_p\equiv \mathcal{O}(-p)\rightarrow S^2$, with the 
gauge field flux quantized as
\bea\label{boltperiod}
\int_{S^2}\frac{F}{2\pi} &=& \frac{n}{2}~.
\eea
In all cases $n\equiv p$ mod 2, which is equivalent to $A$ be a spin$^c$ gauge field, 
as discussed in detail in appendix \ref{spinc}.

For simplicity, we shall consider first the case of uplifting when the internal manifold $Y_7$ is a \emph{regular} Sasaki-Einstein manifold. 
 By definition this means
that $Y_7$ is the total space of a $U(1)$ principal bundle over a K\"ahler-Einstein six-manifold $B_6$ with 
metric $\diff s^2_T$. We may then write $\eta=\diff\Reeb + \sigma$, where standard formulae give $\diff\sigma=\rho/4$ where
$\rho$ is the Ricci-form on $B_6$. The canonical period for $\Reeb$ is then $2\pi/4$, which for a Sasaki-Einstein manifold 
with precisely two Killing spinors is also the smallest 
period compatible with supersymmetry: the Killing spinors on $Y_7$ are charged under the Reeb 
vector $\partial_\Reeb$, and taking $\Reeb$ to have period $2\pi/4m$ for any $m>1$ would lead to 
spinors that are not single-valued. When $\Reeb$ has period $2\pi/4$ $Y_7$ is in fact 
the total space of the $U(1)$ principal bundle associated to the anti-canonical line bundle over $B_6$.
On the other hand, $\Reeb$ can sometimes have \emph{larger} period. 
In fact
$Y_7$ is simply-connected if and only if $\Reeb$ has period $2\pi I/4$ where $I=I(B_6)\in \mathbb{N}$ is 
a positive integer called the \emph{Fano index} of $B_6$ \cite{Gauntlett:2006vf}. In particular, for $B_6=\mathbb{CP}^3$ we have $I(\mathbb{CP}^3)=4$, 
so that for $Y_7=S^7$ we must take $\Reeb$ to have period $2\pi$.\footnote{In this case the discussion 
of Killing spinors is somewhat modified compared with that for a generic Sasaki-Einstein manifold: $S^7/\Z_k$ 
preserves $\mathcal{N}=6$ supersymmetry for $k>2$.  The six Killing spinors here 
are invariant under the $\Z_k$ action for all $k\in\Z$.} In fact $I\in\{1,2,3,4\}$, with $I=4$ only 
for $B_6=\mathbb{CP}^3$.\footnote{For completeness we note that examples exist 
for all values of $I\in\{1,2,3,4\}$: $Y_7=V^{5,2}=SO(5)/SO(3)$ has $I=I(\mathrm{Gr}(5,2))=3$; $Y_7=Q^{1,1,1,}$ has $I=I(\mathbb{CP}^1\times\mathbb{CP}^1\times\mathbb{CP}^1)=2$; $Y_7=M^{3,2}$ has 
$I=I(\mathbb{CP}^1\times \mathbb{CP}^2)=1$.}

We may summarize the previous paragraph, then, by taking $\Reeb$ to have period $2\pi I/4k$, where 
$I=I(B_6)\in\{1,2,3,4\}$ is the Fano index, and the positive integer $k$ must then divide $I$ 
in order that the two $U(1)_R$ charged Killing spinors are single-valued. The number of cases is then very small.

The global restriction on the internal space $Y_7$ in the uplifting ansatz (\ref{uplift}) may then be understood 
by fixing a point in $B_6$ and looking at the corresponding circle bundle over the bolt  $S^2\subset \mathcal{M}^{(4)}$. 
Since $\Reeb$ has period $2\pi I/4k$, it follows from the connection term $\eta+\frac{1}{2}A$ appearing in the metric 
(\ref{uplift}) that we will obtain a well-defined circle bundle only if
\bea\label{fcc}
\frac{4k}{2I}\int_{S^2}\frac{F}{2\pi} &  = & m \ \in \ \mathbb{Z}
\eea
is an integer. Geometrically, this integer $m$ is (minus) the first Chern class of 
the circle bundle, with coordinate $\xi$, integrated over the bolt $S^2$. Recalling 
that $2A$ is a connection on what we called $L\rightarrow \mathcal{M}^{(4)}$, 
we thus see that the eleven-dimensional circle $\xi$ is twisted by 
the line bundle\footnote{Notice that this means the rational number $\lambda$ we alluded to in 
footnotes \ref{newfoot}  and \ref{morefoot} takes the value $\lambda = 2k/I$.} $L^{k/I}=\mathcal{O}(m)$ in general, 
rather than by $L$. When $k=I$ these are the same, which is precisely the case when the internal Sasaki-Einstein
manifold $Y_7$ is the $U(1)$ principal bundle associated  to the anti-canonical bundle over $B_6$.
Given (\ref{boltperiod}), the quantization condition (\ref{fcc}) is equivalent to
\bea\label{integers}
nk &=& m I~.
\eea
This necessary condition is then also sufficient for the eleven-dimensional metric (\ref{uplift}) 
to be globally well-defined. Specifically, the eleven-dimensional spacetime is by construction the total space 
of the circle bundle over $\mathcal{M}_p\times B_6$ with first Chern class 
$c_1=-m\Phi - \frac{k}{I}c_1(B_6)$, where $\Phi$ is the generator of 
$H^2(\mathcal{M}_p,\Z)\cong \Z$. In the $G$-flux in (\ref{uplift}) notice 
that now $\diff\eta$ is no longer an exact form on the eleven-dimensional 
spacetime. In fact its cohomology class is equal to the cohomology class 
of $-\frac{1}{2}F$. But then $*_4F\wedge F$ is proportional to the 
volume form on $\mathcal{M}_p$, which is exact on $\mathcal{M}_p$ 
and thus also is exact on the eleven-dimensional spacetime. 
It follows that there is no quantization condition on $G$. 
In appendix \ref{11spinor} we show that 
if the eleven-dimensional metric is regular then the 
eleven-dimensional geometry is always a spin manifold (for all $p$), and the eleven-dimensional Killing spinors are  
smooth and globally defined.

Taking $k=I$, which leads to the canonical period of $2\pi/4$ for $\Reeb$, we see that the condition (\ref{integers}) is always 
satisfied. Thus \emph{all} Taub-Bolt-AdS solutions can be uplifted for \emph{all} regular Sasaki-Einstein $Y_7$ 
with the canonical period of $2\pi/4$ for $\Reeb$. This is true for any $p$.
 Examples are then $Y_7=S^7/\Z_4$, $Y_7=V^{5,2}/\Z_3$, 
$Y_7=Q^{2,2,2}=Q^{1,1,1}/\Z_2$, and $Y_7=M^{3,2}$. In this case 
the Reeb $U(1)$ principal bundle, with fibre coordinate $\xi$, 
is twisted over the base spacetime $\mathcal{M}^{(4)}$ by the line bundle $L$.

However, more generally (\ref{integers}) leads to restrictions. Consider the case of $Y_7=S^7$, which 
has $I=4$ and $k=1$. It follows from (\ref{integers}) that $n$ is necessarily divisible by $4$. But recall
that $n\equiv p$ mod 2, so we see immediately that \emph{none} of the Taub-Bolt-AdS solutions 
with $p$ odd can be uplifted on the round seven-sphere! In particular, the 1/4 BPS 
Taub-Bolt-AdS solution that fills the squashed $S^3$ cannot be lifted on $S^7$ (nor can it be lifted on 
$S^7/\Z_2$, although from the previous paragraph it \emph{can} be lifted on $S^7/\Z_4$). 
Concretely, this means that \emph{the 1/4 BPS Taub-Bolt-AdS filling of the squashed $S^3$ does not exist for the ABJM theory}. 
We shall discuss this further in section \ref{comments} below. 
Other cases may be analysed similarly. For example, again taking $Y_7=S^7$, which 
has $I=4$ and $k=1$, the 1/2 BPS solutions have $n=\pm p$, which leads to the restriction 
$p=\pm 4m$, so that the 1/2 BPS Taub-Bolt-AdS solutions uplift on $S^7$ only if $p$ is divisible by 4.

Above we have focused on regular Sasaki-Einstein manifolds $Y_7$, but it is straightforward to 
extend this analysis.  \emph{Irregular} Sasaki-Einstein manifolds 
have $\partial_\Reeb$ with generically non-closed orbits. This means that
the coordinate $\Reeb$ is not periodically identified over a dense open 
subset of $Y_7$. On the other hand, the expression $\eta+\frac{1}{2}A$ 
defines a global one-form only if $\Reeb$ is periodically identified 
in $\eta=\diff\Reeb + \sigma$. Thus one can \emph{never}\footnote{However, see footnote \ref{exception}.} 
lift any of these bolt solutions on irregular Sasaki-Einstein manifolds.

Finally, we conclude this section by commenting on an equivalent 
way of seeing the restriction on $Y_7$, that perhaps more directly 
makes contact with the field theory dual description. 
For simplicity, we again take $Y_7$ to be a regular Sasaki-Einstein 
manifold with K\"ahler-Einstein base $B_6$, Fano index $I=I(B_6)$ 
and $\Reeb$ to have period $2\pi I/4k$. It follows that 
$Y_7$ is the unit circle bundle in $\mathcal{L}=\mathcal{K}^{k/I}$, where $\mathcal{K}$ denotes 
the canonical line bundle of $B_6$. In this notation, scalar BPS 
operators arise in the dual field theory from \emph{holomorphic functions} 
on the metric cone over $Y_7$. These correspond to holomorphic 
sections of  $\mathcal{L}^{-t}$, with $t\in \mathbb{N}$ a positive integer.
The R-charge of the holomorphic function is then proportional to $t$, namely 
$R=\lambda t = \frac{2k}{I}t$.
However, because of the twisting in (\ref{uplift}), these holomorphic functions 
 become tensored with sections of a line bundle over the $S^2$ bolt. 
Specifically, in its dependence on $\mathcal{M}_p\times B_6$, 
a holomorphic function with R-charge $\frac{2k}{I}t$ becomes a section of 
$\mathcal{L}^{-t}\otimes \mathcal{O}(tm)\cong \mathcal{L}^{-t}\otimes L^{tk/I}$, where 
the integer $m$ satisfies (\ref{integers}).
In the irregular case the holomorphic functions generically have 
irrational R-charges, which then do not lead to well-defined 
sections over the $S^2$ bolt.

\subsection{Comments on field theory duals}
\label{comments}

We have seen that the 1/4 BPS Taub-Bolt-AdS filling of the biaxially squashed $S^3$ uplifts 
on any regular Sasaki-Einstein manifold with period $2\pi/4$ for $\Reeb$. 
Examples are $Y_7=S^7/\Z_4, V^{5,2}/\Z_3, Q^{2,2,2}$ and $M^{3,2}$. 
Proposals for the corresponding field theory duals 
have been discussed in  \cite{Martelli:2008si, Hanany:2008cd, Martelli:2009ga, Jafferis:2009th, Aganagic:2009zk, Franco:2009sp, Benini:2009qs, Closset:2012ep}.
However,  the solution does \emph{not} lift on the simply-connected covering spaces in the first 
three examples. We begin this section by examining this $p=1$ case, noting that 
all  other Taub-Bolt-AdS solutions fill 
the biaxially squashed Lens spaces $S^3/\Z_p$ with $p>1$, and so far in the literature no one 
has studied $\mathcal{N}=2$ supersymmetric gauge theories in this setting: 
the 1/2 BPS and 1/4 BPS biaxially squashed spheres were studied in \cite{Hama:2011ea}, \cite{Imamura:2011wg}, and \emph{round} Lens spaces $S^3/\Z_p$ \emph{without torsion} gauge fields were studied 
in \cite{Alday:2012au}.

\subsubsection{$S^3$ boundary}

We first note that, thus far in the literature, 
the large $N$ limit of the partition function
 of the field theory models dual to $Q^{2,2,2}$ or $M^{3,2}$  has only been computed using
 an \emph{ad hoc} prescription \cite{Gang:2011jj}. 
The issue is that the proposed field theory duals 
for these Sasaki-Einstein manifolds are \emph{chiral}, meaning 
that the matter representation is not real, and this leads to a more 
complicated matrix model behaviour. In particular, it is possible 
that saddle points exist within these models corresponding to the 
Taub-Bolt-AdS solutions.

The $S^7/\Z_4$ case is also intriguing. Naively one might identify the field theory 
dual in this case with the ABJM model with $k=4$; afterall, the ABJM theory is a 
$U(N)_k\times U(N)_{-k}$ Chern-Simons-matter theory that is dual 
to the case $Y_7=S^7/\Z_k$. However, the problem is quite subtle. 
The central issue is that the $\Z_k\subset U(1)$ quotient 
in the ABJM theory generally leaves $\mathcal{N}=6$ supersymmetry 
unbroken, but the $U(1)$ R-symmetry that is being gauged when 
the theory is put on the squashed sphere corresponds to 
an $\mathcal{N}=2$ subalgebra of this $\mathcal{N}=6$. 
For the Taub-NUT-AdS solutions we may take 
$Y_7=S^7/\Z_k$ and identify the $\Reeb$ circle 
in the uplifting ansatz (\ref{uplift}) with 
a $U(1)_R\subset SO(6)$. Here the $\Z_k$ quotient is 
not contained in this $SO(6)$, where the latter rotates the $\mathcal{N}=6$ supercharges 
in the vector representation. 
We are then gauging the manifest $U(1)_R$ symmetry 
of the ABJM when viewed in $\mathcal{N}=2$ language. 
However, this does \emph{not} work for the 
Taub-Bolt-AdS solutions  on $\mathcal{M}_1=\mathcal{O}(-1)\rightarrow S^2$, because we are forced to 
take $\Reeb$ to have period $2\pi/4$, {\it i.e.} the Taub-Bolt-AdS 
solutions are then defined with internal space 
$Y_7=S^7/\Z_k\times \Z_4$. A dual field theory for the latter is 
then \emph{unknown} (it is not simply an orbifold of the 
ABJM theory). 

Of course, one might instead 
directly identify the $\Z_k$ quotient in the ABJM 
theory with with $\Reeb$ direction in the uplift (\ref{uplift}). 
This then forces $k=4$ for the Taub-Bolt-AdS solutions, 
and we are gauging a $U(1)_R$ symmetry that is \emph{not} 
contained in the manifest $\mathcal{N}=6$ supersymmetry 
of the ABJM theory with $k=4$. This statement might puzzle 
some readers, since in the literature it is claimed that 
the ABJM theory has $\mathcal{N}=6$ supersymmetry for all $k>2$, 
while only $k=1$ and $k=2$ have enhanced $\mathcal{N}=8$ 
supersymmetry. In fact this is incorrect, but subtly so. 
In fact there are 8 Killing spinors on $S^7/\Z_k$ for 
$k=1,2$ \emph{and} $k=4$, but for $k=4$ the 
2 additional Killing spinors are sections of a 
different spin bundle to the $\mathcal{N}=6$ Killing spinors 
that exist on $S^7/\Z_k$ for all $k$. Recall that 
spin bundles on a manifold $\mathcal{M}$ are in general classified by 
$H^1(\mathcal{M},\Z_2)$, and in the case at hand notice that 
$H^1(S^7/\Z_4,\Z_2)\cong \Z_2$. The $\mathcal{N}=6$ spinors 
are sections of one of these two spin bundles, while 
the $\mathcal{N}=2$ Killing spinors that exist 
when $S^7/\Z_4$ is viewed as a regular Sasaki-Einstein manifold 
over $\mathbb{CP}^3$ are sections of the other spin bundle.\footnote{The corresponding 
situation for $S^3/\Z_p$ is discussed at length in appendix \ref{spinc}.} 
Thus although there are 8 Killing spinors, there is \emph{not} 
an $SO(8)$ R-symmetry that rotates them. 
In the field theory we are then gauging this non-manifest 
$\mathcal{N}=2$ $U(1)_R$ symmetry that exists only when $k=4$, which seems rather 
hard to study in practice. 

The conclusion of this is that the internal spaces $Y_7$ for which 
the 1/4 BPS Taub-Bolt-AdS filling of $S^3$ exists do not currently have 
known field theory duals for which the large $N$ partition function computation is under 
good control: either the field theory models are chiral, and the large $N$ limit of the partition function
is correspondingly not well-understood, or no field theory model is currently known, 
or the field theory \emph{is} known and non-chiral, but we are gauging 
a classically non-manifest R-symmetry of that field theory. 

\subsubsection{$S^3/\Z_p$ boundary}

Let us now turn to the  Lens space solutions for $p>1$. Since in general there are 
a number of distinct cases to consider, we shall confine ourselves to commenting 
on what we believe are the more interesting cases/features. 

Let us first discuss the solutions relevant for the 
ABJM model: in this case $Y_7=S^7$ and correspondingly  
we have $I=4$, $k=1$, and hence $\lambda=\frac{1}{2}$. The latter is indeed the value of the R-charge of a chiral field in the ABJM field theory (these fields are usually called $A_1, A_2, B_1, B_2$), 
and the R-charges of gauge-invariant scalar chiral primary operators are $t/2$, where geometrically $t$ is the positive integer of section \ref{liftbolt} (these operators are constructed using monopole operators of zero R-charge). 
Let us focus on the 1/2 BPS class of M-theory solutions. In this case, the Taub-Bolt-AdS solutions 
 have \emph{globally distinct} boundary conditions, as M-theory solutions, from the corresponding Taub-NUT-AdS$/\Z_p$ solution. 
To see this, note that from (\ref{integers}), and using $n=\pm p$, we see that a 1/2 BPS Taub-Bolt-AdS solution uplifts on $S^7$ only if $p=4q$ is divisible by 4. In this case, $S^7$ is fibred over the base $\mathcal{M}_p$ by twisting the Hopf $S^1$ bundle by the line bundle 
$\mathcal{O}(m)=\mathcal{O}(\pm q)$. Alternatively, and equivalently, we may describe the total M-theory spacetime as
the total space of the $U(1)$ principal bundle over $\mathcal{M}_p\times \mathbb{CP}^3$ with first Chern 
class $c_1 = \mp q \Phi - H$, where recall that $\Phi$ generates $H^2(\mathcal{M}_p,\Z)\cong \Z$ 
and $H$ is the hyperplane class generating $H^2(\mathbb{CP}^3,\Z)\cong\Z$. However, since $\pm q\not\equiv 0$ mod $p=4q$, 
this $U(1)$ principal bundle is \emph{also} non-trivially fibred over the boundary Lens space $S^3/\Z_p$.  
On the other hand, the Taub-NUT-AdS$/\Z_p$ solution is always trivially fibred.

To see what this means in terms of the dual boundary field theory,
recall from the discussion at the end of section \ref{liftbolt} that 
the functions on $S^7$ also become non-trivially fibred over $\mathcal{M}_p$ 
via the twisting, and in particular the Kaluza-Klein modes that are dual 
to the four chiral fields of the ABJM (or rather their gauge-invariants 
constructed using monopole operators), become sections of 
$\mathcal{O}(\pm q)$. This implies that for the Taub-Bolt-AdS solutions these basic matter fields 
are \emph{twisted via their R-charge}, becoming sections of $\mathscr{L}^q$ rather than functions. 
We have attempted to study precisely this twisting in the 1/2 BPS case with $s=\frac{1}{2}$, since 
this is then (conjecturally) simply a twisted version of matrix model studied in \cite{Alday:2012au}.\footnote{We 
are very grateful to L. F. Alday for collaboration on this topic.} It is straightforward to see that this 
twisting does indeed preserve supersymmetry, and that localization goes through similarly to the untwisted case.
Our results so far are somewhat inconclusive: the behaviour of the matrix model is now much more involved, 
although interestingly we find that the Wilson loop VEV, discussed in appendix \ref{wilson}, is indeed 
exactly zero, thus agreeing with the gravity prediction. We also find $N^{3/2}$ scaling of the free
energy at large $N$, but with a coefficient that doesn't seem to match the gravity prediction
of section \ref{free1/2}. However, a key issue that affects both this example, and indeed all of the Taub-Bolt-AdS 
examples, is whether the (potential) twisting of the matter fields by their R-charge is the \emph{only} 
effect on the Lagrangian of the untwisted theory, or whether the correct dual field theory is a more complicated deformation. For now 
we leave this issue open.

Having discussed an example  where $I\neq k$, let us conclude this section with the 
class of Taub-Bolt-AdS solutions where the Sasaki-Einstein manifold has $k=I$, which 
then all  uplift to M-theory.
In this case notice that the circle bundle $\xi$ is twisted over the base $\mathcal{M}^{(4)}$ by the line bundle $L$. For the 1/2 BPS 
solutions this has first Chern class $\pm p$ through the bolt, implying 
that $L$ restricted to the boundary $S^3/\Z_p$ is \emph{always trivial} in this 1/2 BPS class. 
This implies that all the 1/2 BPS solutions filling  a fixed squashed $S^3/\Z_p$ in fact have the same global boundary data. 
In turn, in the dual field theory we then  don't have the twisting by the flat R-symmetry Wilson line, discussed in the previous paragraph for the ABJM case. 
If the field theory Lagrangians are exactly the same in all cases,  
one should then compare the free energies of \emph{all} the solutions plotted in Figure \ref{freeplots2}.  
However, to our knowledge all field theories within this class are \emph{chiral models}, 
for which the matrix model is not under good control.


\section{Discussion}
\label{sowhat}

In this paper we have presented all supersymmetric asymptotically locally AdS$_4$ solutions of Euclidean Einstein-Maxwell theory, 
possessing  $SU(2)\times U(1)$ symmetry. We have shown that in general these solutions
have one modulus, which is the squashing parameter $s$ of the Lens space metric at conformal infinity. 
However,  we have also uncovered an intricate moduli space of solutions, comprising  different branches, 
joining at special values of the parameter. Perhaps surprisingly, we found that typically for fixed conformal boundary data there exist  
multiple solutions, with different topologies. 
We studied global aspects of these solutions,  finding a  subtle interplay between  bulk and  boundary spin structures,

We showed that the Taub-Bolt-AdS solutions, despite
being perfectly smooth and globally well-defined in four dimensions, can be uplifted to eleven-dimensional supergravity only for particular
internal Sasaki-Einstein manifolds.\footnote{A caveat here is that in our analysis we have first imposed regularity of the solutions 
in four dimensions, and later checked which of the solutions can be uplifted to eleven dimensions.}
Moreover, we showed that in these solutions the gauge field in the bulk  induces non-zero
gauge field on the boundary,  whose global properties are intimately related to the specific 
Sasaki-Einstein manifold in the eleven-dimensional solution. 
Therefore, generically, the supersymmetric Taub-NUT-AdS solutions (and their 
orbifolds) are the only supersymmetric solutions filling a biaxially squashed Lens space. In particular, there exist only two distinct choices
of instantonic gauge field such that the solutions preserve 1/2 or 1/4 supersymmetry, respectively. Following \cite{Martelli:2011fw, Martelli:2011fu}, 
we have argued  that these correspond to the two different constructions of supersymmetric field theories on a biaxially squashed three-sphere
 discussed in \cite{Imamura:2011wg} and \cite{Hama:2011ea}, respectively.

Nevertheless, there exist many examples where the Taub-Bolt-AdS solutions exist as global, 
smooth supersymmetric solutions of eleven-dimensional  supergravity.  In particular, we have shown that there exist (infinitely many) examples where 
fixed boundary data  can be filled, supersymmetrically, with bulk solutions with different topologies, and with different holographic free energies.  
In order to address the problem of holographic dual field theories systematically, an important problem that remains 
open is the possible existence of \emph{further} M-theory solutions, with the same boundary data as those we have found, but with smaller gravitational free energies.
At present we can't exclude that such solutions exist outside the ansatz that leads to minimal gauged supergravity; 
for example, there could be supersymmetric solutions with scalar ``hair'' in the bulk.

We have argued that at least in certain cases (in particular for the ABJM model),
 the presence of a non-trivial torsion gauge field on the boundary 
should correspond  in the dual field theory to coupling the field theory to a non-trival R-symmetry 
Wilson line.  Although in general this is quite
 complicated, it may be possible to understand the large $N$ matrix model in the special case when the Lens space metric 
becomes round ($s=\tfrac{1}{2}$).   The corresponding matrix model should then be a twisted version of that studied in \cite{Alday:2012au}. 
Comparing our gravity predictions to some field theory calculations would 
be extremely  interesting, and could teach us something new  about supersymmetric field theories on curved manifolds.

\subsection*{Acknowledgments}
We would like to thank Juan Maldacena for  inspiring comments on preliminary notes, and Kostas Skenderis for discussions on holographic renormalization of the 
Einstein-Maxwell theory.  We would also very much like to thank Fernando Alday for collaboration on a project 
addressing a specific question raised in the paper.
D. M. would like to thank the Simons Center for Geometry and Physics 
for hospitality during the 10th Simons Workshop in Mathematics and Physics.
D. M. is supported by an EPSRC Advanced Fellowship EP/D07150X/3 and also acknowledges partial
support from the STFC grant ST/J002798/1. A.P. is supported by an A.G. Leventis Foundation grant, 
an STFC studentship and via the Act ``Scholarship Programme of S.S.F. by 
the procedure of individual assessment, of 2011-12'' by resources of 
the Operational Programme for Education and Lifelong Learning, of the European Social Fund (ESF) 
and of the NSRF, 2007-2013.  J. F. S. is supported by the Royal Society and Oriel College. 

\appendix

\section{Solving the Einstein-Maxwell equations}
\label{app:Einstein}

In this section we find the general solution to Einstein-Maxwell equations (\ref{EOM}) with 
$SU(2)\times U(1)$ symmetry. The ansatz for the  metric and gauge field takes the form 
\bea\label{ansatz}
\diff s^2_4 & = & \alpha^2 (r) \dd r^2 + \beta^2( r) (\sigma_1^2 + \sigma_2^2)  + \gamma^2(r) \sigma_3^2~, \nonumber\\
A & = & h(r) \sigma_3 ~,
\eea 
where $\sigma_1, \sigma_2, \sigma_3$  are left-invariant one-forms for $SU(2)$, given explicitly by (\ref{leftinvariant}).
In the following analysis we will use the local orthonormal frame
\begin{equation}
\begin{aligned}
&\hat{e}^1 \ = \ \beta(r) \diff\theta~,~~
&&\hat{e}^2 \ = \ \beta(r) \sin\theta \diff\phi~,  \\
&\hat{e}^3 \ = \  \gamma(r) (\diff\psi + \cos\theta \diff\phi)~,~~
&&\hat{e}^4 \ = \ \alpha (r) \diff r~,
\end{aligned}
\end{equation}
and introduce frame indices $a,b,c=1,2,3,4$.
The Einstein equations read (with $\ell=1$)
\bea
R_{ab} & = & - 3 \delta_{ab} + 2 T_{ab} ~,
\eea
where $T_{ab} =  F_a{}^c F_{bc} - \frac{1}{4} F^2 \, \delta_{ab}$
is the stress-energy tensor of the gauge field. For the ansatz \eqref{ansatz} we compute
\begin{align}
R_{44} & \ =  \ - \frac{\gamma''}{\alpha^2\gamma} + \frac{\alpha'\gamma'}{\alpha^3\gamma} 
-  \frac{2\beta''}{\alpha^2\beta} + \frac{2\alpha'\beta'}{\alpha^3\beta} ~,\nonumber\\
R_{33} &\ = \ - \frac{\gamma''}{\alpha^2\gamma} + \frac{\alpha'\gamma'}{\alpha^3\gamma} 
-  \frac{2\beta'\gamma'}{\alpha^2\beta\gamma} + \frac{\gamma^2}{2\beta^4} ~, \nonumber \\
R_{11} \ = \ R_{22} & \ = \  - \frac{\beta''}{\alpha^2\beta} + \frac{\alpha'\beta'}{\alpha^3\beta} 
- \frac{\beta'\gamma'}{\alpha^2\beta\gamma} - \frac{\beta'^2}{\alpha^2\beta^2} + \frac{1}{\beta^2} - \frac{\gamma^2}{2\beta^4}  ~, \nonumber \\
T_{11} &\ =\ T_{22} \ =\  - T_{33} \ =\  -T_{44}\  = \ \frac{1}{2} \frac{h^2}{\beta^4} - \frac{1}{2} \frac{h'^2}{\alpha^2\gamma^2}~,
\end{align}
where a prime denotes derivative with respect to $r$. Furthermore, the equation of motion of the gauge field $\diff * F = 0$ becomes
\bea
 - \left(\frac{\beta^2}{\alpha\gamma} h' \right)' + \frac{\alpha\gamma}{\beta^2}h &= & 0 ~.
\eea

By considering the difference $R_{44} - R_{33}$ we obtain the equation
\bea
-  \frac{2\beta''}{\alpha^2\beta} + \frac{2\beta'}{\alpha^2\beta}\left(\frac{\alpha'}{\alpha} + \frac{\gamma'}{\gamma}\right)  - \frac{\gamma^2}{2\beta^4} & =&  0~,
\eea
and by an appropriate reparametrization of $r$ we can take
\bea
\alpha\gamma &=&  2s ~,~~~~ \beta^2 \ = \ r^2 - s^2~.
\eea
The equation of motion for the gauge field then becomes an ordinary differential equation for $h(r)$:
\bea\label{heqn}
- ((r^2-s^2)h')'  + \frac{4s^2}{r^2-s^2}h & = & 0~.
\eea
The general solution to (\ref{heqn}) is easily found to be
\bea
h(r) &=& P\frac{r^2+s^2}{r^2-s^2}-Q\frac{2rs}{r^2-s^2}~,
\eea
where $P$ and $Q$ are integration constants. 
Substituting  this back into the $33$-component of the Einstein equation gives a second 
order ODE for the metric function $\gamma(r)$. The  general solution to this is
\bea
\gamma^2(r)&=& \frac{4s^2}{r^2-s^2}\left[P^2-Q^2-2Mr + r^2(r^2-3s^2)+C\left(1+\frac{r^2}{s^2}\right)\right]~,
\eea
where $C$ and $M$ are two new integration constants. Substituting this into the $11$-component of the Einstein equations then constrains
\bea
C &=& s^2(1-3s^2)~.
\eea
This is precisely an  analytic continuation the Reissner-Nordstr\"om-Taub-NUT-AdS (RN-TN-AdS) solution in \cite{AlonsoAlberca:2000cs}. Hence we have proven that 
this is the most general solution to the Einstein-Maxwell equations  with $SU(2)\times U(1)$ symmetry.

\section{Integrability and BPS equations}
\label{app:BPS}

In this appendix we compute the  general integrability conditions for supersymmetry
for the Euclidean RN-TN-AdS solutions derived in appendix \ref{app:Einstein}. 
An analysis for Lorentzian solutions was performed in \cite{AlonsoAlberca:2000cs}. 

The  Euclidean RN-TN-AdS solutions are given by (\ref{solution}), (\ref{poly}). 
In this section we use the orthonormal frame $e^a$ in (\ref{frame}), which 
we note is \emph{different} to the orthonormal frame $\hat{e}^a$ used in 
appendix  \ref{app:Einstein}, and take the basis of gamma matrices 
(\ref{gamma}). The integrability condition for the Killing spinor equation \eqref{KSE} reads\footnote{We use frame indices $a,b,c,\dots$ and set $\ell=1$.}
\bea\label{int}
\mathcal{I}_{ab} \, \epsilon \  = \ 0~,
\eea
where
\bea
\mathcal{I}_{ab} &\equiv & \frac{1}{4}R_{ab}^{\ \ \ cd}\Gamma_{cd} + \frac{1}{2}\Gamma_{ab}- 
\ii F_{ab} \mathbb{I}_4 + \frac{\ii}{2}\nabla_{[a }F_{|\, cd}\Gamma^{cd}_{\ \ |}\Gamma_{b]}+ \frac{\ii}{4}\Gamma_{[a }F_{|\, cd}\Gamma^{cd}_{\ \ |}\Gamma_{a]}\nonumber\\
&&- \frac{1}{16}\left[F_{cd}\Gamma^{cd}\Gamma_a,F_{cd}\Gamma^{cd}\Gamma_b\right]+\frac{\ii}{4}F_{cd}\Gamma^{cd}\Gamma_{ab}~,
\eea
is a two-form with values in the Clifford algebra.

A necessary condition to have a non-trivial solution to (\ref{int}) is that 
\bea
\det{\! }_\mathrm{Cliff}\, \mathcal{I}_{ab} & =& 0~,
\eea
holds for all $a,b$. We compute
\bea\label{remarkable}
\det{\! }_\mathrm{Cliff}\,  \mathcal{I}_{ab} &=& \frac{-B_+B_-+D(B_+-B_-)r+D^2r^2}{(r^2-s^2)^6}W_{ab}~,
\eea
where 
\bea
D &\equiv & 2\left[MP-sQ(1-4s^2)\right]~,\nonumber\\
B_\pm & \equiv & (M\pm sQ)^2 - s^2(1\pm P-4s^2)^2 - (1\pm 2P-5s^2)(P^2-Q^2)~,
\eea
and
\bea
\left(W_{ab}\right) & \equiv & \left(\begin{array}{cccc}0 & 1 & \frac{1}{16} & \frac{1}{16} \\ 1 & 0 & \frac{1}{16} & \frac{1}{16} \\ 
\frac{1}{16} & \frac{1}{16} & 0 & 1 \\ \frac{1}{16} & \frac{1}{16} & 1 & 0\end{array}\right)~.
\eea

We thus conclude that a \emph{necessary} condition to have a supersymmetric solution is that the numerator in (\ref{remarkable}) is zero, which is equivalent to
\bea\label{BPSapp}
D \ = \ 0~, \qquad B_+B_- \ = \ 0~.
\eea
These can also be obtained from an analytic continuation \cite{Martelli:2011fu} of the integrability 
conditions in  \cite{AlonsoAlberca:2000cs}, but here we have derived the equations from first principles. We study the general solutions to (\ref{BPSapp}) in 
section \ref{bpsses}.

\section{Class III and supersymmetry}
\label{classIII}

In this appendix we show that the condition $P = \pm Q$ characterizing $\mbox{Class III}$ is not sufficient for supersymmetry, but rather the existence of a Killing spinor requires in addition
\bea\label{extra}
P \ &=& \ -\frac{1}{2}(4s^2-1)~, \ \ \ \ \text{or} \nn \\
P \ &=& \ -s\sqrt{4s^2-1} ~.
\eea
In order to prove this we look at the boundary Killing spinor equation, which can be derived from \eqref{KSE} upon expanding in powers of $1/r$. At lowest order we find 
\bea\label{KSE3}
(\nabla^{(3)}_\alpha - \ii A^{(3)}_\alpha)\chi - \frac{\ii s}{2}\gamma_\alpha \chi + \ii V_\beta \gamma_\alpha\gamma^\beta \chi \ = \ 0 ~.
\eea
Here $\nabla^{(3)}$ denotes the spin connection for the three-metric
\bea\label{3metricapp}
\diff s^2_3 &=& \sigma_1^2+ \sigma_2^2 + 4s^2\sigma_3^2~,
\eea
with $\gamma_{\alpha}$, $\alpha=1,2,3$ generating the corresponding 
Cliff$(3,0)$ algebra, and $\chi$ is a two-component spinor. Furthermore
\bea
A^{(3)} &=& \lim_{r\rightarrow\infty} A  \ = \ P\sigma_3~, \nn \\
V &=& \frac{s^2(4s^2-1)}{Q} \sigma_3 ~.
\eea

The integrability condition for \eqref{KSE3} reads
\bea\label{int3}
\mathcal{I}^{(3)}_{\alpha\beta} \chi \ = \ 0 ~,
\eea
where
\begin{equation}
\begin{split}
\mathcal{I}^{(3)}_{\alpha\beta} \  \equiv \ 
&\frac{1}{4} R^{(3)}_{\alpha\beta}{}^{\alpha_1\alpha_2} \gamma_{\alpha_1\alpha_2} 
  - \ii F^{(3)}_{\alpha\beta} 
  - \frac{s^2}{2}\gamma_{\alpha\beta}  
  - 2\ii \nabla_{[\alpha|} V_{\alpha_1} \gamma_{|\beta]} \gamma^{\alpha_1}  \\
&- 2s\gamma_{[\alpha} V_{\beta]}
  + 2 V^{\alpha_1} V_{\alpha_1} \gamma_{\alpha\beta} - 4 V_{\alpha_1} \gamma_{[\alpha}V_{\beta]}\gamma^{\alpha_1} ~.
\end{split}
\end{equation}
A necessary condition to have a non-trivial solution to \eqref{int3} is that 
\bea
\det{\! }_\mathrm{Cliff}\, \mathcal{I}^{(3)}_{\alpha\beta} &=& 0~.
\eea
Taking into account $P=\pm Q$ we find that this is equivalent to
\bea
\frac{[(1-4s^2)^2-4Q^2][Q^2 +s^2(1-4s^2)]^2}{4Q^4} \ = \ 0 ~,
\eea
and hence \eqref{extra} must hold.

\section{Spin$^c$ structures on bolt solutions}\label{spinc}

In this appendix we discuss in detail the spin$^c$ structures, in the bulk and on the conformal boundary, 
for the bolt-type solutions. This is a little subtle, because for $p$ odd the bolt solutions 
are not spin manifolds (but nevertheless are supersymmetric and admit Killing spinors).
Correlated with this, the four-dimensional graviphoton in the bulk is in general a spin$^c$ connection, 
meaning that when $p$ is odd it is not a gauge field in the usual sense. 
We begin in section \ref{top} with a general 
 topological discussion, and then in section \ref{explicit} give some 
 more explicit details in the cases of interest. Section \ref{11spinor}
 contains a brief discussion of lifting these spinors to eleven dimensions.

\subsection{Topological discussion}\label{top}

In general, recall 
that on an orientable four-manifold $\mathcal{M}^{(4)}$ the spin bundle $\mathcal{S}=\mathcal{S}_+\oplus 
\mathcal{S}_-$
exists if and only if the second Stiefel-Whitney class is zero, so $w_2(\mathcal{M}^{(4)})=0\in 
H^2(\mathcal{M}^{(4)},\Z_2)$. However, it is also true that on \emph{every} four-manifold 
the spin$^c$ bundles $\mathcal{S}_\pm\otimes L^{1/2}$ exist, where 
$L$ is a line bundle satisfying
\bea\label{c1w2}
c_1(L) & \equiv & w_2(\mathcal{M}^{(4)}) \ \ \mbox{mod}\ 2~.
\eea
A spin$^c$ gauge field then has the property that
$2A$ is a connection on $L$, so that (formally) 
$A$ is a connection on $L^{1/2}$.

Recall that the bolt-type solutions all have the topology $\mathcal{M}^{(4)}=\mathcal{M}_p=$ total 
space of $\mathcal{O}(-p)\rightarrow S^2$. A simple computation shows that 
$w_2(\mathcal{M}_p)$ is zero for $p$ even, while for $p$ odd 
$w_2(\mathcal{M}_p)$ generates the cohomology group $H^2(\mathcal{M}_p,\Z_2)\cong \Z_2$. 
We assume that the gauge field has field 
strength $F$ satisfying
\bea\label{Fflux}
\int_{S^2}\frac{F}{2\pi} &=& \frac{n}{2}~,
\eea
where $S^2\subset \mathcal{M}_p$ denotes the bolt/zero-section, so that 
$c_1(L)=n\in H^2(\mathcal{M}_p,\Z)\cong \Z$. 
Then via (\ref{c1w2}), we see that $A$  is a spin$^c$ gauge field if and only if $n\equiv p$ mod 2. 
Notice that for all the solutions discussed in the main text regularity of the metric 
fixes the gauge field, and that $n\equiv p$ mod 2 was then indeed found to hold automatically 
for this gauge field. This is a necessary condition for supersymmetry.

In this section we would like to describe the spin$^c$ bundles $\mathcal{S}_\pm\otimes L^{1/2}$ more explicitly. We begin by noting that, although the metrics on $\mathcal{M}_p$ are not K\"ahler, 
nevertheless $\mathcal{M}_p$ admits a K\"ahler structure. We may then use the 
fact that on a K\"ahler four-manifold the spin bundles are (formally)
\bea\label{spinbundles}
\mathcal{S}_+ &=& K^{1/2}\oplus K^{-1/2}~,\nn\\
\mathcal{S}_- &=& K^{1/2}\otimes \Omega^{0,1}~.
\eea
Here $K$ denotes the canonical line bundle, while $\Omega^{0,1}$ denotes 
the holomorphic tangent bundle. The spin bundles (\ref{spinbundles}) exist if and only if 
the square root $K^{1/2}$ exists. A \emph{natural} choice for $L$ on a K\"ahler manifold is thus $L=K^{-1}$. 
If we denote $\pi:\mathcal{M}_p\rightarrow S^2$ as the projection onto the bolt/zero-section, 
then for the natural complex structure on $\mathcal{M}_p$ implied by our notation we have
\bea
K &=& \pi^*\mathcal{O}(p-2)~.
\eea
We thus see that $K^{1/2}$ indeed exists if and only if $p$ is even. The spinor bundles 
are (formally when $p$ is odd) hence
\bea\label{spins}
\mathcal{S}_+ &=& \pi^*\left[\mathcal{O}(\tfrac{p}{2}-1)\oplus
\mathcal{O}(-\tfrac{p}{2}+1)\right]~,\nn\\
\mathcal{S}_- &=& \pi^*\left[\mathcal{O}(-\tfrac{p}{2}-1)\oplus
\mathcal{O}(\tfrac{p}{2}+1)\right]~.
\eea
 Since 
$L=\pi^*\mathcal{O}(n)$ by definition, we thus compute the spin$^c$ bundles
\bea\label{spincs}
\mathcal{S}_+\otimes L^{1/2} &=& \pi^*\left[\mathcal{O}(\tfrac{n+p}{2}-1)\oplus
\mathcal{O}(\tfrac{n-p}{2}+1)\right]~,\nn\\
\mathcal{S}_-\otimes L^{1/2} &=& \pi^*\left[\mathcal{O}(\tfrac{n-p}{2}-1)\oplus
\mathcal{O}(\tfrac{n+p}{2}+1)\right]~.
\eea
In particular, notice that since $n\equiv p$ mod 2, these bundles always exist on 
$\mathcal{M}_p$, as advertised. The Dirac spinors on our bolt solutions are globally sections of the bundles 
$\mathcal{S}\otimes L^{1/2}=\left(\mathcal{S}_+\otimes L^{1/2}\right)\oplus \left(\mathcal{S}_-\otimes L^{1/2}\right)$, where 
the factors are given by 
(\ref{spincs}) and $n$ is the flux number given by (\ref{Fflux}). Notice 
we have made use of (\ref{spincs}) in the main text, for example to 
deduce (\ref{nigel}).

Now we consider how these spinors restrict to the conformal boundary $S^3/\Z_p=\partial\mathcal{M}_p$. 
Denote the inclusion of this boundary as $\iota:S^3/\Z_p\hookrightarrow \mathcal{M}_p$. 
Then $H^2(S^3/\Z_p,\Z)\cong H_1(S^3/\Z_p,\Z)\cong \Z_p$, and the map
\bea\label{tormap}
\Z\ \cong \  H^2(\mathcal{M}_p,\Z)\ \stackrel{\iota^*}{\longrightarrow}\ H^2(S^3/\Z_p,\Z) \ \cong \ \Z_p
\eea
is simply reduction mod $p$. Let us denote the torsion line bundle that generates 
$H^2(S^3/\Z_p,\Z)\cong \Z_p$ by $\mathscr{L}$, so that $c_1(\mathscr{L})=1\in \Z_p$.
Then using (\ref{tormap}) we can determine that the restriction of either 
spin$^c$ bundle to the conformal boundary is
\bea\label{boundaryspinc}
\mbox{boundary spin$^c$ bundle} \ = \ \iota^*\mathcal{S}_\pm\otimes L^{1/2} &=& \mathscr{L}^{\frac{n+p}{2}}\otimes\left(\mathscr{L}\oplus\mathscr{L}^{-1}\right)~.
\eea
Here it is important to note that $\mathscr{L}^p=1$ is a trivial line bundle, so that 
$\mathscr{L}^{\frac{n+p}{2}}=\mathscr{L}^{\frac{n-p}{2}}$. 
Thus the boundary spinors are typically sections of a non-trivial bundle. 

Recall that every orientable three-manifold is spin, so a spin bundle of 
$S^3/\Z_p$ certainly exists. However, an important subtley here is that for $p$ 
odd there is a \emph{unique} spin bundle, namely
\bea\label{spin3}
\mathscr{S} &=& \mathscr{L}\oplus\mathscr{L}^{-1}~,
\eea
while for $p$ even there are \emph{two} inequivalent spin bundles, namely
\bea\label{spin01}
\mathscr{S}_0 &=& \mathscr{L}\oplus\mathscr{L}^{-1}~, \qquad 
\mathscr{S}_1 \ =\  \mathscr{L}^{\frac{p}{2}+1}\oplus\mathscr{L}^{-\frac{p}{2}-1}~.
\eea
This arises from the fact that, quite generally, inequivalent spin bundles 
correspond to elements of $H^1(\mathcal{M},\Z_2)$, and in the case 
at hand using the universal coefficient theorem one can compute $H^1(S^3/\Z_p,\Z_2) \cong \Z_{\mathrm{gcd}(p,2)}$. 
Thus for $p$ odd this group is trivial, while for $p$ even it is isomorphic to $\Z_2$.
Concretely, when $p$ is even the two spinor bundles in (\ref{spin01}) differ in that the spinors 
differ by a sign on going once around the Hopf fibre. We have then explicitly shown 
that the spin bundle $\mathscr{S}_1$ extends to either of the unique chiral spin bundles $\mathcal{S}_\pm$ 
over $\mathcal{M}_p$ in (\ref{spins}), while $\mathscr{S}_0$ extends instead to 
a particular spin$^c$ bundle on $\mathcal{M}_p$.\footnote{The reader might be more familiar 
with this in the case of spinors on the circle $S^1$: there are two spin structures, periodic 
and anti-periodic. Only the anti-periodic choice extends to the spin structure on $\R^2$. It is similar here: 
it is the ``anti-periodic'' spinor bundle $\mathscr{S}_1$ that extends to a spinor bundle on $\mathcal{M}_p$.} 

The above 
discussion implies that a section of the spin bundle $\mathscr{S}_0$ is the same 
thing as a section of the spin$^c$ bundle $\mathscr{S}_1\otimes \mathscr{L}^{\frac{p}{2}}$. 
This isomorphism is important for understanding the 
Killing spinors. Recall that in the 1/2 BPS case we always have 
$n=\pm p$. When $p$ is even the spinor bundles $\mathcal{S}_\pm$ 
restrict to $\mathscr{S}_1$ on the boundary, and it is precisely 
the flux $n=\pm p$  that turns this into the spinor bundle $\mathscr{S}_0$, as is clear from (\ref{boundaryspinc}). 
 At the level of the Killing spinor equation itself, 
the difference in the global form of the 
spin connection for $\mathscr{S}_0$ and $\mathscr{S}_1$ is equivalent to the
difference between having no flat connection and the specific flat connection 
on $\mathscr{L}^{\frac{p}{2}}$.  The reader might re-examine 
the (essentially local) discussion of the explicit spinors in section \ref{KSsection} in light 
of this global point.
 The 1/4 BPS case involves an additional 
subtlety, that we address in the next subsection \ref{explicit}.

Finally, let us explain \emph{why} (\ref{spin3}), (\ref{spin01}) are in fact spinor bundles 
for $S^3/\Z_p$! If we view $S^3/\Z_p$ as a $p$th power of the Hopf fibration over $S^2$, then 
this naturally leads to the tangent bundle being
\bea\label{tangent}
T(S^3/\Z_p) &=& \R\oplus \mathscr{L}^2~,
\eea
where we have used that the tangent bundle for $S^2$ is $\mathcal{O}(2)$, and pulled this back 
to $S^3/\Z_p$ to obtain $\mathscr{L}^2$. The factor of $\R$ in (\ref{tangent}) is tangent to the vector field 
$\partial_\psi$, generating the $S^1$ fibres. Given that the spinor bundle is a $\C^2$ vector bundle with structure group $SU(2)$, 
combined with the constraint that $\mathbb{P}(\mathscr{S})=T_{\mathrm{unit}}$, relating 
the projectivized spinor bundle to the bundle of unit tangent vectors, this 
implies that $\mathscr{S}$ must be of the form $P\oplus P^{-1}$ where 
$P$ is a line bundle satisfying $P^2=\mathscr{L}^2$. This leads directly 
to (\ref{spin3}) as the unique solution when $p$ is odd, and to 
the two solutions (\ref{spin01}) when $p$ is even. 

\subsection{Explicit computations}\label{explicit}

Guided by the above discussion, we may now look more closely at the local solutions to the Killing spinor 
equations in section \ref{KSsection}. 

\subsubsection{Flat connections}\label{connect4}

We first look more closely at the gauge field on $\mathcal{M}_p$, and in particular its 
global structure on the boundary.
Suppose we have a gauge field on 
$\mathcal{M}_p$ given by
\bea\label{localA}
A &=& \kappa(r)(\diff\psi+\cos\theta\diff\varphi)~,
\eea
where $\psi$ has period $4\pi/p$ and the bolt is at $r=r_0$. Flux quantization through this bolt gives
\bea
\int_{S^2_{r=r_0}}\frac{F}{2\pi} &=& -2\kappa(r_0) \ \equiv \ q~.
\eea
Then $A$ is a connection on the line bundle $\mathcal{O}(q)\rightarrow \mathcal{M}_p$, 
where we are for now assuming that $q\in\Z$ is an integer, so that this makes sense. 
The expression (\ref{localA}) is ill-defined at $r=r_0$, where the vector field 
$\partial_\psi$ is zero. This is because $A$ cannot have an 
expression in terms of a global one-form on $\mathcal{M}_p$ when $q\neq 0$.

We remedy this as follows. Let $\theta$ and $\varphi$ be the standard coordinates on $S^2$, and cover this $S^2$ with 
coordinate patches $U_\pm$, in which $U_+$ excludes the south pole at $\theta=\pi$, and 
$U_-$ excludes the north pole at $\theta=0$. On the products $U_\pm \times S^1_\pm$ 
we may define the one-forms
\bea\label{e3}
D\nu_\pm &\equiv & \diff\nu_\pm + \frac{p}{2}(\cos\theta \mp 1)\diff\varphi~.
\eea
Here $\nu_\pm$ are coordinates on $S^1_\pm$, respectively, each with period $2\pi$. 
In order to form $S^3/\Z_p$, which are the constant $r>r_0$ surfaces, we then glue these together on the overlap via
\bea
\nu_+ - \nu_- &=& p\varphi~.
\eea
Here the transition function $g:(0,\pi)\times S^1\rightarrow U(1)$ is 
$g(\theta,\varphi)=\ex^{\ii p \varphi}$. This has winding number $p\in\mathbb{Z}$, 
and defines the principal $U(1)$ bundle over $S^2$ with first Chern class 
$p\in \Z\cong H^2(S^2,\Z)$.
Then on the overlap $D\nu_+=D\nu_-$, and (\ref{e3}) defines the global angular form for the 
principal $U(1)$ bundle. Notice then that, in terms of the Euler angles used in the main text,
\bea
\nu_\pm &=& \frac{p}{2}\psi_\pm~,
\eea
and the globally defined one-form defined by (\ref{e3}) is simply $\frac{p}{2}\sigma_3$.

We may then cover our manifold $\mathcal{M}_p$ by the two coordinate patches 
$\R_{\geq 0}\times U_\pm \times S^1_\pm$, where $r-r_0$ is a coordinate 
on $\R_{\geq 0}$. Then in these two patches we define
\bea\label{Apm}
A_\pm &=& \frac{q}{2}\diff\psi_\pm+\kappa(r)(\diff\psi_\pm + (\cos\theta \mp 1)\diff\varphi)~.
\eea
This is the correct non-singular form of (\ref{localA}) in each coordinate patch. 
Moreover, on the overlap in $\R_{>0}\times S^3/\Z_p=\{r>r_0\}$ (notice it is crucial here that we exclude the bolt at $r=r_0$) we have
\bea\label{tran}
A_+-A_- &=& q\diff\varphi~.
\eea
It follows that on the complement of the bolt $\R_{>0}\times S^3/\Z_p$ we may write
\bea
A &=& \kappa(r)\sigma_3 + A^{(3)}_{\mathrm{flat}}~,
\eea
where $A^{(3)}_{\mathrm{flat}}$ is a \emph{flat connection} on 
$\mathscr{L}^q$, where $\mathscr{L}$ has first Chern class $c_1(\mathscr{L})=1\in \Z_p\cong H^2(S^3/\Z_p,\Z)$. This is 
defined in the two patches
\bea\label{Apatches}
A^{(3)}_{\mathrm{flat}} &=& \begin{cases}\  \frac{q}{2}\diff\psi_+ & \mbox{in}\  U_+\times S^1_+\\ \ \frac{q}{2}\diff\psi_- & \mbox{in}\  U_-\times S^1_-\end{cases}~.
\eea
This is manifestly flat, and on the overlap we have
\bea\label{Apatch}
A^{(3)}_{\mathrm{flat}, +} - A^{(3)}_{\mathrm{flat}, -} &=& q\diff\varphi~,
\eea
which is indeed precisely the transition function that defines $\mathscr{L}^q$. The holonomy 
of this connection around any $S^1$ fibre in $S^3/\Z_p$ is 
\bea\label{Ken}
\exp\left(\ii \int_{S^1_{\mathrm{fibre}}} A^{(3)}_{\mathrm{flat}}\right)&=& \exp\left(\frac{2\pi \ii q}{p}\right)~,
\eea
which is the observable Wilson line of this non-trivial connection. 

What we have shown here, very explicitly, is that if the gauge field is a connection on 
$\mathcal{O}(q)\rightarrow\mathcal{M}_p$, which has first Chern class 
$c_1(\mathcal{O}(q))=q\in \Z\cong H^2(\mathcal{M}_p,\Z)$, then 
the restriction of this first Chern class  to the boundary $S^3/\Z_p$ is simply $q$ mod $p$ 
in $H^2(S^3/\Z_p,\Z)\cong \Z_p$. Topologically this is clear, since the 
natural map 
\bea
\Z \ \cong \ H^2(\mathcal{M}_p,\Z) \ \rightarrow \ H^2(S^3/\Z_p,\Z) \ \cong \ \Z_p~,
\eea
is just reduction mod $p$. 

When $q$ is half-integer, which happens 
when $p$ is odd and $A$ is a spin$^c$ connection, the above discussion
cannot be applied directly. For example, for the 1/2 BPS solutions we have $q=\pm\frac{p}{2}$.  
In particular, the transition function (\ref{tran}) 
is not a single-valued $U(1)$ gauge transformation in this case.
One might proceed in this case by multiplying the gauge field by 2, and note 
that $2q=\pm p=0$ mod $p$, and then that 
when $p$ is odd the only solution to $2q= 0$ mod $p$ is $q=0$. 
Thus the boundary torsion is zero in this case. Although slightly 
indirect, this is a perfectly valid argument to reach this 
conclusion, which we have then used in the main text. 
A more direct proof, using coordinate patches, requires 
a more involved explicit treatment than we have given 
above.

\subsubsection{Boundary spinors}

With this in hand, we can return to the explicit boundary Killing spinors in section \ref{KSsection}. Beginning with the 1/2 BPS case, the explicit solution to the 
Killing spinor equation is (\ref{chi}). We first note that the frame 
$\tilde{e}^a$ in (\ref{tildeframe}) is not invariant under $\mathcal{L}_{\partial_\psi}$, but rather
\bea\label{merry}
\left(\begin{array}{c} \tilde{e}^1 \\ \tilde{e}^2 \\ \tilde{e}^3\end{array}\right) &=& \left(\begin{array}{ccc}\cos \psi & \sin\psi & 0 \\ -\sin\psi & \cos\psi & 0 \\ 0 & 0 & 1 \end{array}\right)\left(\begin{array}{c} \diff\theta \\ \sin\theta\diff\varphi\\ 2s\sigma_3\end{array}\right)~.
\eea
Here $\sigma_3$ is globally defined on $S^3/\Z_p$, being $\frac{2}{p}D\nu_\pm$ 
in each patch given by (\ref{e3}). The $SO(3)$ rotation above corresponds 
to the $SU(2)=\mathrm{Spin}(3)$ rotation
\bea\label{spin3so3}
 \left(\begin{array}{ccc}\cos \psi & \sin\psi & 0 \\ -\sin\psi & \cos\psi & 0 \\ 0 & 0 & 1 \end{array}\right) & \sim & \left(\begin{array}{cc}\ex^{\ii \psi/2} & 0 \\ 0 & \ex^{-\ii\psi/2}\end{array}\right)~,
\eea
so that in the frame $\check{e}^1=\diff\theta$, $\check{e}^2=\sin\theta\diff\varphi$, $\check{e}^3=2s\sigma_3$ the spinor 
(\ref{chi}) reads
\bea\label{checkchi}
\check{\chi} & =& \left(\begin{array}{cc}\cos\tfrac{\theta}{2}\ex^{\ii\varphi/2} & 
-\sin\tfrac{\theta}{2}\ex^{-\ii\varphi/2} \\ \gamma\sin\tfrac{\theta}{2}\ex^{\ii\varphi/2} & \gamma\cos\tfrac{\theta}{2}\ex^{-\ii\varphi/2} \end{array}\right)\chi_{(0)}~.
\eea
This is independent of $\psi$, as claimed. However, the frame $\check{e}^a$ is 
\emph{singular} at the poles $\theta=0$, $\theta=\pi$, which are coordinate 
singularities. In the patch $U_+\times S^1_+$, which recall excludes 
the south pole $\theta=\pi$, we may further rotate the frame to
\bea\label{rotp}
\left(\begin{array}{c} {e}_+^1 \\ {e}_+^2 \\ {e}_+^3\end{array}\right) &\equiv & \left(\begin{array}{ccc}\cos \varphi & -\sin\varphi & 0 \\ \sin\varphi & \cos\varphi & 0 \\ 0 & 0 & 1 \end{array}\right)\left(\begin{array}{c} \check{e}^1 \\ \check{e}^2 \\ \check{e}^3\end{array}\right) \ \sim \ \left(\begin{array}{c} \diff x_+ \\ \diff y_+ \\ 2s\sigma_3\end{array}\right) ~,
\eea
where we have defined $x_+=\theta\cos\varphi$, $y_+=\theta\sin\varphi$, and the last 
equality is true to leading order near to $\theta=0$. 
Near to $\theta=0$, these are standard Cartesian coordinates on 
$\R^2$, with $\theta$ playing the role of the usual radial coordinate. 
Thus the frame $e_+^a$ is non-singular in the patch $U_+\times S^1_+$, 
and the corresponding spinor rotates similarly to (\ref{spin3so3}) to give\footnote{Here $\chi_+$ 
denotes the spinor $\chi$ in the patch $U_+\times S^1_+$, and is not to be confused with 
the use of $\pm$ in section \ref{KSsection} to denote chirality!}
\bea\label{chip}
\chi_+ &=& \left(\begin{array}{cc}\cos\tfrac{\theta}{2} & 
-\sin\tfrac{\theta}{2}\ex^{-\ii\varphi} \\ \gamma\sin\tfrac{\theta}{2}\ex^{\ii\varphi} & \gamma\cos\tfrac{\theta}{2} \end{array}\right)\chi_{(0)}~.
\eea
We see that this is indeed smooth in this patch, the point being that 
the terms $\ex^{-\ii\varphi}$, which are ill-defined at $\theta=0$, 
have coefficients which vanish as $\mathcal{O}(\theta)$ at $\theta=0$. 

A similar argument now works in the south patch $U_-\times S^1_-$, 
with $x_-=-(\pi-\theta)\cos\varphi$, $y_-=(\pi-\theta)\sin\varphi$. The 
rotation then has the \emph{opposite} sign to (\ref{rotp}),
\bea\label{rotm}
\left(\begin{array}{c} {e}_-^1 \\ {e}_-^2 \\ {e}_-^3\end{array}\right) &\equiv & \left(\begin{array}{ccc}\cos \varphi & \sin\varphi & 0 \\ -\sin\varphi & \cos\varphi & 0 \\ 0 & 0 & 1 \end{array}\right)\left(\begin{array}{c} \check{e}^1 \\ \check{e}^2 \\ \check{e}^3\end{array}\right) \ \sim \ \left(\begin{array}{c} \diff x_- \\ \diff y_- \\ 2s\sigma_3\end{array}\right) ~,
\eea
 leading to the
corresponding spinor in the corresponding smooth frame $e_-^a$ 
\bea\label{chim}
\chi_- &=& \left(\begin{array}{cc}\cos\tfrac{\theta}{2}\ex^{\ii\varphi} & 
-\sin\tfrac{\theta}{2} \\ \gamma\sin\tfrac{\theta}{2} & \gamma\cos\tfrac{\theta}{2}\ex^{-\ii\varphi} \end{array}\right)\chi_{(0)}~.
\eea
This is then smooth in the patch $U_-\times S^1_-$.

Our spinor is thus smooth in each coordinate patch of $S^3/\Z_p$, and on the overlap
region they are related by the $U(1)\subset SU(2)\cong \mathrm{Spin}(3)$ transformation
\bea
\chi_- &=& \left(\begin{array}{cc}\ex^{\ii\varphi} & 0
\\0 & \ex^{-\ii\varphi} \end{array}\right)\chi_+~.
\eea
This precisely means that, globally, the spinors are sections of 
$\mathscr{L}\oplus\mathscr{L}^{-1}$, precisely as we claimed using 
more abstract reasoning in section  \ref{top}. We have thus checked
that the 1/2 BPS spinors are globally well-defined and smooth on the 
constant $r>r_0$ surfaces $S^3/\Z_p$, 
and sections of the bundle $\mathscr{S}$ in (\ref{spin3}) and $\mathscr{S}_0$ in (\ref{spin01}), when $p$ is odd and even, respectively.

The story for the 1/4 BPS spinors is very similar, with just one important 
difference. Although the spinor (\ref{1/4BPSchi}) is simply constant in the 
frame $\tilde{e}^a$, because the latter depends on $\psi$ as in (\ref{merry})
in fact the 1/4 BPS spinors are charged under $\partial_\psi$. Specifically, 
(\ref{1/4BPSchi}) satisfies
\bea\label{1/4rot}
\mathcal{L}_{\partial_\psi} \chi &=& \frac{\ii}{2}\chi~,
\eea
implying an overall phase dependence of
$\ex^{\ii \psi/2}$.
This would then seem problematic if one tries to 
take $\psi$ to have period $4\pi/p$ for general $p>1$. However, 
we emphasized in section \ref{KSsection} that the computation 
was only valid \emph{locally}, and indeed for the 1/4 BPS 
Quaternionic-Eguchi-Hanson solutions in section \ref{selfdual} (and of course the more general 1/4 BPS solutions in section \ref{1/4}), 
the gauge field flux (\ref{QEHflux}) implies that 
on $S^3/\Z_p$ we have an additional flat connection on 
$\mathscr{L}^{-1}$. This flat connection is given explicitly in coordinate patches by (\ref{Apatches}), (\ref{Apatch}), with $q=-1$. If one \emph{includes} this gauge field 
when solving for the 1/4 BPS Killing spinors in each patch, then 
one obtains an \emph{additional} phase dependence of $\ex^{-\ii \psi_\pm/2}$. 
This phase then \emph{cancels} the phase arising from (\ref{1/4rot}), 
and the upshot is that the global 1/4 BPS spinor is in fact \emph{independent of 
$\psi$}. We thus see that the $-1$ factor in the quantized flux (\ref{QEHflux})
and (\ref{fluxy1/4}) is crucial for supersymmetry for general $p>1$. 

Including this flat connection, then in the frame $\breve{e}^1=\diff\theta$, 
$\breve{e}^2=\sin\theta\diff\varphi$, $\breve{e}^3_{\pm} = 
2s(\diff\psi_\pm + (\cos\theta \mp 1)\diff\varphi)$, one find that the 1/4 BPS spinors in the two 
patches are  explicitly
\bea
\breve{\chi}_\pm &=& \ex^{\mp\ii\varphi/2}\left(\begin{array}{c} 0 \\ \chi_{(0)}^{(-)}\end{array}\right)~.
\eea
Rotating as in (\ref{rotp}) and (\ref{rotm}) in each patch, to give 
smooth frames $e_\pm^a$ as before, one then sees that these 1/4 BPS spinors on constant $r>r_0$ surfaces $S^3/\Z_p$  are 
smooth sections of $(\mathscr{L}\oplus\mathscr{L}^{-1})\otimes \mathscr{L}^{-1}$.

\subsubsection{Regularity at the bolt}\label{spinbolt}

The above discussion guarantees that the spinors are well-defined and smooth on $\{r>r_0\}$, 
where the bolt $S^2$ is at $r=r_0$. For completeness, we should also verify that the spin$^c$ spinors in 
section \ref{KSsection} are smooth at the bolt itself. 

This is easily checked along the lines of the previous subsection. We first note that the 
four-frame (\ref{frame}) is \emph{singular} at the bolt $r=r_0$ itself, and moreover the 
gauge field in (\ref{solution}) is also singular at the bolt. Thus the spinors in section
\ref{KSsection} are in a singular frame, in a singular gauge! However, this is easily 
rectified by making an appropriate frame rotation and gauge transformation, respectively. 

If we denote by $\rho$ the geodesic distance from the bolt at $r=r_0$, then to leading order near $\rho=0$ the frame (\ref{frame}) reads
\bea
e^1 &\sim & \sqrt{r_0^2-s^2}\sigma_1~, \quad 
e^2 \ \sim \ \sqrt{r_0^2-s^2}\sigma_2~, \quad e^3 \ \sim \ \rho\left[\diff\left(\frac{p\psi}{2}\right)+\frac{p}{2}\cos\theta\diff\varphi\right]~, \nonumber\\
e^4 &\sim& \diff \rho~,
\eea
as in equation (\ref{nearthebolt}). The $e^3$ and $e^4$ directions suffer the same polar coordinate 
type singularity at $\rho=0$ as the frame $\check{e}^a$ suffered at $\theta=0$, $\theta=\pi$ in the previous subsection. 
If we rotate
\bea\label{rotbolt}
\left(\begin{array}{c} {e}_0^1 \\ {e}_0^2 \\ {e}_0^3 \\ e_0^4\end{array}\right) &\equiv & \left(\begin{array}{cccc}1 & 0 & 0 & 0\\ 0 & 1 & 0& 0 \\ 0 & 0 & \cos\frac{p\psi}{2} & \sin\frac{p\psi}{2} \\ 0 & 0 & -\sin\frac{p\psi}{2} & \cos\frac{p\psi}{2} \end{array}\right)\left(\begin{array}{c} {e}^1 \\ {e}^2 \\ {e}^3 \\ e^4\end{array}\right)~,
\eea
the $e_0^3$, $e_0^4$ are now smooth near the bolt. The corresponding action on the Dirac spinors 
may be deduced from the four-dimensional gamma matrices (\ref{gamma}), 
and is 
\bea
\mathrm{diag}(\ex^{-\ii {p\psi}/{4}}, \ex^{\ii {p\psi}/{4}}, \ex^{\ii {p\psi}/{4}}, \ex^{-\ii {p\psi}/{4}})\in \mathrm{Spin}(4)\, \cong \,
SU(2)\times SU(2)~.\eea
Of course, we should again introduce coordinate patches $U_\pm$ on the $S^2$ bolt, 
and rotate the $e_0^1$ and $e_0^2$ directions precisely as we did in the previous section, 
{\it i.e.} we apply the rotation (\ref{merry}) so that the frame is invariant under $\mathcal{L}_{\partial_\psi}$, 
and the rotations (\ref{rotp}), (\ref{rotm}) in the $U_+$ and $U_-$ patches, respectively. 
In this way we obtain four-frames $e_\pm^a$, $a=1,2,3,4$, in patches 
$U_\pm \times S^1_\pm \times \R_{\geq 0}$ which cover a neighbourhood of the bolt. Here 
$\rho\in\R_{\geq 0}$ is geodesic distance from the bolt. 
In this frame, the 1/2 BPS spinors (\ref{spinor1/2}) read
\bea
\epsilon &=& \left(\begin{array}{c} \sqrt{\frac{(r-r_3)(r-r_4)}{r-s}} \chi^{(+)} \ex^{-\ii {p\psi}/{4}}\\ \sqrt{\frac{(r-r_1)(r-r_2)}{r-s}}\chi^{(-)} \ex^{\ii {p\psi}/{4}}\\ \ii\sqrt{\frac{(r-r_1)(r-r_2)}{r+s}}  \chi^{(+)}\ex^{\ii {p\psi}/{4}} \\ \ii\sqrt{\frac{(r-r_3)(r-r_4)}{r+s}} \chi^{(-)} \ex^{-\ii {p\psi}/{4}} \end{array}\right)~,
\eea
where $\chi^{(\pm)}$ are the two components of $\chi$ in (\ref{chip}) and (\ref{chim}), in the two patches respectively.
Similarly, one should understand $\psi=\psi_\pm$ in the two patches, respectively.

Finally, recall that the gauge for the spin$^c$ gauge field $A$ is singular at the bolt, as discussed in section \ref{connect4}. 
For the positive/negative branch 1/2 BPS solutions, the singular gauge field is to leading order
\bea
A & \sim & \mp \frac{p}{4}(\diff\psi + \cos\theta\diff\varphi)~,
\eea
near the bolt, respectively. This follows directly from (\ref{fluxy}). Thus 
for the positive/negative branch solutions we must make a gauge transformation 
$A\rightarrow A \pm \frac{p}{4}\diff\psi$ (in each patch appropriately) in order 
that $A$ is well-defined at the bolt (where the azimuthal coordinate $\psi$ is not defined). 
Doing so, we obtain the following form of the spinors for the positive branch solutions 
\bea\label{positivespinor}
\epsilon_{\mathrm{positive\, branch}} & = &  \left(\begin{array}{c} \sqrt{\frac{(r-r_3)(r-r_4)}{r-s}} \chi^{(+)}\\ \sqrt{\frac{(r-r_1)(r-r_2)}{r-s}}\chi^{(-)}  \ex^{\ii {p\psi}/{2}} \\ \ii\sqrt{\frac{(r-r_1)(r-r_2)}{r+s}}  \chi^{(+)} \ex^{\ii {p\psi}/{2}} \\ \ii\sqrt{\frac{(r-r_3)(r-r_4)}{r+s}} \chi^{(-)}  \end{array}\right)~,
\eea
while the negative branch spinors are
\bea\label{negativespinor}
\epsilon_{\mathrm{negative\, branch}} &=& \left(\begin{array}{c} \sqrt{\frac{(r-r_3)(r-r_4)}{r-s}} \chi^{(+)} \ex^{-\ii {p\psi}/{2}}\\ \sqrt{\frac{(r-r_1)(r-r_2)}{r-s}}\chi^{(-)} \\ \ii\sqrt{\frac{(r-r_1)(r-r_2)}{r+s}}  \chi^{(+)} \\ \ii\sqrt{\frac{(r-r_3)(r-r_4)}{r+s}} \chi^{(-)} \ex^{-\ii {p\psi}/{2}} \end{array}\right)~.\eea
These spinors are now in a non-singular frame and gauge at the bolt, and we indeed see that they are smooth.
Here  one must recall that for the positive branch the bolt is at $r_0=r_2$, while for the negative branch 
instead $r_0=r_4$. In both cases $r_0$ is the largest root, so $r>s$ for all $r$ while $r>r_i$ provided
 $r_i$ is not the root $r_0$. The key point is that for the positive branch spinor (\ref{positivespinor}), 
the components that depend on $\psi$ tend to zero at the bolt $r=r_2$, with 
a corresponding statement holding for (\ref{negativespinor}). Indeed, notice that 
$p\psi/2$ has the canonical period $2\pi$, with geodesic distance $\rho\propto\sqrt{r-r_0}$ near the bolt,
so that the spinors tend to zero near the bolt 
in the same way as they tend to zero near the poles $\theta=0$, $\theta=\pi$ in (\ref{chip}), (\ref{chim}), 
respectively. This proves that the 1/2 BPS spin$^c$ spinors are smooth and well-defined everywhere, 
for both positive and negative branches.

The discussion for the 1/4 BPS case is essentially identical (although here notice that our labelling 
of roots $r_4\leftrightarrow r_2$ for the two types of branch is interchanged relative to the 1/2 BPS case).

\subsection{Eleven-dimensional spinors}\label{11spinor}

In this appendix we briefly consider the eleven-dimensional spinors for the bolt 
solutions. Even though the four-dimensional Taub-Bolt-AdS solutions are not spin manifolds 
for $p$ odd, we will see that the eleven-dimensional Euclidean space is always spin, 
and that the eleven-dimensional spinors 
are indeed globally well-defined whenever the metric is. We follow the notation of section \ref{liftingit}.

We consider the case of lifting a Taub-Bolt-AdS solution, with topology 
$\mathcal{M}_p=\mathcal{O}(-p)\rightarrow S^2$, on a regular Sasaki-Einstein manifold $Y_7$  
with K\"ahler-Einstein base $B_6$, Fano index $I=I(B_6)$, and for simplicity we take 
$k=I$ so that $Y_7$ is the total space of the $U(1)$ principal bundle associated to the 
canonical bundle of $B_6$. In this case from (\ref{uplift}) we see that the eleven-dimensional 
geometry is the total space of a $U(1)$ principal bundle over $\mathcal{M}_p\times B_6$, 
with global angular form $\eta+\tfrac{1}{2}A$. We denote the corresponding line bundle 
by $\mathcal{V}$. We will show that the total space $Z$ of $\mathcal{V}$ (which is twelve-dimensional) is always a spin manifold. Since $Z$ deformation retracts onto its zero section, 
it is sufficient to compute the restriction of $w_2(Z)$ to the zero section $\mathcal{M}_p\times B_6$.
In turn, we note that $Z$ has a natural complex structure (with $\mathcal{M}_p$ having the 
complex structure of section \ref{top}), and then $w_2(Z)$ is the mod 2 reduction 
of the first Chern class $c_1(Z)$.  We then compute $c_1(\mathcal{M}_p\times B_6) = 
(2-p)\Phi+c_1(B_6)$, where $\Phi$ denotes the generator of $H^2(\mathcal{M}_p,\Z)\cong\Z$.\footnote{That is, $\int_{S^2_{\mathrm{bolt}}}\Phi = 1$.} Then the connection term 
$\eta+\tfrac{1}{2}A$ implies that $c_1(\mathcal{V})=-n\Phi-c_1(B_6)$.  
The Whitney product formula then gives $c_1(Z)=(2-p-n)\Phi$ (with $\Phi$ understood as appropriately 
pulled back). Since $p\equiv n$  mod 2, we see that $c_1(Z)=0$ mod 2, which implies 
that ${Z}$ is indeed a spin manifold. Its eleven-dimensional boundary, which 
is our spacetime, is thus also spin.

The connection term $\eta+\frac{1}{2}A$ is thus precisely ensuring that the eleven-dimensional 
spacetime is a spin manifold, even though the base four-dimensional spacetime in general is not. 
This term also plays an important role in ensuring that the eleven-dimensional spinor 
is indeed a spinor, rather than a section of a spin$^c$ bundle.  The eleven-dimensional spinor 
is a tensor product $\epsilon\otimes \beta$, where $\epsilon$ is the Dirac spin$^c$ spinor  on
$\mathcal{M}_p$, and $\beta$ is a spinor on the internal space $Y_7$. In particular, 
$\epsilon$ is coupled to the spin$^c$ line bundle $L^{1/2}$, with (formal) connection $A$. 
However, because of the connection term $\eta+\frac{1}{2}A$ the spinor $\beta$ is 
also fibred over $\mathcal{M}_p$. To see this, note that on a Sasaki-Einstein seven-manifold the 
Killing spinor has charge $2$ under $\partial_\Reeb$, where recall $\eta=\diff\Reeb+\sigma$. 
Thus the additional connection term in  $\eta+\frac{1}{2}A$ implies that 
$\beta$ has charge $-1$ under $A$. Thus $\beta$ is a spinor on $Y_7$, but 
also valued in $L^{-1/2}$. Altogether, we see that the dependence on $L$ cancels in the tensor product $\epsilon\otimes\beta$, 
which precisely ensures that this is then an eleven-dimensional spinor, rather than a spin$^c$ spinor. 
As we have seen in the previous paragraph, this is then guaranteed to be globally defined.

\section{Holographic free energy and one-point functions}

\label{app:FreeEnergy}

In this appendix we present  some further details of the computation of the holographic free energy/Euclidean action of the 
solutions described in the main text.  We also collect expressions for simple holographic one-point functions in these backgrounds.

\subsection{Free energy}

We begin by writing the supergravity action 
\bea
I \ \equiv \ I^{\text{grav}}_{\text{bulk}} + I^F \  = \  -\frac{1}{16\pi G_4}\int \diff^4x\sqrt{g}\left(R + 6\right) +\frac{1}{16\pi G_4}\int \diff^4x\sqrt{g} \, F^2 ~.
\eea
This action diverges as $r\to\infty$ and in order to obtain a finite value we apply the standard technique of holographic renormalization   \cite{Skenderis:2002wp,Papadimitriou:2005ii}. 
We introduce a cut-off at $r=\rcut$ and consider the hypersurface $\mathcal{S}_\rcut$ of constant $r=\rcut$ with induced metric
\bea
\g_{\mu\nu} = g_{\mu\nu} - \hat n_\mu \hat n_\nu~,
\eea
where $\hat n$ is the unit vector normal to $\mathcal{S}_\rcut$. As $\rcut\to \infty$, $\mathcal{S}_\rcut$ becomes the (conformal) boundary and $\gamma_{\mu\nu}$ the boundary metric.
We regularize the action by adding the following term
\bea
I^{\text{grav}}_{\text{ct}} + I^{\text{grav}}_{\text{bdry}} &=&  \frac{1}{8\pi G_4} \int_{\mathcal{S}_\rcut} \diff^3x \sqrt{\gamma} \left( 2 + \tfrac{1}{2} R(\gamma) - K\right) ~.
\eea
Here $R(\gamma)$ is the Ricci scalar of $\gamma_{\mu\nu}$, and $K$ is the trace of the second fundamental form of $\mathcal{S}_\rcut$, the latter being the Gibbons-Hawking boundary term. 

For the metric \eqref{solution} we compute
\bea
I^{\text{grav}}_{\text{bulk}} & =&  \frac{1}{8\pi G_4} \int \diff^3x  \int_{r_0}^\rcut \diff r \, 6 s (r^2-s^2)~.
\eea
Here $\int \diff^3x$ is an integral over the Euler angular variables $\theta,\varphi,\psi$, and $r_0$ is the largest root of $\Omega(r)$ where the metric closes off. We have also used
\bea
R &=& -12 ~, \\
\sqrt{g} &=& 2s(r^2-s^2)\sin\theta~.
\eea
In addition  we find
\bea
R(\gamma) &=& \frac{2}{\rcut^2-s^2} - \frac{2s^2\Omega(\rcut)}{(\rcut^2-s^2)^3}~, \nonumber\\
\sqrt{\gamma} &=& 2s\sin\theta \sqrt{(\rcut^2-s^2)\Omega(\rcut)}~, \nonumber \\
\sqrt{\gamma} K &=& \mathcal{L}_{\hat n} \sqrt{\gamma} \quad = \quad \frac{s \sin\theta}{\rcut^2-s^2} \left. \frac{\diff((r^2-s^2) \Omega(r))}{\diff r}\right|_{r=\rcut}~,
\eea
where $\mathcal{L}$ denotes the Lie derivative and 
\bea
\hat n & = & \sqrt{\frac{\Omega(r)}{r^2-s^2}} \frac{\partial}{\partial r}~.
\eea

\subsection{Energy-momentum tensor and $U(1)_R$ current}

For any of our solutions the holographic one-point functions  of the energy momentum tensor $\langle T_{\mu\nu}\rangle $,
and of the $U(1)_R$ current $\langle J_\mu \rangle$, may be computed using the standard methods of holographic renormalization in asymptotically locally AdS solutions. In general, the holographic energy-momentum  tensor is given by the expression
\bea
\langle T_{\mu\nu} \rangle  & = &  - \frac{1}{8\pi G_4} \lim_{\rcut\to \infty} \, r \left( K_{\mu\nu} - K h_{\mu\nu} + 2 h_{\mu\nu} - R_{\mu\nu}[h] + \frac{1}{2} R[h] h_{\mu\nu} \right)~,
\eea
where $h_{\mu\nu} = g_{\mu\nu} - \hat{n}_\mu\hat{n}_\nu$ is the induced metric on a 
surface $\mathcal{S}_\rcut$ of constant $r=\rcut$, and $\hat n^\mu$ is the unit vector normal to $\mathcal{S}_\rcut$.
Here the metric is  
\bea
\diff s^2_3 &=&   h_{\mu\nu} \diff x^\mu \diff x^\nu \ =  \ (\rcut^2-\s^2)(\sigma_1^2 + \sigma_2^2) + \frac{4s^2\Omega(\rcut)}{\rcut^2-\s^2}\sigma_3^2 ~,
\eea
and computing the extrinsic curvature $K_{\mu\nu} = \nabla_{(\mu} \hat{n}_{\nu)}$  we obtain
\bea
 \langle T_{\mu\nu} \rangle \diff x^\mu \diff x^\nu  & =  &  \frac{M}{8\pi G_4}\left [ \sigma^2_1 + \sigma^2_2 - 8s^2 \sigma^2_3 \right] ~,
\eea
where 
\bea
M & = & \left\{ \begin{array}{ll}
Q_{1/2}\sqrt{4s^2-1} & \quad 1/2~ \mathrm{BPS} ~,\\
 2sQ_{1/4}  & \quad 1/4 ~\mathrm{BPS}~.\\
\end{array}
\right.
\eea
Note that $\langle T_{\mu\nu}\rangle $ is always traceless with respect to the conformal boundary metric $\gamma_{\mu\nu}$, 
namely $\gamma^{\mu\nu}\langle T_{\mu\nu}\rangle =0$.  Similarly, in general 
the VEV of the $U(1)_R$ current can be extracted from the expansion of the 
the gauge field in the bulk
\bea\label{expansion}
A &=& P\sigma_3 + A^{(3)}_{\mathrm{flat}}  - \frac{1}{r} 2Qs\, \sigma_3+ \mathcal{O}\left(\frac{1}{r^2}\right)~,
\eea
and  we obtain
\bea
\langle J_\mu \rangle \diff x^\mu   & = & \frac{sQ }{2\pi G_4} \, \sigma_3 ~.
\eea

Recall that  regularity fixes $Q=Q(s)$, and then different 
solutions have different values of the parameter $Q(s)$, for fixed $s$. 
Thus the VEVs $\langle T_{\mu\nu}\rangle $ and $\langle J_\mu \rangle$  are different 
for the Taub-NUT-AdS and the inequivalent branches of Taub-Bolt-AdS solutions. 
We note that the  obvious  Ward identity
\bea\label{iden}
\frac{\diff}{\diff s} I  & = &   \int \vol_\gamma \left( \frac{\diff \gamma^{\mu\nu}}{\diff s} \frac{1}{2} \langle T_{\mu\nu} \rangle +  \frac{\diff A^{(3)}{}^\mu} {\diff s}  \langle J_\mu  \rangle \right)
\eea
must hold in the field theory, as a  consequence of the chain rule. Here $\gamma^{\mu\nu}$ is the inverse boundary metric
and $\vol_\gamma$ denotes the  volume form. Indeed, using that 
\bea
\frac{\diff\gamma^{\mu\nu}}{\diff s} \frac{\de}{\de x^\mu }\frac{\de}{\de x^\nu} &=&  - \frac{1}{s^3} \frac{\de^2}{\de \psi^2}~,
\eea
one can check that (\ref{iden})  is satisfied  by the holographic one-point functions in all cases.  

\section{Holographic Wilson loops}
\label{wilson}

In this appendix we present an argument showing that the Taub-NUT-AdS and Taub-Bolt-AdS solutions behave \emph{qualitatively} differently 
with respect to the holographic computation of the VEV of a BPS Wilson loop. Given a specific dual field theory Lagrangian,  
the latter is in principle computable (at finite $N$) using localization methods.

We consider an M2-brane that wraps the M-theory circle together with a  copy of $\R^2\subset \mathcal{M}^{(4)}$ that 
has boundary an $S^1\subset S^3$ at conformal infinity. This naturally 
corresponds to a Wilson loop in the boundary gauge theory. Notice that, from the IIA 
point of view, this is a fundamental string wrapping the copy of $\R^2$.
Taking the $S^1\subset S^3$ to be a Hopf fibre/great circle, 
which in our coordinate system is coordinatized by the Euler angle $\psi$, and the $\R^2$ to 
be this together with the radial direction coordinatized by $r$ at $\theta=0$, we conjecture that the
wrapped string should be BPS, as it is in AdS$_4$.\footnote{For $Y_7=S^7$, or 
$S^7/\Z_4$ as appropriate for a Taub-Bolt-AdS solution, the string can be at any point 
on the $\mathbb{CP}^3$ base. More generally, it will sit at a point in the IIA base $M_6$ in such a way that 
the M-theory circle fibre above it is calibrated and hence BPS.} For a Taub-Bolt-AdS 
solution, notice this is a copy of the fibre of $\mathcal{M}^{(4)}=\mathcal{O}(-1)\rightarrow S^2$.

The action of the M2-brane/fundamental string should compute the VEV of the corresponding 
BPS Wilson loop in the holographically dual supersymmetric gauge theory, to leading 
order in the large $N$ limit. It is easy enough to compute this action in any particular example. 
The VEV of a BPS Wilson loop can also be computed \emph{exactly} via localization in the gauge theory.

However, there is an important subtlety in this computation, for which the Taub-NUT-AdS 
space and Taub-Bolt-AdS space behave very differently. 
This was first pointed out, in a similar but non-supersymmetric context, in 
 \cite{Witten:1998zw}. The point is that the type IIA string has a coupling 
$\exp \left( \ii\int_{\Sigma} B\right)$. When we insert this  string into our 
string theory path integral, we should include this coupling in the computation
of the action. Moreover, in the supergravity partition function we should remember to 
sum over flat $B$-fields. Adding a closed $B$-field does not affect the supergravity equations 
of motion, but different closed $B$-fields can be gauge inequivalent, and should be summed/integrated over. 
This is a key point.

In the present situation, with boundary conditions fixed at infinity, we should sum over 
$B$-fields in spacetime that are zero at infinity, modulo shifts $B\rightarrow B+\diff\Lambda$, 
where $\Lambda$ is also zero at infinity. This means that physically distinct $B$-fields, 
with fixed boundary condition at infinity, are measured by $H^2_{\mathrm{cpt}}(\mathcal{M}^{(4)})$. 
In fact including large gauge transformations this becomes $H^2_{\mathrm{cpt}}(\mathcal{M}^{(4)},U(1))$.
The key point is that for the Taub-NUT-AdS spacetime this group is zero, so there are 
no flat $B$-fields to sum over. But for the Taub-Bolt-AdS spacetimes, because 
of the $S^2$ bolt in fact $H^2_{\mathrm{cpt}}(\mathcal{M}^{(4)},\R)\cong \R$, and is generated 
by a closed two-form that integrates to 1 over the fibre of $\mathcal{M}^{(4)}=\mathcal{M}_1=\mathcal{O}(-1)\rightarrow S^2$, 
and has rapid decay up the fibre. Including large gauge transformations, this means 
there is an $S^1$ moduli space of $B$-fields to integrate over, and the supergravity 
saddle point approximation for the path integral with the type IIA string inserted should be
\bea
\langle \, \mathrm{string} \, \rangle_{\mathrm{Bolt}} & =& \int_{\vartheta=0}^{2\pi} \exp\left[-A_{\mathrm{string}}+\ii \vartheta\right] \ = \ 0~.
\eea
Here $A_{\mathrm{string}}$ is the area of the string (its action), while $\vartheta$ parametrizes the different 
$B$-fields integrated over the fibre. For the Taub-NUT-AdS solution, there is no 
such integral, and the VEV is just given by the classical area, in the large $N$ limit.
On the other hand, this argument shows that the VEV of the Wilson loop in the Taub-Bolt-AdS backgrounds is 
\emph{identically zero}.

\section{Proof that $I_\mathrm{sing}=\frac{n^2 \pi }{8pG_4}$}
\label{proofising}

The space Taub-NUT-AdS$/\Z_p$ is a singular orbifold for $p>1$. 
We have seen in our explicit examples that 
Taub-NUT-AdS$/\Z_p$ solutions can arise, with specific squashing parameters, 
as limits of Taub-Bolt-AdS solutions. When this happens, the 
\emph{singularity} of the Taub-NUT-AdS$/\Z_p$ can effectively contribute to the free energy. 
This is because, in these limits, the Taub-NUT-AdS$/\Z_p$ solution has an additional flat 
gauge field turned on, which can be understood as originating from ``trapped flux'' at the 
bolt which has collapsed to zero size.
In this appendix we attempt to understand this phenomenon more generally.
We
argue that 
the singularity can contribute to the free energy via 
\bea\label{sing}
I_\mathrm{sing} & \equiv &  \frac{n^2}{8p}\cdot \frac{\pi}{G_4}~.
\eea
The basic physical idea here is that the singularity can have $\frac{n}{2}$ units of flux 
``trapped'' in it, so that
\bea\label{trapped}
\int_{\mathrm{collapsed}\, \mathrm{cycle} }\frac{F}{2\pi} &=& \frac{n}{2}~.
\eea
This flux then induces a corresponding 
torsion line bundle on Taub-NUT-AdS$/\Z_p$ minus the singularity, 
which has topology $\R_{>0}\times S^3/\Z_p$.
In practice, we compute this singular contribution by choosing a one-parameter family of 
resolutions of  the orbifold singularity 
to $\mathcal{M}_p=\mathcal{O}(-p)\rightarrow S^2$, and then calculating 
the free energy of an appropriate gauge field satisfying (\ref{trapped}), 
where the collapsed cycle is resolved to the $S^2$ bolt/zero-section. 
This one-parameter family, depending on $\varepsilon>0$, will be such 
that in the $\varepsilon\rightarrow 0$ limit we recover the 
Taub-NUT-AdS$/\Z_p$ metric with a \emph{flat} torsion gauge field, with 
line bundle depending on $n$.
The result will end up being a topological invariant, provided we make certain natural assumptions.\footnote{Notice that the naive contribution of a flat torsion gauge field to the free energy is zero 
(because $F=0$).}

We begin by first choosing an explicit resolution of the metric and  appropriate 
gauge field, which will lead to (\ref{sing}). Having done this, we will then discuss 
to what extent the result is independent of these choices, and why (\ref{sing}) 
may then be interpreted as a topological invariant.

Recall that the self-dual Einstein metric on the Taub-NUT-AdS space can be written as
\bea
\diff s^2_4 &=& \frac{r^2-s^2}{\Omega(r)}\diff r^2 + (r^2-s^2)(\sigma_1^2 + \sigma_2^2)+\frac{4s^2\Omega(r)}{r^2-s^2} \sigma_3^2~,
\label{heremetric}
\eea
where
\bea
\sigma_1 + \ii \sigma_2 & = & \ex^{-\ii\psi}(\diff\theta + \ii \sin\theta \diff\varphi)~,~~~ \sigma_3 \ = \ \diff\psi + \cos\theta\diff\varphi~.
\eea
Here $\Omega(r) = (r-s)^2(r-r_1)(r-r_2)$, where 
\bea\label{Einsteinrootsmore}
\left\{\begin{array}{c}r_2 \\ r_1\end{array}\right\} &=& \left\{\begin{array}{c} - s + \sqrt{4s^2-1}\\ - s - \sqrt{4s^2-1} \end{array}\right\} ~.
\eea
Taking  $\theta\in [0,\pi]$ and the periodicities   $\varphi \in [0,2\pi), \psi \in [0, 4\pi)$ 
this space is topologically $\R^4$.  Taking instead $\psi \in [0, \frac{4\pi}{p})$ this becomes topologically the orbifold 
$\R^4/\Z_p$, with a (NUT) orbifold singularity located at $r=s$.  

To compute $I_\mathrm{sing}$ we resolve
the singularity, replacing it with an $S^2_\varepsilon$ of radius proportional to a small
 parameter $\varepsilon$, for any value of $s$.  Obviously, we cannot do this while preserving supersymmetry 
 and $SU(2)\times U(1)$ isometry in general, otherwise we would have found this metric within some class of BPS solutions.
 However, it is straightforward to write a metric  on the resolved space that has the same isometry group  and with same conformal boundary.  A simple example of such a  metric is  obtained by replacing\footnote{We have checked that for any 
 choice of parameters in $\Omega^\varepsilon (r) = (r-s-\varepsilon)(r - s  -a \varepsilon)(r- r_1- b \varepsilon )(r-r_2  -c\varepsilon)~$ the resulting metric is not Einstein. However, this is not an issue, as will become apparent.}
  $\Omega(r)$ with 
 \bea
 \Omega^\varepsilon (r) & = & (r-s-\varepsilon)(r - s -  a \varepsilon)(r- r_1-\varepsilon )(r-r_2 -\varepsilon)~,
 \eea
 where we assume that $\varepsilon \geq 0$.
 Notice that the roots are now all distinct, with the largest root being  $r^\varepsilon = s + \varepsilon$, provided that $a<1$.
  Using the method described in the text, it is straightforward to check that taking 
 \bea
 a   & = & 1-p +  {\cal O} (\varepsilon) ~,
 \eea
 this gives a \emph{smooth} metric on the space  $\mathcal{M}_p^\varepsilon= \mathcal{O}(-p)\rightarrow S^2$, 
 for any value of $s$ and sufficiently small $\varepsilon>0$.  Notice that  indeed $a<1$ for any $p$, thus $r^\varepsilon$ is the largest root. Then $\Omega^\varepsilon (r)$ reduces smoothly 
 to the Taub-NUT-AdS metric function for $\varepsilon\to 0$, where two roots coalesce. 

In order to compute the contribution to the free energy of the trapped flux, we will 
choose a one-parameter family of gauge fields on this resolved space $\mathcal{M}_p^\varepsilon$ 
 with self-dual field strength $F^\varepsilon$.   Recall that locally 
 the most general (anti-)self-dual gauge field preserving the isometry of the 
metric (\ref{heremetric}) is given by 
 $A^\pm   =  C_\pm  f_\pm (r) \sigma_3$,
where $C_\pm$ are constants and 
\bea
f_\pm(r) &=& \frac{r\mp s}{r\pm s}~.
\eea
It turns out that choosing the (local) gauge field 
\bea\label{Achoice}
A^\varepsilon & = & -\frac{n\varepsilon}{4(2s+\varepsilon)} \, \frac{r + s}{r -  s} \, \sigma_3~,
\eea
 the flux through the  $S^2_\varepsilon \subset \mathcal{M}_p^\varepsilon$ at $r=s+\varepsilon$ is the desired one, namely
\bea
\int_{S^2_\varepsilon} \frac{F^\varepsilon}{2\pi} &=&  \frac{n}{2}~,
\eea
 again independently of  $s$ and $\varepsilon$. Moreover $F^\varepsilon \to 0$ for $\varepsilon \to 0$ implying that,
 \emph{globally},  $A^\varepsilon $ becomes a flat torsion gauge field in the limit.
Finally, it is straightforward to compute the contribution to the action/free energy
 \bea
 \frac{1}{16 \pi G_4} \int_{\mathcal{M}_p} (F^\varepsilon)^2  &= &  - \frac{1}{8 \pi G_4} \int_{\mathcal{M}_p} F^\varepsilon \wedge F^\varepsilon \nn\\
 & = & \frac{1}{8 \pi G_4}  \frac{n^2 \varepsilon^2}{16 (2s+\varepsilon)^2 }[f_- (r=r^\varepsilon)^2 - f_-(r=\infty)^2] \int \diff \psi \sin\theta \diff \theta \diff \varphi \nn\\[2mm]
 & = &    \frac{n^2}{8p}\cdot \frac{\pi}{G_4} + {\cal O}(\varepsilon^2) ~,
 \eea
 where in the last equality we used  the fact that $\psi \in [0,\frac{4\pi}{p})$.
We have thus derived (\ref{sing}), as advertised.

Although this result depends {\it a priori} on the choice of resolved metric and gauge field we picked, 
we will now explain to what extent it is in fact \emph{independent} of these choices. 
Having resolved the singularity to $\mathcal{M}_p=\mathcal{O}(-p)\rightarrow S^2$, 
more generally we may consider any one-parameter family of gauge fields 
on this space, depending on $\varepsilon>0$, which satisfy the following properties: 
(i) the curvature $F^\varepsilon$ has finite action, (ii) $\frac{F^\varepsilon}{2\pi}$ has period 
$\frac{n}{2}$ through the $S^2$ bolt/zero-section, (iii) 
the curvature tends to zero in the Taub-NUT-AdS$/\Z_p$ space
as $\varepsilon\rightarrow 0$ (say, $\mathcal{O}(\varepsilon)$). These are all clearly necessary (or at least reasonable) assumptions.
In order to compute the contribution of this gauge field to the free energy, we will 
also assume that (iv) $F^\varepsilon$ satisfies the gauge field equation of motion.  
Of course, all these conditions are satisfied in our computation above.

With these assumptions in place, the integral $\int_{\mathcal{M}_p}F^\varepsilon\wedge *F^\varepsilon$ is 
in fact independent of the cohomology class of $F^\epsilon$, to leading order ({\it i.e.} ignoring 
$\mathcal{O}(\varepsilon)$ corrections). This follows by taking $F^\varepsilon\rightarrow F^\varepsilon+\diff\Lambda$, 
where $\Lambda$ is any closed form, using the equation of motion for $F^\epsilon$, Stokes' Theorem, 
and the fact that the curvature is $\mathcal{O}(\varepsilon)$ at infinity. We may thus, 
without loss of generality, pick the particular representation (\ref{Achoice}) for this cohomology 
class. So far we have not specified what metric we are using to define the Hodge dual, 
but notice that since the one-parameter family of metrics is required to 
tend to the Taub-NUT-AdS metric as $\varepsilon\rightarrow 0$, and since 
our choice of gauge field (\ref{Achoice}) becomes (anti-)self-dual 
in this limit, without loss of generality we may pick an (anti-)self-dual gauge field 
for all $\varepsilon$. Essentially, any other choice will simply change only the $\mathcal{O}(\varepsilon)$ 
corrections to the final action/free energy integral.

The advantage of choosing an anti-self-dual field strength is that this makes it clear why 
the final result (\ref{sing}) is a topological invariant (even though the above argument shows 
that picking an anti-self-dual field strength is not necessary). We have, as before,
\bea
\frac{1}{16 \pi G_4} \int_{\mathcal{M}_p} (F^\varepsilon)^2  &= &  - \frac{1}{8 \pi G_4} \int_{\mathcal{M}_p} F^\varepsilon \wedge F^\varepsilon~.\eea
The right hand side may then be understood  topologically, to leading order in $\varepsilon$,  as the 
pairing $H^2_{\mathrm{cpt}}(\mathcal{M}_p,\R)\times H^2(\mathcal{M}_p,\R)\rightarrow \R$. 
Although $F^\varepsilon$ is \emph{not} necessarily compactly supported (and is not in our 
example computation), it \emph{is} to leading order in $\varepsilon$. We have 
$H^2_{\mathrm{cpt}}(\mathcal{M}_p,\Z)\cong \Z$, and the generator $\Psi$ 
has unit integral over a fibre of $\mathcal{M}_p=\mathcal{O}(-p)\rightarrow S^2$ 
(it is the Thom class of this bundle). It is then a standard fact that
$\int_{S^2}\Psi = -p$, the latter being the Euler class of the bundle
$\mathcal{O}(-p)\rightarrow S^2$, so that $\int_{\mathcal{M}_p}\Psi\wedge \Psi =
-p$ (integrating first over the fibre, and then over the bolt).
 Thus the cohomology class $[F^\varepsilon]=-\frac{\pi n}{p}\Psi$, 
and we hence compute
\bea
\frac{1}{16 \pi G_4} \int_{\mathcal{M}_p} (F^\varepsilon)^2  &= &  - \frac{1}{8 \pi G_4} \left(\frac{\pi n}{p}\right)^2\int_{\mathcal{M}_p} \Psi\wedge \Psi~,\nonumber\\
&=&   \frac{n^2}{8p}\cdot \frac{\pi}{G_4}~.
\eea
Here each equality should be understood as up to $\mathcal{O}(\varepsilon)$. 
This explains why (\ref{sing}) may be understood as a topological invariant.


\begin{thebibliography}{99}

\bibitem{Pestun:2007rz} 
  V.~Pestun,
  ``Localization of gauge theory on a four-sphere and supersymmetric
Wilson loops,''
  Commun.\ Math.\ Phys.\  {\bf 313}, 71 (2012)
  [arXiv:0712.2824 [hep-th]].

\bibitem{Kapustin:2009kz} 
  A.~Kapustin, B.~Willett and I.~Yaakov,
  ``Exact Results for Wilson Loops in Superconformal Chern-Simons Theories with Matter,''
  JHEP {\bf 1003}, 089 (2010)
  [arXiv:0909.4559 [hep-th]].

\bibitem{Jafferis:2010un} 
  D.~L.~Jafferis,
  ``The Exact Superconformal R-Symmetry Extremizes Z,''
  JHEP {\bf 1205}, 159 (2012)
  [arXiv:1012.3210 [hep-th]].
  
\bibitem{Festuccia:2011ws}
  G.~Festuccia, N.~Seiberg,
  ``Rigid Supersymmetric Theories in Curved Superspace,''
  JHEP {\bf 1106}, 114 (2011).
  [arXiv:1105.0689 [hep-th]].

\bibitem{Samtleben:2012gy} 
  H.~Samtleben and D.~Tsimpis,
  ``Rigid supersymmetric theories in 4d Riemannian space,''
  JHEP {\bf 1205}, 132 (2012)
  [arXiv:1203.3420 [hep-th]].

\bibitem{Klare:2012gn} 
  C.~Klare, A.~Tomasiello and A.~Zaffaroni,
  ``Supersymmetry on Curved Spaces and Holography,''
  JHEP {\bf 1208}, 061 (2012)
  [arXiv:1205.1062 [hep-th]].

\bibitem{Dumitrescu:2012ha} 
  T.~T.~Dumitrescu, G.~Festuccia and N.~Seiberg,
  ``Exploring Curved Superspace,''
  JHEP {\bf 1208}, 141 (2012)
  [arXiv:1205.1115 [hep-th]].

\bibitem{Cassani:2012ri} 
  D.~Cassani, C.~Klare, D.~Martelli, A.~Tomasiello and A.~Zaffaroni,
  ``Supersymmetry in Lorentzian Curved Spaces and Holography,''
  arXiv:1207.2181 [hep-th].

\bibitem{Liu:2012bi} 
  J.~T.~Liu, L.~A.~Pando Zayas and D.~Reichmann,
  ``Rigid Supersymmetric Backgrounds of Minimal Off-Shell Supergravity,''
  JHEP {\bf 1210}, 034 (2012)
  [arXiv: 1207.2785 [hep-th]].

\bibitem{deMedeiros:2012sb} 
  P.~de Medeiros,
  ``Rigid supersymmetry, conformal coupling and twistor spinors,''
  arXiv:1209.4043 [hep-th].

\bibitem{Dumitrescu:2012at} 
  T.~T.~Dumitrescu and G.~Festuccia,
  ``Exploring Curved Superspace (II),''
  JHEP {\bf 1301}, 072 (2013)
  [arXiv:1209.5408 [hep-th]].

\bibitem{Kehagias:2012fh} 
  A.~Kehagias and J.~G.~Russo,
  ``Global Supersymmetry on Curved Spaces in Various Dimensions,''
  arXiv:1211.1367 [hep-th].

\bibitem{newcyril} 
  C.~Closset, T.~T.~Dumitrescu, G.~Festuccia and Z.~Komargodski,
  ``Supersymmetric Field Theories on Three-Manifolds,''
  arXiv:1212.3388 [hep-th].
  
\bibitem{Martelli:2011fu} 
  D.~Martelli, A.~Passias and J.~Sparks,
  ``The gravity dual of supersymmetric gauge theories on a squashed three-sphere,''
  Nucl.\ Phys.\ B {\bf 864}, 840 (2012)
  [arXiv:1110.6400 [hep-th]].

\bibitem{Martelli:2011fw} 
  D.~Martelli and J.~Sparks,
  ``The gravity dual of supersymmetric gauge theories on a biaxially squashed three-sphere,''
  Nucl.\ Phys.\ B {\bf 866}, 72 (2013)
  [arXiv:1111.6930 [hep-th]].

\bibitem{Hama:2011ea}
  N.~Hama, K.~Hosomichi and S.~Lee,
  ``SUSY Gauge Theories on Squashed Three-Spheres,''
  JHEP {\bf 1105}, 014 (2011)
  [arXiv:1102.4716 [hep-th]].

\bibitem{Imamura:2011wg} 
  Y.~Imamura and D.~Yokoyama,
   ``${\cal N}=2$ supersymmetric theories on squashed three-sphere,''
  Phys.\ Rev.\ D {\bf 85}, 025015 (2012)
  [arXiv:1109.4734 [hep-th]].
  
\bibitem{anderson} Michael T. Anderson, ``Extension of symmetries on Einstein manifolds with boundary,'' 
  arXiv:0704.3373.

\bibitem{Papadimitriou:2005ii} 
  I.~Papadimitriou and K.~Skenderis,
  ``Thermodynamics of asymptotically locally AdS spacetimes,''
  JHEP {\bf 0508}, 004 (2005)
  [hep-th/0505190].

\bibitem{Hawking:1982dh} 
  S.~W.~Hawking and D.~N.~Page,
  ``Thermodynamics of Black Holes in anti-De Sitter Space,''
  Commun.\ Math.\ Phys.\  {\bf 87}, 577 (1983).

\bibitem{Witten:1998zw} 
  E.~Witten,
  ``Anti-de Sitter space, thermal phase transition, and confinement in gauge theories,''
  Adv.\ Theor.\ Math.\ Phys.\  {\bf 2}, 505 (1998)
  [hep-th/9803131].

\bibitem{Hawking:1998ct} 
  S.~W.~Hawking, C.~J.~Hunter and D.~N.~Page,
  ``Nut charge, anti-de Sitter space and entropy,''
  Phys.\ Rev.\ D {\bf 59}, 044033 (1999)
  [hep-th/9809035].

\bibitem{Chamblin:1998pz} 
  A.~Chamblin, R.~Emparan, C.~V.~Johnson and R.~C.~Myers,
  ``Large N phases, gravitational instantons and the nuts and bolts of AdS holography,''
  Phys.\ Rev.\ D {\bf 59}, 064010 (1999)
  [hep-th/9808177].
  
  \bibitem{Dunajski:2010zp} 
  M.~Dunajski, J.~Gutowski, W.~Sabra and P.~Tod,
  ``Cosmological Einstein-Maxwell Instantons and Euclidean Supersymmetry: Anti-Self-Dual Solutions,''
  Class.\ Quant.\ Grav.\  {\bf 28}, 025007 (2011)
  [arXiv:1006.5149 [hep-th]].

\bibitem{Freedman:1976aw}
  D.~Z.~Freedman, A.~K.~Das,
  ``Gauge Internal Symmetry in Extended Supergravity,''
  Nucl.\ Phys.\  {\bf B120}, 221 (1977).

\bibitem{Gauntlett:2007ma}
  J.~P.~Gauntlett, O.~Varela,
  ``Consistent Kaluza-Klein reductions for general supersymmetric AdS solutions,''
  Phys.\ Rev.\  {\bf D76}, 126007 (2007).
  [arXiv:0707.2315 [hep-th]].
  
\bibitem{Gauntlett:2009zw}
  J.~P.~Gauntlett, S.~Kim, O.~Varela, D.~Waldram,
  ``Consistent supersymmetric Kaluza-Klein truncations with massive modes,''
  JHEP {\bf 0904}, 102 (2009).
  [arXiv:0901.0676 [hep-th]].  
 
\bibitem{Page:1985bq} 
  D.~N.~Page and C.~N.~Pope,
  ``Inhomogeneous Einstein Metrics On Complex Line Bundles,''
  Class.\ Quant.\ Grav.\  {\bf 4}, 213 (1987).

\bibitem{Gibbons:1979zt} 
  G.~W.~Gibbons and S.~W.~Hawking,
  ``Gravitational Multi - Instantons,''
  Phys.\ Lett.\ B {\bf 78}, 430 (1978).
  
 
\bibitem{Carter:1968ks} 
  B.~Carter,
  ``Hamilton-Jacobi and Schrodinger separable solutions of Einstein's equations,''
  Commun.\ Math.\ Phys.\  {\bf 10}, 280 (1968).
  
  \bibitem{plebe}
  J.~ F.~ Plebanski, ``A class of solutions of Einstein-Maxwell equations,'' 
  Annals \ Phys. \ {\bf 90}, 196 (1975).

  
  
\bibitem{AlonsoAlberca:2000cs}
  N.~Alonso-Alberca, P.~Meessen, T.~Ortin,
  ``Supersymmetry of topological Kerr-Newman-Taub-NUT-AdS space-times,''
  Class.\ Quant.\ Grav.\  {\bf 17}, 2783-2798 (2000).
  [hep-th/0003071].

\bibitem{Caldarelli:2003pb} 
  M.~M.~Caldarelli and D.~Klemm,
  ``All supersymmetric solutions of N=2, D = 4 gauged supergravity,''
  JHEP {\bf 0309}, 019 (2003)
  [hep-th/0307022].

\bibitem{peder}
H. ~Pedersen, 
``Einstein metrics, spinning top motions and monopoles,''
Math. Ann. 274 (1986) 35 - 39.


\bibitem{lebru}
C. ~R. ~LeBrun, 
``Counterexamples to the generalized positive action conjecture,''
 Comm. Math. Phys. 118 (1988) 591 - 596.


\bibitem{hitch}
N. ~J. ~Hitchin, 
``Twistor spaces, Einstein metrics and isomonodromic deformations,''
 J. Diff. Geom. 42 (1995) 30 - 112.



\bibitem{Alday:2012au} 
  L.~F.~Alday, M.~Fluder and J.~Sparks,
  ``The Large N limit of M2-branes on Lens spaces,''
  JHEP {\bf 1210}, 057 (2012)
  [arXiv:1204.1280 [hep-th]].

\bibitem{Emparan:1999pm} 
  R.~Emparan, C.~V.~Johnson and R.~C.~Myers,
  ``Surface terms as counterterms in the AdS / CFT correspondence,''
  Phys.\ Rev.\ D {\bf 60}, 104001 (1999)
  [hep-th/9903238].

\bibitem{Skenderis:2002wp} 
  K.~Skenderis,
  ``Lecture notes on holographic renormalization,''
  Class.\ Quant.\ Grav.\  {\bf 19}, 5849 (2002)
  [hep-th/0209067].
  
\bibitem{Gabella:2012rc} 
  M.~Gabella, D.~Martelli, A.~Passias and J.~Sparks,
  ``${\cal N}=2$ supersymmetric AdS$_4$ solutions of M-theory,''
  arXiv:1207.3082 [hep-th].

\bibitem{Gabella:2011sg} 
  M.~Gabella, D.~Martelli, A.~Passias and J.~Sparks,
  ``The free energy of ${\cal N}=2$ supersymmetric AdS$_4$ solutions of M-theory,''
  JHEP {\bf 1110}, 039 (2011)
  [arXiv: 1107.5035 [hep-th]].

\bibitem{Gauntlett:2006vf} 
  J.~P.~Gauntlett, D.~Martelli, J.~Sparks and S.~-T.~Yau,
  ``Obstructions to the existence of Sasaki-Einstein metrics,''
  Commun.\ Math.\ Phys.\  {\bf 273}, 803 (2007)
  [hep-th/0607080].
  
\bibitem{Martelli:2008si} 
  D.~Martelli and J.~Sparks,
  ``Moduli spaces of Chern-Simons quiver gauge theories and AdS(4)/CFT(3),''
  Phys.\ Rev.\ D {\bf 78}, 126005 (2008)
  [arXiv:0808.0912 [hep-th]].
  
\bibitem{Hanany:2008cd} 
  A.~Hanany and A.~Zaffaroni,
  ``Tilings, Chern-Simons Theories and M2 Branes,''
  JHEP {\bf 0810}, 111 (2008)
  [arXiv:0808.1244 [hep-th]].

\bibitem{Martelli:2009ga} 
  D.~Martelli and J.~Sparks,
  ``AdS(4) / CFT(3) duals from M2-branes at hypersurface singularities and their deformations,''
  JHEP {\bf 0912}, 017 (2009)
  [arXiv: 0909.2036 [hep-th]].
  
\bibitem{Jafferis:2009th} 
  D.~L.~Jafferis,
  ``Quantum corrections to N=2 Chern-Simons theories with flavor and their AdS(4) duals,''
  arXiv:0911.4324 [hep-th].

\bibitem{Aganagic:2009zk} 
  M.~Aganagic,
  ``A Stringy Origin of M2 Brane Chern-Simons Theories,''
  Nucl.\ Phys.\ B {\bf 835}, 1 (2010)
  [arXiv:0905.3415 [hep-th]].

\bibitem{Franco:2009sp} 
  S.~Franco, I.~R.~Klebanov and D.~Rodriguez-Gomez,
  ``M2-branes on Orbifolds of the Cone over Q**1,1,1,''
  JHEP {\bf 0908}, 033 (2009)
  [arXiv:0903.3231 [hep-th]].

\bibitem{Benini:2009qs} 
  F.~Benini, C.~Closset and S.~Cremonesi,
  ``Chiral flavors and M2-branes at toric CY4 singularities,''
  JHEP {\bf 1002}, 036 (2010)
  [arXiv:0911.4127 [hep-th]].

\bibitem{Closset:2012ep} 
  C.~Closset and S.~Cremonesi,
  ``Toric Fano varieties and Chern-Simons quivers,''
  JHEP {\bf 1205}, 060 (2012)
  [arXiv:1201.2431 [hep-th]].
 
\bibitem{Gang:2011jj} 
  D.~Gang, C.~Hwang, S.~Kim and J.~Park,
  ``Tests of AdS$_4$/CFT$_3$ correspondence for $\mathcal{N}=2$ chiral-like theory,''
  JHEP {\bf 1202}, 079 (2012)
  [arXiv:1111.4529 [hep-th]].
 
\bibitem{Cheon:2011vi} 
  S.~Cheon, H.~Kim and N.~Kim,
  ``Calculating the partition function of N=2 Gauge theories on $S^3$ and AdS/CFT correspondence,''
  JHEP {\bf 1105}, 134 (2011)
  [arXiv:1102.5565 [hep-th]].

\bibitem{Jafferis:2011zi} 
  D.~L.~Jafferis, I.~R.~Klebanov, S.~S.~Pufu and B.~R.~Safdi,
  ``Towards the F-Theorem: N=2 Field Theories on the Three-Sphere,''
  JHEP {\bf 1106}, 102 (2011)
  [arXiv:1103.1181 [hep-th]].

\end{thebibliography}
\end{document}